\documentclass[twocolumn]{aastex6}
\usepackage{graphicx}
\usepackage{amssymb,amsfonts,amsmath,amstext,amsgen,amsopn,amsxtra,indentfirst,times}

\begin{document}

\title{\texttt{HELIOS-Retrieval}: An Open-source, Nested Sampling Atmospheric Retrieval Code, Application to the HR 8799 Exoplanets and Inferred Constraints for Planet Formation}
\author{Baptiste Lavie\altaffilmark{1,2,4}, Jo\~{a}o M. Mendon\c{c}a\altaffilmark{1}, Christoph Mordasini\altaffilmark{2}, Matej Malik\altaffilmark{1}, Micka\"{e}l Bonnefoy\altaffilmark{3}, Brice-Olivier Demory\altaffilmark{1}, Maria Oreshenko\altaffilmark{2}, Simon L. Grimm\altaffilmark{1}, David Ehrenreich\altaffilmark{4}, Kevin Heng\altaffilmark{2}}
\altaffiltext{1}{University of Bern, Space Research and Planetary Sciences, Sidlerstrasse 5, CH-3012, Bern, Switzerland Email: baptiste.lavie@space.unibe.ch}
\altaffiltext{2}{University of Bern, Center for Space and Habitability, Sidlerstrasse 5, CH-3012, Bern, Switzerland.  Email: kevin.heng@csh.unibe.ch}
\altaffiltext{3}{Universit\'{e} Grenoble Alpes, IPAG, 38000, Grenoble, France; CNRS, IPAG, 38000, Grenoble, France}
\altaffiltext{4}{Observatoire de l'Universit\'{e} de Gen\`{e}ve, 51 chemin des Maillettes, 1290, Sauverny, Switzerland}

\begin{abstract}
We present an open-source retrieval code named \texttt{HELIOS-RETRIEVAL}, designed to obtain chemical abundances and temperature-pressure profiles from inverting the measured spectra of exoplanetary atmospheres. We use an exact solution of the radiative transfer equation, in the pure absorption limit, in our forward model, which allows us to analytically integrate over all of the outgoing rays. Two chemistry models are considered: unconstrained chemistry and equilibrium chemistry (enforced via analytical formulae). The nested sampling algorithm allows us to formally implement Occam's Razor based on a comparison of the Bayesian evidence between models. We perform a retrieval analysis on the measured spectra of the four HR 8799 directly imaged exoplanets. Chemical equilibrium is disfavored for HR 8799b and c. We find supersolar C/H and O/H values for the outer HR 8799b and c exoplanets, while the inner HR 8799d and e exoplanets have a range of C/H and O/H values.  The C/O values range from being superstellar for HR 8799b to being consistent with stellar for HR 8799c and being substellar for HR8799d and e.  If these retrieved properties are representative of the bulk compositions of the exoplanets, then they are inconsistent with formation via gravitational instability (without late-time accretion) and consistent with a core accretion scenario in which late-time accretion of ices occurred differently for the inner and outer exoplanets.  For HR 8799e, we find that spectroscopy in the K band is crucial for constraining C/O and C/H.  \texttt{HELIOS-RETRIEVAL} is publicly available as part of the Exoclimes Simulation Platform (ESP; \texttt{www.exoclime.org}).
\end{abstract}

\section{Introduction}

\subsection{Motivation}

Traditionally, the masses and radii of brown dwarfs and substellar objects have been inferred from applying evolutionary tracks to measurements of their luminosities and ages (e.g., \citealt{burrows97,chabrier00,baraffe02}).  On rare occasions, brown dwarfs and low-mass stars may transit their binary companions and allow for their other properties to be studied (see \citealt{burrows11} and references therein).  A particularly important study was conducted by \cite{konopacky10}, who were able to obtain dynamical masses for 15 brown dwarfs residing in binaries.  By comparing the dynamical and photometric masses, \cite{konopacky10} showed that both the \cite{burrows97} and \cite{chabrier00} models underpredicted the masses of M and L dwarfs and overpredicted the mass of the lone T dwarf in their sample by $\sim 10\%$ (tens of percent).  From studying a sample of 46 L dwarfs, \cite{hira16} suggested that a dust haze of sub-micron-sized particles exist in their upper atmospheres, which are neglected by the standard evolutionary tracks.  

Taken together, these results suggest that the traditional approach of using self-consistent evolutionary tracks may be incomplete and motivates alternative and complementary ways of interpreting the spectra of brown dwarfs and substellar objects.  We expect this train of thought to apply to the recently discovered directly imaged exoplanets as well, since the interpretation of their photometry and spectroscopy is typically performed using the evolutionary tracks computed for brown dwarfs (e.g., \citealt{bonnefoy16}).

\subsection{Theoretical Improvements}

\begin{figure*}%[!h]
\begin{center}
%\vspace{0.2in}
\includegraphics[width=1.5\columnwidth]{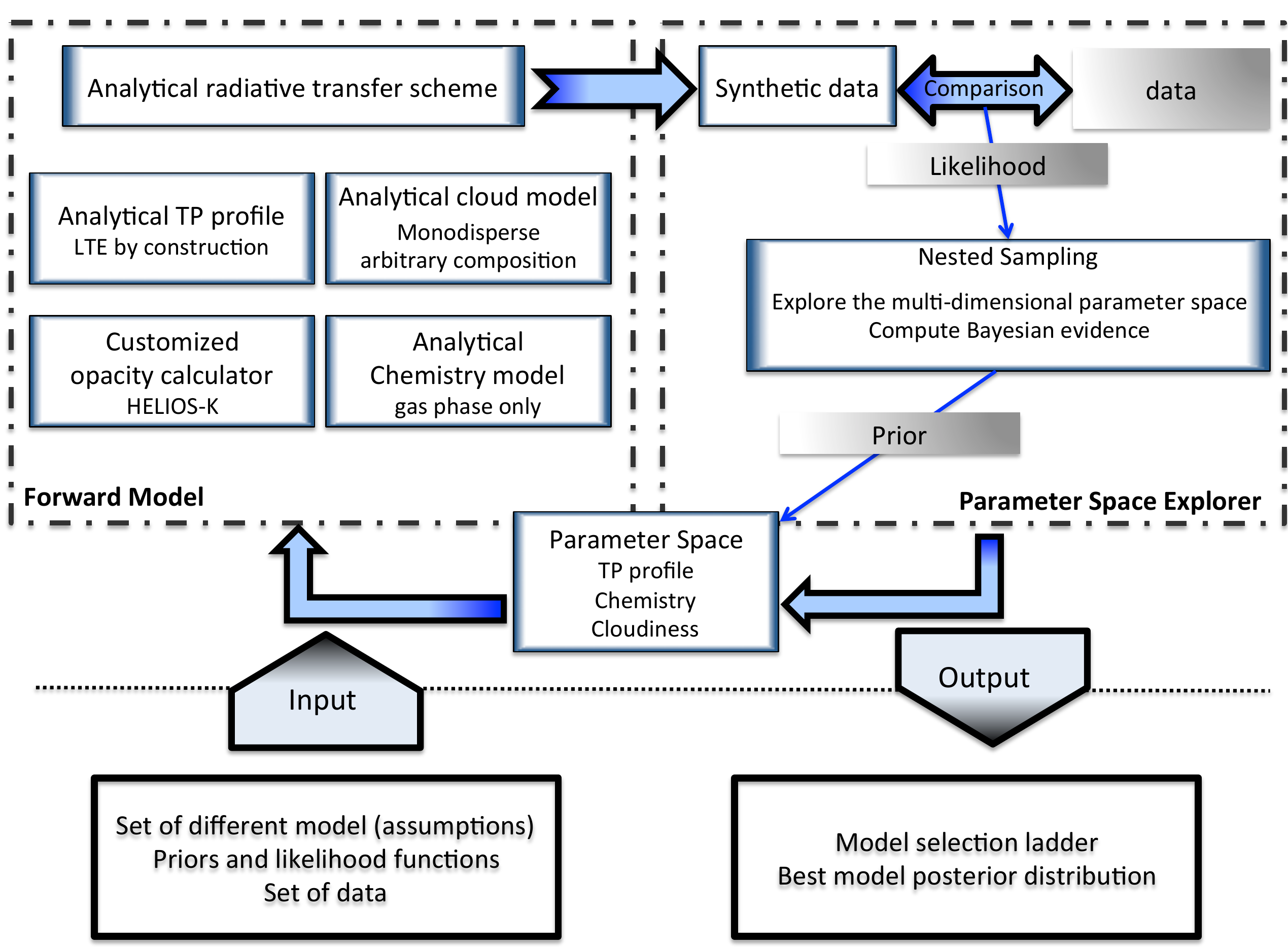}
\end{center}
\vspace{0.01in}
\caption{Flow chart for \texttt{HELIOS-R} and a description of its main components.  Note that enforcing equilibrium chemistry is optional and our approach for distinguishing between equilibrium and non-equilibrium chemistry is completely data-driven.}
%\vspace{0.1in}
\label{fig:schematic}
\end{figure*}

Self-consistent forward modeling starts with a set of assumptions and computes forward to predict the temperature-pressure profile and synthetic spectrum of an object.  Atmospheric retrieval is a complementary approach borrowed from the Earth remote sensing community, where one applies an inversion method to obtain the temperature-pressure profile and chemical abundances from finding the best-fit solution to the measured spectrum (e.g., \citealt{ms09,bs13,lee13,line13,line16,barstow15,waldmann15}).  It sacrifices self-consistency and sophistication for simplicity, which allows for a more thorough exploration of parameter space.  Atmospheric retrieval is particularly well-suited for addressing questions regarding planet formation, since it allows for the posterior distributions of the carbon-to-oxygen ratio (C/O) and the elemental abundances of carbon (C/H) and oxygen (O/H) to be computed.

The HR 8799 system hosts four exoplanets \citep{marois08,marois10}, whose formation mechanisms remain an enigma \citep{kratter10}.  Spectra with resolutions of about 30 to 4000 have been obtained by, e.g., \cite{barman11,barman15}, \cite{konopacky13}, \cite{oppen13}, \cite{ingra14} and \cite{zurlo16}.  Since these spectra have resolutions that are considerably higher than those obtained for hot Jupiters using WFC3 on the Hubble Space Telescope (e.g., \citealt{deming13,mandell13,kreidberg14,stevenson14}), they present an opportunity for performing remote sensing of exoplanetary atmospheres that is similar to what planetary scientists had to work with a few decades ago, before the advent of probes.  A key difference is that the radii and masses of these directly imaged exoplanets are unknown\footnote{Meaning they are typically not directly measured, but rather inferred using evolutionary models, which means the radii and masses are model-dependent.}, unlike for transiting exoplanets.  A recent review of directly imaged exoplanets, which includes the HR 8799 system, may be found in \cite{bowler16}.

The first atmospheric retrieval analysis of directly imaged exoplanets was performed by \cite{lee13}, who studied only the HR 8799b exoplanet.  In the current study, we collect all of the published spectra of the HR 8799b, c, d and e exoplanets and subject them to the same retrieval method with the intention of using the retrieved chemistry to constrain planet formation scenarios.  

Besides the novelty of our analysis, the current study is also a method paper for our new atmospheric retrieval code named \texttt{HELIOS-R}, which we constructed from scratch to study exoplanetary atmospheres.  \texttt{HELIOS-R} is part of the \texttt{HELIOS} radiation package of the Exoclimes Simulation Platform\footnote{\texttt{http://www.exoclime.org}} and has the following features (Figure \ref{fig:schematic}):
\begin{itemize}

\item We have implemented a nested sampling algorithm to explore the multi-dimensional parameter space \citep{skilling06,feroz09,bs13,waldmann15,line16}.  Unlike other approaches (e.g., Markov Chain Monte Carlo, non-linear optimal estimation), nested sampling allows for the Bayesian evidence to be directly calculated, which in turn allows for models with different parametrizations (and number of parameters) to be compared on an equal footing. Models with extra complexity are penalized, which allows for Occam's Razor\footnote{Whether Occam's Razor always yields the correct answer is another matter.  In the current study, we are guided by Occam's Razor in the limit of sparse data.} to be formally enforced. For example, our retrieval analysis allows us to formally determine if chemical equilibrium is favored or disfavored in an atmosphere in a completely data-driven manner.  As another example, it allows us to determine the number and types of molecules to be included in the retrieval.

\item Our temperature-pressure profile is taken from \cite{hml14}, who generalized the work of \cite{guillot10} and \cite{hhps12} to include non-isotropic scattering and non-constant opacities.  When stellar irradiation and scattering are omitted, the temperature-pressure profile reduces to the classical solution of Milne for self-luminous objects \citep{mihalas70}.  By construction, it conserves energy in an analytical and exact sense.  

\item Our atmospheric cross sections are computed using our customized opacity calculator named \texttt{HELIOS-K}, which was previously published by \cite{gh15}.  

\item To combine the cross sections of different molecules, one needs to have a chemistry model that calculates their relative abundances.  We use the analytical solutions of \cite{hlt16}, \cite{hl16} and \cite{ht16}, which have been shown to be accurate at the $\sim 1\%$ level (or better) when benchmarked against numerical solutions using Gibbs free energy minimization.  These analytical solutions allow for fast computation if one wishes to enforce chemical equilibrium.

\item Our radiative transfer scheme, which translates cross sections and temperatures into fluxes (and hence allows us to compute the synthetic spectrum), uses the exact analytical solution in the limit of isothermal model layers and pure absorption \citep{hml14}.  It allows us to analytically integrate over all of the incoming and outgoing angles associated with every ray.  

\item Our cloud model is based on the basic principles of Mie theory (e.g., \citealt{pierrehumbert}).  It assumes a monodisperse set of particles, which may be interpreted as the dominant size in a size distribution of particles (e.g., \citealt{burrows11}).  It includes a dimensionless parameter that is a proxy for the cloud composition.  When the particles are small compared to the wavelength, it reproduces Rayleigh scattering.  By contrast, models that implement a constant cloud-top pressure implicitly assume the cloud particles to be large (compared to the wavelength observed) and preclude Rayleigh scattering by construction.

\end{itemize}

While each component of \texttt{HELIOS-R} may not be novel by itself, the assembly of all of these components into a single code and retrieval tool is a novel endeavor.  Furthermore, we have designed \texttt{HELIOS-R} to run on Graphics Processing Units (GPUs), which affords speed-ups of at least a factor of several compared to the CPU version.  With a UCrg model (see Table 1) retrieval performed on the HR 8799b dataset, the GPU version is 5 times faster than the CPU version on a Macbook Pro laptop equipped with a NVIDIA GeForce GT 750M GPU card and an Intel Core i7 2.5 GHz CPU.  For this analysis, we used our GPU cluster of NVIDIA K20 cards; it takes $10^{-2}$ seconds to evaluate one likelihood of this UCrg model.

In \S\ref{sect:methods}, we provide a detailed description of each component or ingredient of \texttt{HELIOS-R}.  In \S\ref{sect:tests}, we subject \texttt{HELIOS-R} to several tests before applying it to the measured spectra of the HR 8799b, c, d and e directly imaged exoplanets.  In \S\ref{sect:results}, we present our retrieval results of the HR 8799 system.  In \S\ref{sect:discussion}, we compare our study to previous work and describe opportunities for future work.  Table 1 shows the suite of models tested in the current study.  Table 2 states the priors used for our fitting parameters.  Table 3 summarizes our retrieval results.  Appendix \ref{append:e1} states our fast analytical formulae for evaluating the exponential integral of the first order. Appendix \ref{append:full} includes, for completeness, the full posterior distributions of the best models for the atmospheres of HR 8799b, c, d and e.

\begin{table}
\label{tab:models}
\begin{center}
\caption{Shorthand Notation for the Suite of Models Tested in this Study}
\begin{tabular}{lc}
\hline
\hline
Notation & Meaning \\
\hline
U & Unconstrained chemistry \\
E & Equilibrium chemistry \\
B & Cloudfree (``blue sky") \\
C & Cloudy \\
1 & H$_2$O is included in retrieval \\
2 & CO$_2$ is included in retrieval \\
5 & CO is included in retrieval \\
6 & CH$_4$ is included in retrieval \\
r & planet radius $R$ is included in retrieval \\
g &planet surface gravity $g$ is included in retrieval \\
d & distance of the system $d$ is included in retrieval \\
\hline
\hline
\end{tabular}\\
%\vspace{0.05in}
\end{center}
\scriptsize
Note: ``1", ``2", ``5" and ``6" refer to the HITRAN/HITEMP labels for these molecules.  When no number is specified, it means that all four molecules are included in the retrieval.  Example: UBrg16 is a cloudfree model with unconstrained chemistry, where the mixing ratios of water and methane, as well as the planetary radius and surface gravity, are included as fitting parameters.  By contrast, the UB model includes all four molecules in the retrieval, but fixes the planetary radius and surface gravity to user-specified values.
\vspace{0.2in}
\end{table}

\begin{table}
\label{tab:priors}
\begin{center}
\caption{Priors Used in This Study}
\begin{tabular}{lcc}
\hline
\hline
Symbol & Prior Used & Value \\
\hline
$R$ & Gaussian & $R = 1.2 \pm 0.1 R_{\rm J}$ \\
$g$ & Gaussian & $\log{g} = 4.1 \pm 0.3$ (cgs) \\
$X_i$ & Log-uniform & $10^{-20}$ to $10^{-1}$ \\
$\kappa_0$ & Log-uniform & $\log{\kappa_0} = 10^{-15}$ to 10 (mks) \\
$T_{\rm int}$ & Uniform & 10 to 1500 K \\
$Q_0$ & Uniform & 1 to 100 \\
$r_{\rm c}$ & Log-uniform & $10^{-7}$ to $10^{-3}$ m \\
$f_{\rm cloud}$ & Log-uniform & $10^{-30}$ to $10^{-4}$ \\
$d$ & Gaussian & $39.4 \pm 1.0$ pc \\
\hline
\hline
\end{tabular}\\
%\vspace{0.05in}
\end{center}
\scriptsize
cgs: centimeters, grams and seconds. \\
mks: meters, kilograms and seconds. \\
\vspace{0.2in}
\end{table}

\section{Methodology}
\label{sect:methods}

The executive summary is that each model of the retrieval contains up to 11 parameters: the radius, the surface gravity, 2 for the temperature-pressure profile, 2 or 4 for the chemistry (depending on whether one adopts equilibrium or unconstrained chemistry) and 3 for the cloud model.  The mean molecular weight is not a parameter and is constructed from the mixing ratios.  Each HR 8799 exoplanet typically has between 40 and 120 data points for its measured spectrum: 68 for b, 105 for c, 115 for d and 48 for e.

To construct an atmospheric retrieval model, we need a forward model.  By ``forward model", we refer to the temperature-pressure profile, atmospheric opacities, chemistry model, radiative transfer scheme and cloud model. We also need a method to scan the vast multi-dimensional parameter space of our forward model to locate the highest likelihood region, i.e., the best solution that fits the data (e.g., for a review, see \citealt{nr}).

\subsection{Nested Sampling}

We use a nested sampling algorithm \citep{skilling06} to scan the diverse, multi-dimensional parameter space describing our one-dimensional model atmospheres.  \cite{bs13} previously gave a detailed overview of the nested sampling method.  \cite{waldmann15} and \cite{line16} also used nested sampling.  Here, we provide a concise description of our implementation.

Consider a model with a set of parameters $\vec{\theta} = \{ \theta_1, \theta_2, ..., \theta_{N_\theta}\}$, where $N_\theta$ is the number of parameters.  Consider a set of models labeled by the index $i$: ${\cal M}_i$.  The probability density function (PDF) on the parameters for a given model is ${\cal P}(\vec{\theta} \vert {\cal M}_i)$, which is also known as the ``prior".

Discussions of any Bayesian method necessarily start with Bayes's rule, which states that the PDF of a model given the data (denoted by $\vec{D}$) is (e.g., \citealt{skilling06})
\begin{equation}
{\cal P} \left( \vec{\theta} \vert \vec{D}, {\cal M}_i \right) = \frac{{\cal P}\left(\vec{\theta} \vert {\cal M}_i \right) {\cal L}\left(\vec{D} \vert \vec{\theta}, {\cal M}_i \right)}{{\cal Z} \left( \vec{D} \vert {\cal M}_i \right)}.
\end{equation}
The quantity ${\cal L}(\vec{D} \vert \vec{\theta}, {\cal M}_i )$ is the ``likelihood".  We assume ${\cal L}(\vec{D} \vert \vec{\theta}, {\cal M}_i )$ to be the same Gaussian function as equation (5) of \cite{bs13}.

We will term ${\cal P} ( \vec{\theta} \vert \vec{D}, {\cal M}_i )$ the ``posterior".  Since it normalizes to unity, the Bayesian evidence is given by the multi-dimensional integral,
\begin{equation}
{\cal Z} \left( \vec{D} \vert {\cal M}_i \right) = \int {\cal P}\left(\vec{\theta} \vert {\cal M}_i \right) {\cal L}\left(\vec{D} \vert \vec{\theta}, {\cal M}_i \right) ~d\vec{\theta}.
\end{equation}

Fitting a model to a measured spectrum is an exercise in which a better fit is obtained when more free parameters (e.g., more molecules) are introduced.  Model selection is essentially the enforcing of Occam's Razor, meaning that we select the model that has a level of sophistication or complexity that is commensurate with the quality of data available.  It prevents the over-fitting of data by a model that is too complex.  For example, \cite{hansen14} find that for some of the exoplanets the photometric data of Spitzer alone may be fitted with a Planck function and a more complex model is unnecessary.  As the data quality improves, so does the complexity of the best model.

The essence of nested sampling is to reduce the computation of the Bayesian evidence to a one-dimensional integral \citep{skilling06},
\begin{equation}
{\cal Z} \left( \vec{D} \vert {\cal M}_i \right) = \int^1_0 {\cal L}^\prime \left( {\cal X} \right) ~d{\cal X},
\end{equation}
where the likelihood now only depends on a single variable and is denoted by ${\cal L}^\prime$.  This variable ${\cal X}$ is termed the ``prior mass" and is bounded between 0 and 1.  A visualization of the prior mass and its relationship to the Bayesian evidence is given in Figure 3 of \cite{skilling06} and Figure 1 of \cite{bs13}. Numerically, we use the trapezoid rule to compute the Bayesian evidence as a finite sum,
\begin{equation}
{\cal Z} \left( \vec{D} \vert {\cal M}_i \right) = \sum_j \frac{{\cal X}_{j+1} - {\cal X}_j}{2} \left( {\cal L}^\prime_{j+1} + {\cal L}^\prime_j \right).
\label{eq:evidence}
\end{equation}

We begin by randomly drawing $N_{\rm live}$ points from the parameter space($\theta$) subjected to the constraint of the chosen prior.  We use either Gaussian (radius, logarithm of gravity, distance), log-uniform (mixing ratios, mean opacity, cloud particle radius, cloud mixing ratio) or uniform (temperature, cloud composition parameter) priors.  For a set of points drawn, we compute their likelihood values.  At each step of the algorithm, we discard the worst point and replace it with a newly drawn point until the convergence criteria is met (see \citealt{skilling06}). This newly drawn point needs to have a higher likelihood than the worst point just discarded.  Specifically, we use the open-source software named \texttt{PyMultiNest}\footnote{\texttt{https://github.com/JohannesBuchner/PyMultiNest/}} \citep{buchner14}, which is a \texttt{Python} wrapper for the open-source \texttt{MultiNest}\footnote{\texttt{https://ccpforge.cse.rl.ac.uk/gf/project/multinest/}} program written in \texttt{Fortran 90} \citep{feroz08,feroz09,feroz13}.  
For each model, we run the nested sampling algorithm using 40 000 living points parallelised into 20 runs of 2000 ``living points" each. For comparison, \cite{waldmann15} uses $N_{\rm live} = 4000$ living points.  \cite{bs13} use between  $N_{\rm live} = 50$ and 10,000 living points.  \cite{line16} do not specify the number of living points used.  Equation (\ref{eq:evidence}) is used to compute the Bayesian evidence.  As a by-product of this procedure, one also obtains posterior-distribution samples of the model parameters.

For the purpose of comparing two models, which we denote by ${\cal M}_i$ and ${\cal M}_{i+1}$, it is useful to define a quantity known as the Bayes factor,  which is the ratio of the Bayesian evidences \citep{trotta08},
\begin{equation}
{\cal B} = \frac{{\cal Z} \left( \vec{D} \vert {\cal M}_i \right)}{{\cal Z} \left( \vec{D} \vert {\cal M}_{i+1} \right)}.
\end{equation}
The Bayes factor is equal to the posterior odds when both models are considered equally likely.  As shown in Table 2 of \cite{trotta08}, which is reproduced in Table 2 of \cite{bs13}, there is a relationship between the Bayes factor, the $p$-value of the frequentists and the significance in terms of the number of standard deviations. We use the Jeffreys scale \citep{kr95} to evaluate model significances. Weak, moderate and strong evidence for favoring the $i$-th model over the $(i+1)$-th model correspond to $\ln{\cal B}=1$, 2.5 and 5, respectively.

\subsection{Temperature-Pressure Profile}

For the temperature-pressure profile, we assume a one-dimensional, plane-parallel model atmosphere.  Its layers are evenly spaced in the logarithm of pressure between 1 $\mu$bar and 1 kbar.  We implement equation (126) of \cite{hml14}, who previously generalized the work of \cite{guillot10} (pure absorption limit and constant opacities) and \cite{hhps12} (isotropic scattering, constant shortwave/optical opacity) to include non-isotropic scattering and a non-constant shortwave/optical opacity.  Since the HR 8799 exoplanets are non-irradiated, we essentially use a reduced version of equation (126) of \cite{hml14},
\begin{equation}
T^4 = \frac{T^4_{\rm int}}{4} \left[ \frac{8}{3} + \frac{3\tilde{m}}{\beta_{\rm L}^2} \left( \kappa_0 + \frac{\kappa_{\rm CIA} \tilde{m}}{2\tilde{m}_0} \right) \right],
\label{eq:tp}
\end{equation}
where $T_{\rm int}$ is the internal/interior temperature, $\beta_{\rm L}$ is the longwave/infrared scattering parameter, $\kappa_0$ is the constant component of the longwave/infrared opacity and $\kappa_{\rm CIA}$ is the opacity associated with collision-induced absorption (CIA).  The column mass is denoted by $\tilde{m}$, while $\tilde{m}_0$ is the column mass referenced to the bottom of the model atmosphere.  We set $P_0 = \tilde{m}_0 g = 1$ kbar, where $g$ is the surface gravity.

Equation (\ref{eq:tp}) is essentially a generalization of the classical Milne's solution \citep{mihalas70} to include scattering and CIA.  In the limit of pure absorption ($\beta_{\rm L}=1$) and in the absence of CIA, we obtain $T=T_{\rm int}$ when $\kappa_0 \tilde{m} = 4/9$, somewhat different from the classical Milne value of 2/3.  It is worth emphasizing that equation (\ref{eq:tp}) is, \textit{by construction,} a temperature-pressure profile in radiative equilibrium, which implies that both local and global energy conservation are guaranteed in an exact, analytical sense \citep{hml14,hl16}.  By contrast, the versatile fitting function used by \cite{ms09} does not, by construction, obey energy conservation and this has to be enforced as a separate numerical condition.  However, in using a mean opacity equation (\ref{eq:tp}) sacrifices accuracy for simplicity, which makes the temperature-pressure profile more isothermal, at high altitudes, than if a more realistic radiative transfer calculation was performed.

In principle, $\kappa_0$ and $\kappa_{\rm CIA}$ are mean opacities that may be calculated directly from the spectroscopic line lists.  However, in deriving these analytical temperature-pressure profiles \cite{guillot10}, \cite{hhps12} and \cite{hml14} have assumed that the absorption, flux, Planck and Rosseland mean opacities are equal, which makes it unclear how to exactly compute $\kappa_0$ and $\kappa_{\rm CIA}$.  Therefore, we opt to use $\kappa_0$ and $\kappa_{\rm CIA}$ as fitting parameters instead.  In other words, our temperature-pressure profile is not self-consistent with the atmospheric opacities used.

We find that using $\kappa_{\rm CIA}$ and $\beta_{\rm L}$ as fitting parameters have a negligible effect on our results (not shown).  In practice, the use of equation (\ref{eq:tp}) with only $T_{\rm int}$ and $\kappa_0$ as fitting parameters (i.e., setting $\beta_{\rm L}=1$ and $\kappa_{\rm CIA}=0$) is sufficient for our retrieval calculations.

We use a constant value of the surface gravity, as we are sensing at most 6 orders of magnitude in pressure, which corresponds to 13.8 scale heights.  This means that the region of the atmosphere being sensed is only several percent of the radius of the exoplanet.  A constant surface gravity is thus not unreasonable.

\subsection{Atmospheric Cross Sections}

\begin{table}
\vspace{-0.55in}
\label{tab:results}
\begin{center}
\caption{Summary of Retrieved Results}
\begin{tabular}{lcc}
\hline
\hline
Property & Exoplanet & Value \\
\hline
$X_{\rm H_2O}$ & HR8799b & $-2.89_{-0.09}^{+0.09}$   \\ 
$X_{\rm H_2O}$ & HR8799c & $-2.60_{-0.05}^{+0.12}$   \\ 
$X_{\rm H_2O}$ & HR8799d & $[-2.29]$   \\ 
$X_{\rm H_2O}$ & HR8799e & $[-1.84]$   \\ 

\hline
$X_{\rm CO_2}$ & HR8799b & $-6.70_{-6.33}^{+1.52}$   \\ 
$X_{\rm CO_2}$ & HR8799c & $-4.63_{-0.11}^{+0.13}$   \\ 
$X_{\rm CO_2}$ & HR8799d & $[-18.84]$   \\ 
$X_{\rm CO_2}$ & HR8799e & $[-19.13]$   \\ 

\hline
$X_{\rm CO}$ & HR8799b & $-1.86_{-0.09}^{+0.10}$   \\ 
$X_{\rm CO}$ & HR8799c & $-2.48_{-0.20}^{+0.14}$   \\ 
$X_{\rm CO}$ & HR8799d & $[-16.32]$   \\ 
$X_{\rm CO}$ & HR8799e & $[-17.36]$   \\ 

\hline
$X_{\rm CH_4}$ & HR8799b & $-5.03_{-0.16}^{+0.14}$   \\ 
$X_{\rm CH_4}$ & HR8799c & $-5.03_{-0.18}^{+0.17}$   \\ 
$X_{\rm CH_4}$ & HR8799d & $[-28.11]$   \\ 
$X_{\rm CH_4}$ & HR8799e & $[-27.59]$   \\ 

\hline
$\mu$ & HR8799b & $2.18_{-0.00}^{+0.00}$   \\ 
$\mu$ & HR8799c & $2.19_{-0.00}^{+0.00}$   \\ 
$\mu$ & HR8799d & $[2.28]$   \\ 
$\mu$ & HR8799e & $[2.42]$   \\ 

\hline
C/O & HR8799b & $0.92_{-0.01}^{+0.01}$   \\ 
C/O & HR8799c & $0.55_{-0.12}^{+0.10}$   \\ 
C/O & HR8799d & $0.00_{-0.00}^{+0.00}$   \\ 
C/O & HR8799e & $0.00_{-0.00}^{+0.00}$   \\ 

\hline
C/H & HR8799b & $-2.11_{-0.09}^{+0.10}$   \\ 
C/H & HR8799c & $-2.73_{-0.20}^{+0.14}$   \\ 
C/H & HR8799d & $-16.62_{-2.04}^{+4.08}$   \\ 
C/H & HR8799e & $-11.93_{-4.64}^{+4.62}$   \\ 

\hline
O/H & HR8799b & $-2.07_{-0.09}^{+0.09}$   \\ 
O/H & HR8799c & $-2.47_{-0.11}^{+0.09}$   \\ 
O/H & HR8799d & $-3.20_{-0.15}^{+0.19}$   \\ 
O/H & HR8799e & $-2.75_{-0.57}^{+0.57}$   \\ 

\hline
$Q_0$ & HR8799b & $1.21_{-0.73}^{+0.49}$   \\ 
$Q_0$ & HR8799c & $0.79_{-0.26}^{+0.25}$   \\ 
$Q_0$ & HR8799d & $1.39_{-0.34}^{+0.27}$   \\ 
$Q_0$ & HR8799e & $0.95_{-0.52}^{+0.53}$   \\ 

\hline
$r_{c}$ & HR8799b & $-4.37_{-0.46}^{+0.49}$   \\ 
$r_{c}$ & HR8799c & $-4.44_{-0.20}^{+0.25}$   \\ 
$r_{c}$ & HR8799d & $-6.68_{-0.18}^{+0.18}$   \\ 
$r_{c}$ & HR8799e & $-4.69_{-0.76}^{+0.79}$   \\ 

\hline
$X_{c}$ & HR8799b & $-21.22_{-0.85}^{+0.80}$   \\ 
$X_{c}$ & HR8799c & $-20.55_{-0.49}^{+0.41}$   \\ 
$X_{c}$ & HR8799d & $-15.96_{-0.95}^{+1.20}$   \\ 
$X_{c}$ & HR8799e & $-20.56_{-1.37}^{+1.55}$   \\ 

\hline
$$d$$ & HR8799b & $40.30_{-0.79}^{+0.66}$   \\ 
$$d$$ & HR8799c & $39.73_{-0.20}^{+0.22}$   \\ 
$$d$$ & HR8799d & $40.81_{-0.50}^{+0.42}$   \\ 
$$d$$ & HR8799e & $39.40_{-0.78}^{+0.70}$   \\ 
\hline
\end{tabular}\\
\vspace{-0.1in}
\end{center}
\scriptsize
Note: we have listed the $1\sigma$ uncertainties, which were computed by locating the 15.87th and 84.13th percentile points on the horizontal axis.  In the limit of a symmetric Gaussian function, these would yield the full-width at half-maximum of the Gaussian. For planet d and e, the molecules abundances and the mean molecular weight are given at 1 bar.  Values are in $\log_{10}$ (except for C/O) and dimensionless (except for $r_{\rm c}$ which is in meters).\\
\end{table}

We first distinguish between our use of the terms ``cross section" and ``opacity".  The former has units of area.  The latter is the cross section per unit mass.  We previously designed and wrote an open-source opacity calculator \citep{gh15}, based on implementing Algorithm 916 \citep{za12} to perform fast computations of the Voigt profile by recasting it as a Faddeeva function.  Typically, \texttt{HELIOS-K} is able to compute an opacity or cross section function with $\sim 10^5$ spectral lines in $\sim 1$ s on a NVIDIA K20 GPU. In principle, it is agnostic about the spectroscopic line list being used and is able to take any line list as an input.  The details of how to take the inputs of a line list and use them to compute the integrated line strengths and line shapes have previously been summarized in \cite{gh15} and we will not repeat them here.

We restrict ourselves to only four molecules: carbon monoxide (CO), carbon dioxide (CO$_2$), water (H$_2$O) and methane (CH$_4$).   For CO and CO$_2$, we use the HITEMP database \citep{rothman10}.  For H$_2$O and CH$_4$, we use the ExoMol line list \citep{barber06,yt14}.  Acetylene, ammonia, ethylene and hydrogen cyanide have been omitted, because they are subdominant at the photospheric temperatures of the HR 8799 exoplanets \citep{madhu12,ht16,moses16}.  In particular, see Figure 10 of \cite{moses16}.

In the current study, we choose to deal with cross sections instead of opacities.  For our \texttt{HELIOS} self-consistent radiative transfer code, we chose to use opacities instead \citep{malik17}.  There are various strategies to construct the cross section function of the atmosphere.  By ``cross section function", we refer to the function that depends on temperature, pressure, wavenumber and type of molecule.  The cross section function is a theoretical construction: it may be defined continuously or be sampled at an arbitrary number of discrete points.  We consider the way in which the cross section function is sampled as an issue of implementation, which we will now discuss.  Regardless of the approach used to construct and sample the cross section function, the end goal is the same: to use them to construct transmission functions and ultimately integrate fluxes over a waveband.

The first approach is to use the ``k-distribution method", which resamples the highly erratic cross section function into a monotonically increasing cumulative distribution function \citep{lo91,fu92,gh15}.  Since the k-distribution method is only exact for a homogeneous atmosphere with one molecule \citep{gh15}, one has to apply the ``correlated-k approximation" as well, which assumes that the spectral lines are perfectly correlated (see Chapter 4.4.5 of \citealt{pierrehumbert}).  

The second approach is to use ``opacity sampling", which is to discretely sample the opacity function, typically at a smaller number of points than there are lines.  In our context, it is perhaps more accurate to term it ``cross section sampling". 

The ``line-by-line" limit occurs when the integrated fluxes over a waveband is exact (to machine precision).  It is essentially the second approach, but where the cross section function is sampled at more wavenumber points than there are lines.  Since there are $\sim 10^9$ (or more) lines for the water molecule alone, this is a formidable computational challenge and is currently infeasible for any retrieval code dealing with hot exoplanetary atmospheres.  We note that a cross section function that includes all of the lines of a given line list does not qualify it as being ``line by line", if the sampling is not fine enough to resolve each line profile.

In the current study, we adopt the second approach, which is also used by \cite{ms09}, \cite{bs13}, \cite{line13} and \cite{waldmann15}.  Our spectral resolution used is 1 cm$^{-1}$, evenly sampled across wavenumber.  We note that \cite{line13,line15} and \cite{waldmann15} also used a spectral resolution of 1 cm$^{-1}$.  Some authors do not specify the spectral resolution of their atmospheric cross section function (e.g., \citealt{ms09,bs13,barstow15,line16}).  We precompute our cross sections on a grid across wavenumber, pressure and temperature: 100 to 2900 K (in increments of 200 K) and 1 $\mu$bar to 1 kbar (with two points per dex in pressure) for CO, CO$_2$, CH$_4$ and H$_2$O.   The grid is then interpolated to obtain values of the cross sections for any temperature and pressure within the stated ranges.

A lingering issue, which stems from an unsolved physics problem, is that the far line wings of Voigt profiles do not accurately represent the wings of real lines.  Various groups have adopted different ad hoc approaches to truncating the Voigt profiles (see \citealt{gh15} and references therein).  \cite{hm16} discuss this issue, but do not provide any solution for it.  In the current study, we adopt a 100 cm$^{-1}$ cutoff.

\subsection{Chemistry}

\begin{figure}%[!h]
\begin{center}
%\vspace{-0.2in}
\includegraphics[width=\columnwidth]{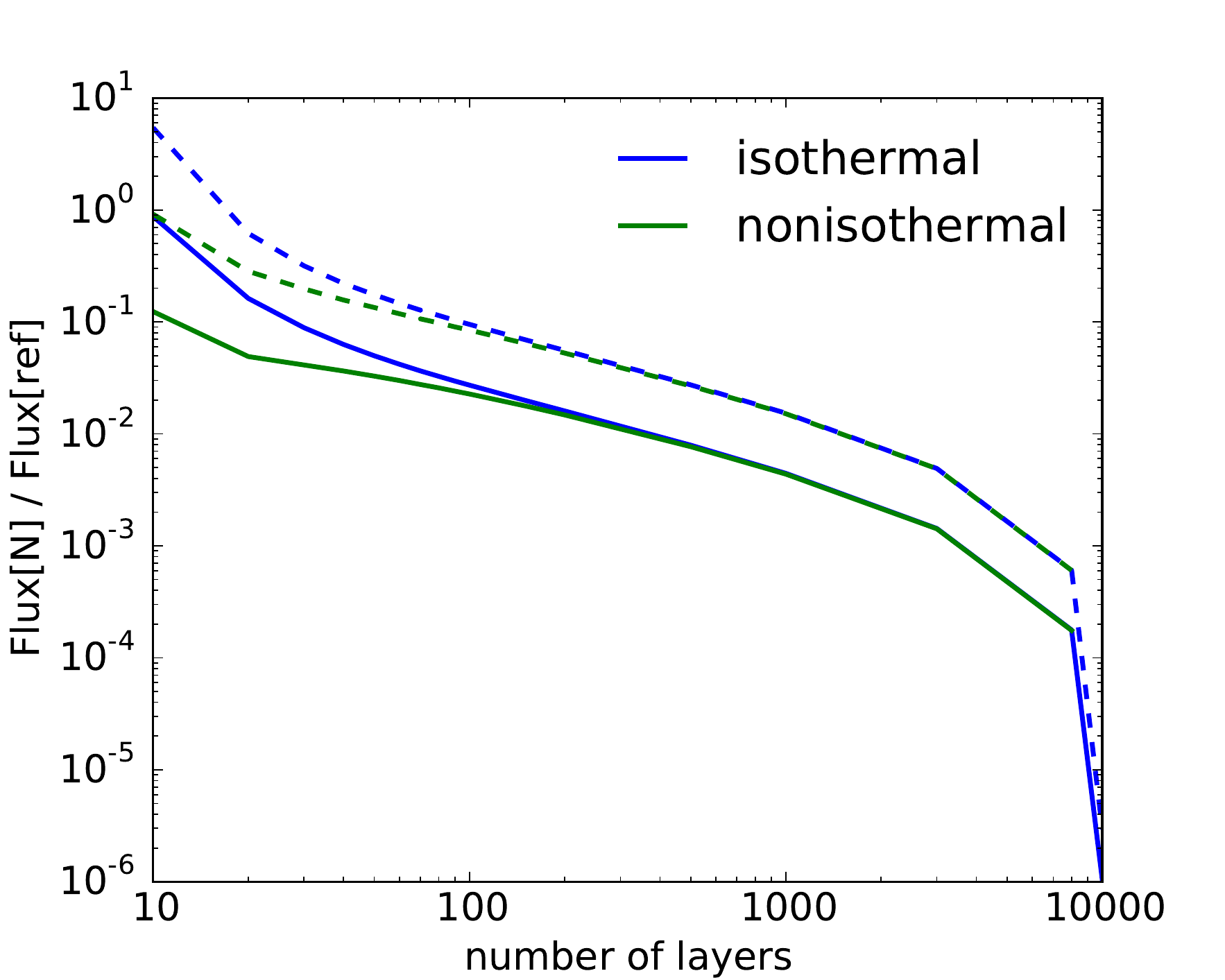}
\end{center}
%\vspace{-0.2in}
\caption{Mean (solid curves) and maximum (dashed curves) errors in the synthetic spectrum as a function of the number of model layers used, computed by performing retrievals on the measured spectrum of HR 8799b.  The reference used is the retrieval with 10,000 model layers (see text).  When about 100 layers are used, the models with isothermal and non-isothermal layers yield the same answers.}
%\vspace{0.1in}
\label{fig:layers}
\end{figure}

\begin{figure}%[!h]q
\begin{center}
%\vspace{0.2in}
\includegraphics[width=\columnwidth]{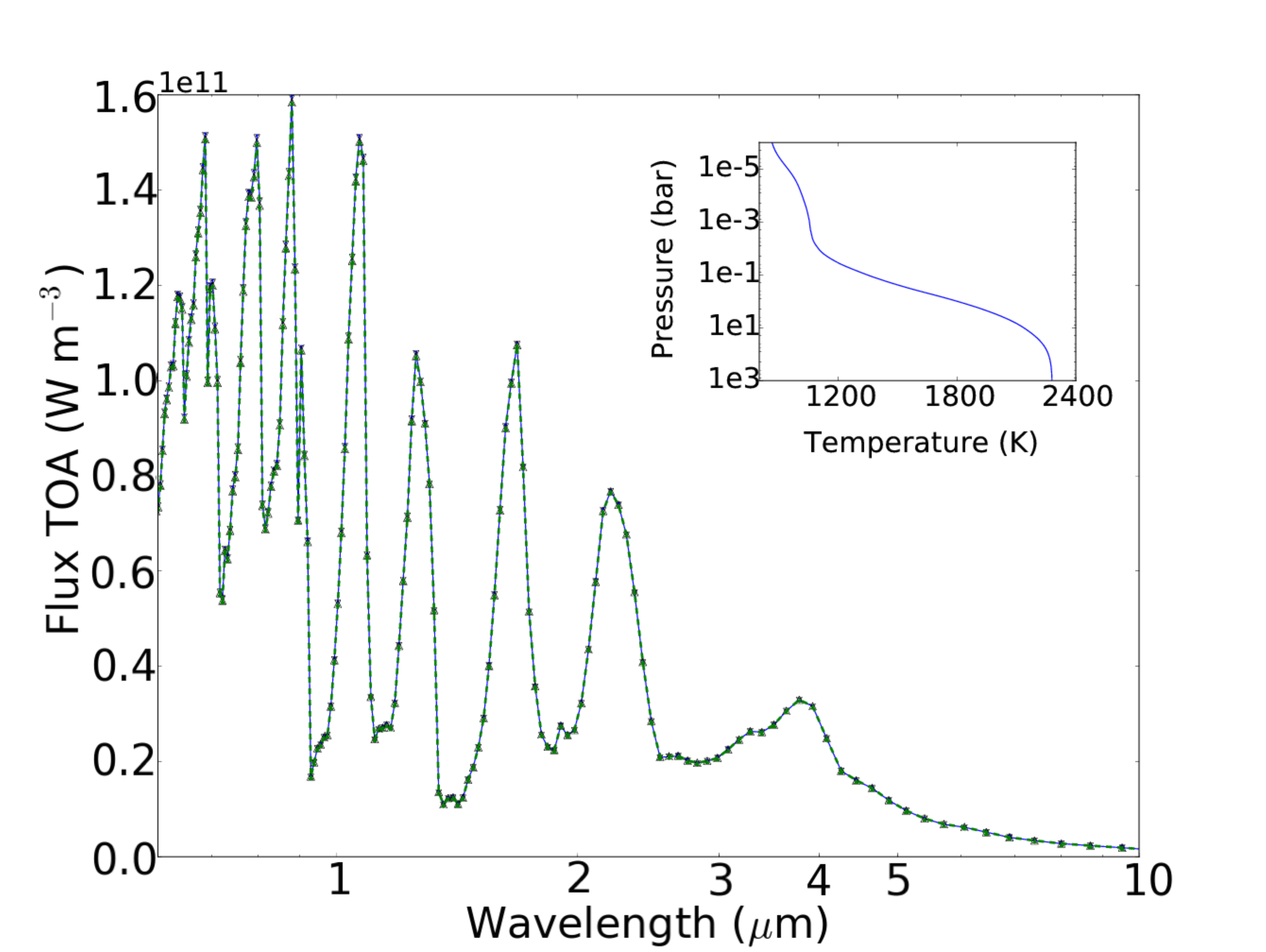}
\end{center}
%\vspace{-0.2in}
\caption{Validation of our \texttt{HELIOS-R} forward model (green dashed curve and upward-facing triangles) against that used in the \texttt{HELIOS} self-consistent radiative transfer code (blue solid curve and downward-facing triangles).  The insert shows the temperature-pressure profile used as an input.}
%\vspace{0.1in}
\label{fig:validation}
\end{figure}

Once the cross sections have been computed, they may be used to compute the optical depth, of each model layer, for all of the molecules,
\begin{equation}
\Delta \tau = \sum_i \frac{X_i \sigma_i}{\bar{m} g} ~\Delta P,
\end{equation}
where $X_i$ and $\sigma_i$ are the mixing ratio and cross section of the $i$-th molecule, respectively.  $\Delta P$ is the thickness of the layer in terms of the difference in pressure.  The mean molecular mass is given by $\bar{m}= \mu m_{\rm u}$, where $\mu$ is the mean molecular weight and $m_{\rm u}$ is the atomic mass unit.  The preceding expression assumes hydrostatic equilibrium, isothermal layers and that the surface gravity is constant throughout our model atmosphere.

Generally, the mixing ratios of molecular hydrogen and helium only show up via CIA as a contribution to the continuum of a spectrum, which implies that they cannot be as definitively determined as that of the molecules.  Our CIA opacities are obtained from \cite{richard12}.  In the range of $X_{\rm H_2} \approx 0.8$--0.9, the effect on the continuum of the synthetic spectrum is very similar (not shown).  The effects of H$_2$-He CIA are even more subtle.  As such, we adjust $X_{\rm H_2}$ to render the sum of the mixing ratios unity,
\begin{equation}
1.1 X_{\rm H_2} + \sum_i X_i = 1,
\end{equation}
where we have assumed that $X_{\rm He} = 0.1 X_{\rm H_2}$ to reflect cosmic abundance.  By denoting the mass of the $i$-th molecule by $m_i$, the corresponding mean molecular weight is calculated using
\begin{equation}
\mu = 2 X_{\rm H_2} + 4 X_{\rm He} + \sum_i \frac{m_i X_i}{m_{\rm u}}.
\end{equation}
For example, if we have $X_{\rm H_2} = 0.85$, $X_{\rm He}=0.085$ and $X_{\rm CO}=0.065$, then we have $\mu = 3.86$.  In models with equilibrium chemistry, the mean molecular weight changes slightly for each layer, because the molecular abundances vary from layer to layer even for the same metallicity.

In the current study, we consider two chemistry models.  ``Unconstrained chemistry" refers to using each $X_i$ as a fitting parameter in the retrieval.  ``Equilibrium chemistry" means that the $X_i$ may be determined using only the elemental abundances of carbon ($f_{\rm C}$) and oxygen ($f_{\rm O}$), if C-H-O gaseous chemistry is considered.  In this case, the 4-parameter system of unconstrained chemistry reduces to 2 parameters.  To compute the four $X_i$ values given $f_{\rm C}$ and $f_{\rm O}$, we use the validated analytical formulae of \cite{hlt16}, \cite{hl16} and \cite{ht16}.  Specifically, we implement equations (12), (20) and (21) of \cite{hl16} for gaseous C-H-O chemistry.  The benchmarking of these formulae against calculations of Gibbs free energy minimization was previously performed by \cite{ht16}, who showed that they are accurate at the $\sim 1\%$ level or better.  Further validation of these formulae comes from matching the trends found by \cite{madhu12} and \cite{moses13}.

For unconstrained chemistry, the carbon-to-oxygen ratio is computed using
\begin{equation}
\mbox{C/O} = \frac{X_{\rm CO} + X_{\rm CO_2} + X_{\rm CH_4}}{X_{\rm CO} + 2 X_{\rm CO_2} + X_{\rm H_2O}}.
\end{equation}
The elemental abundances are inferred using
\begin{equation}
\begin{split}
\mbox{C/H} &= \frac{X_{\rm CO} + X_{\rm CO_2} + X_{\rm CH_4}}{2 X_{\rm H_2} + 4 X_{\rm CH_4} + 2 X_{\rm H_2O}}, \\
\mbox{O/H} &= \frac{X_{\rm CO} + 2 X_{\rm CO_2} + X_{\rm H_2O}}{2 X_{\rm H_2} + 4 X_{\rm CH_4} + 2 X_{\rm H_2O}}.
\end{split}
\end{equation}
Each mixing ratio is assumed to be constant over the entire model atmosphere.  The alternative, which is to have a different value of the mixing ratio for each of the 100 model layers we assume, would result in 400 free parameters.  This is unwarranted given the sparseness of the data, i.e., we have less than 400 data points.

For equilibrium chemistry, the carbon-to-oxygen ratio is simply
\begin{equation}
\mbox{C/O} = \frac{f_{\rm C}}{f_{\rm O}},
\end{equation}
and $f_{\rm C} \equiv \mbox{C/H}$ and $f_{\rm O} \equiv \mbox{O/H}$ are directly the fitting parameters of the retrieval.  Since the mixing ratios of all of the molecules can be exactly specified for each layer, which has its own temperature and pressure, the assumption of constant mixing ratios across height/pressure is unnecessary for the models with equilibrium chemistry.  The 400 values of the mixing ratios are specified by just two free parameters: $f_{\rm C}$ and $f_{\rm O}$.

Unlike in previous studies, we do not manually decide whether to pick unconstrained or equilibrium chemistry.  Rather, we compute both of these models and select between them based on the Bayesian evidence computed.  

We note that, as part of the ESP, we have previously developed a chemical kinetics solver named \texttt{VULCAN} \citep{tsai17}.

\begin{figure*}%[!h]
\begin{center}
\vspace{-0.2in}
\includegraphics[width=0.65\columnwidth]{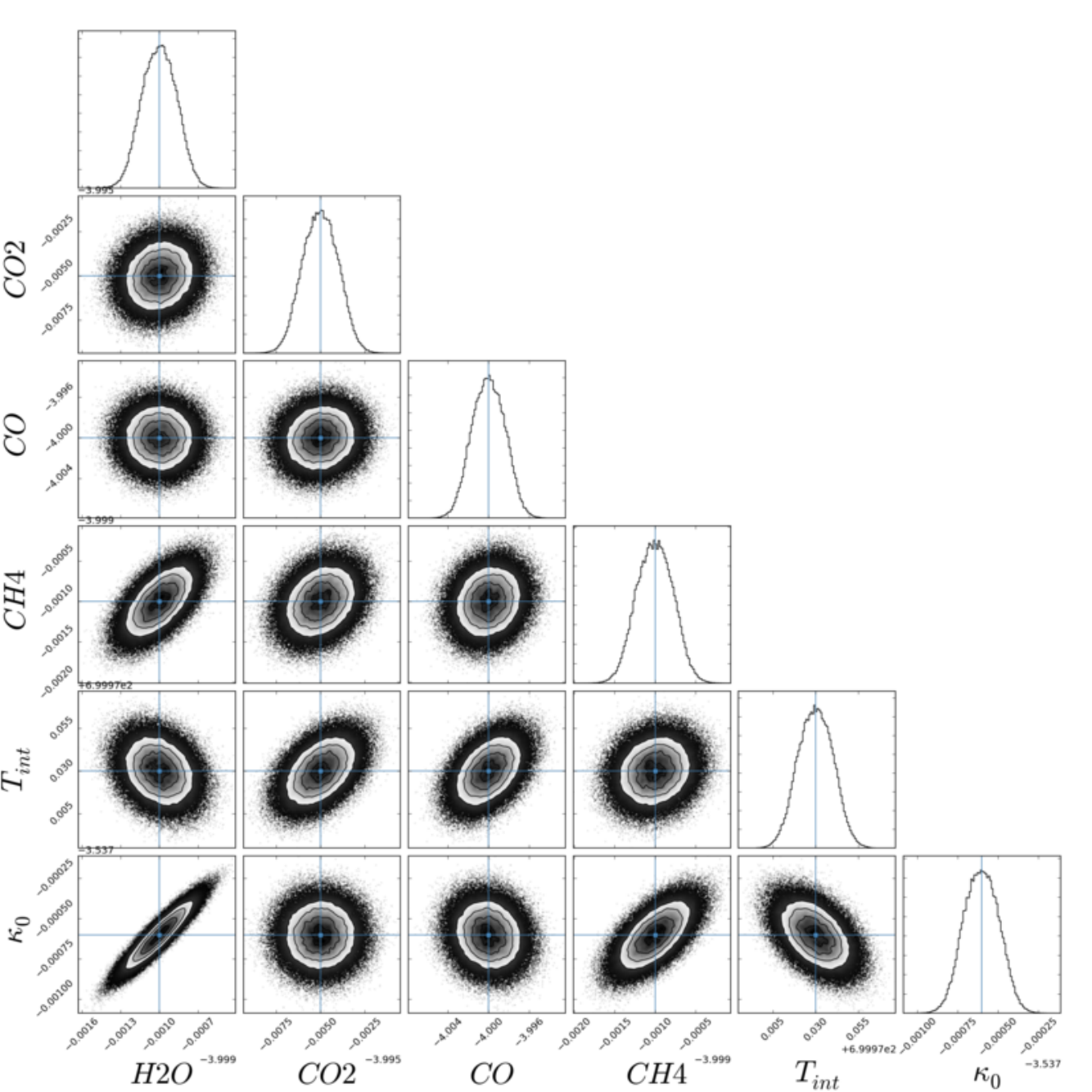}
\includegraphics[width=0.65\columnwidth]{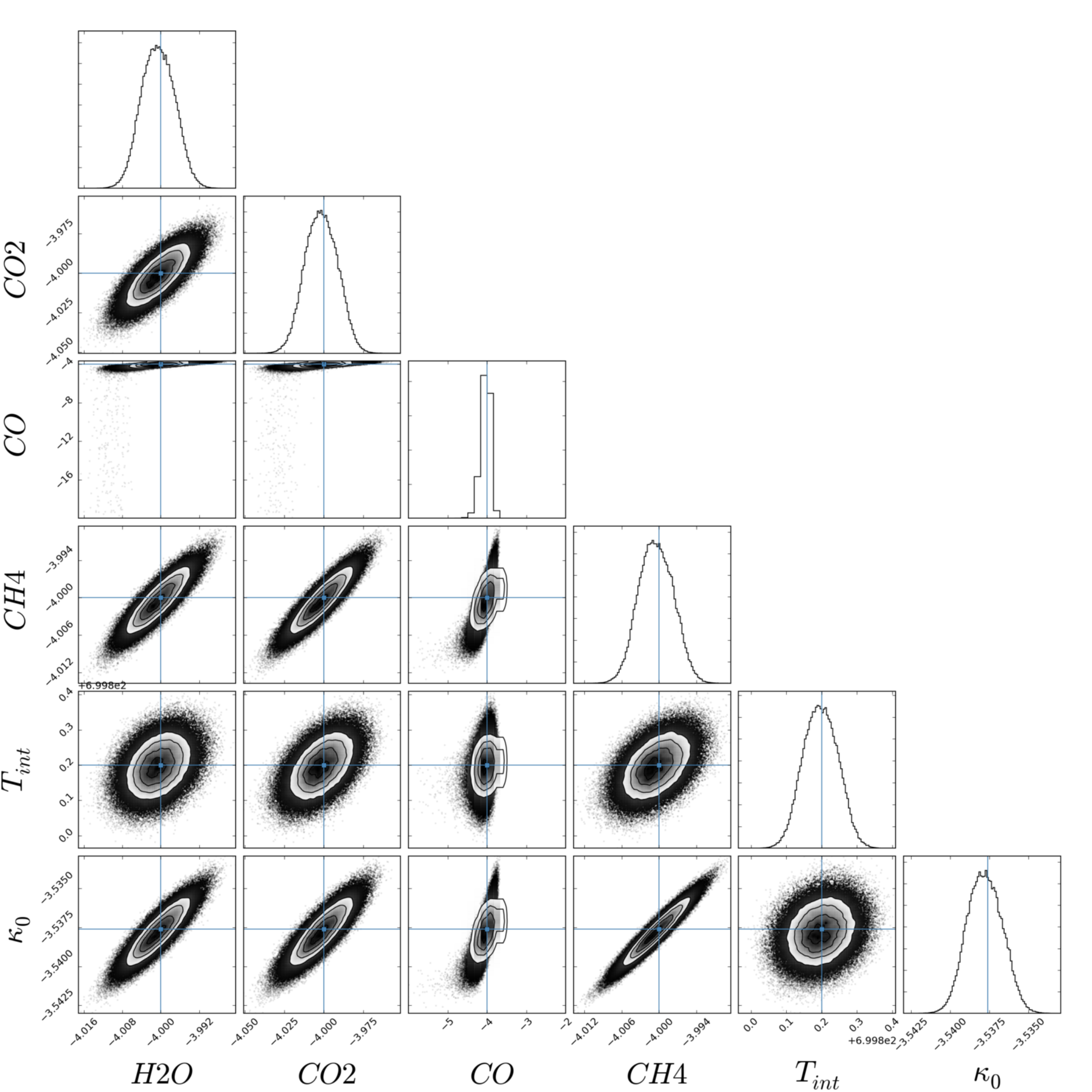}
\includegraphics[width=0.65\columnwidth]{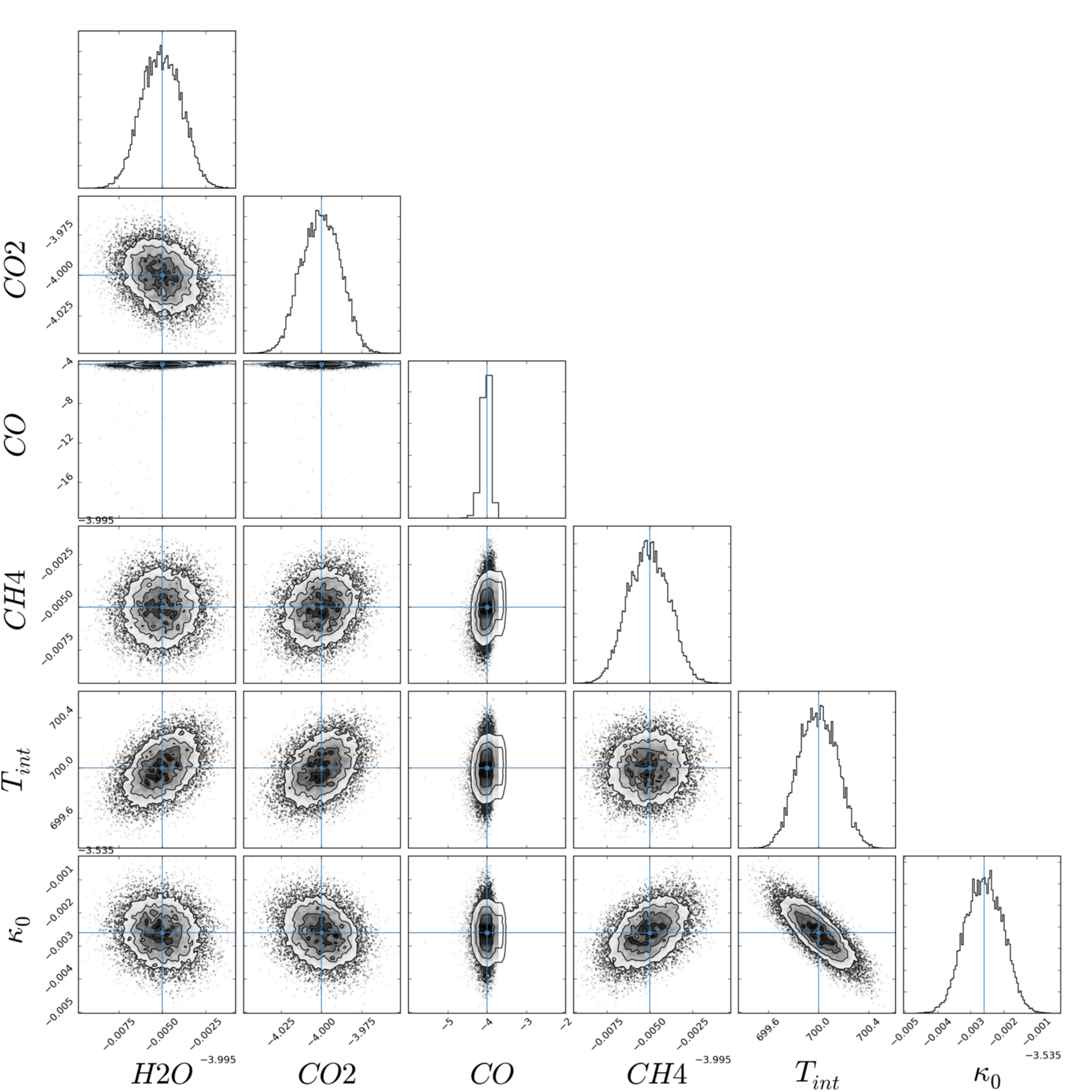}
\includegraphics[width=0.65\columnwidth]{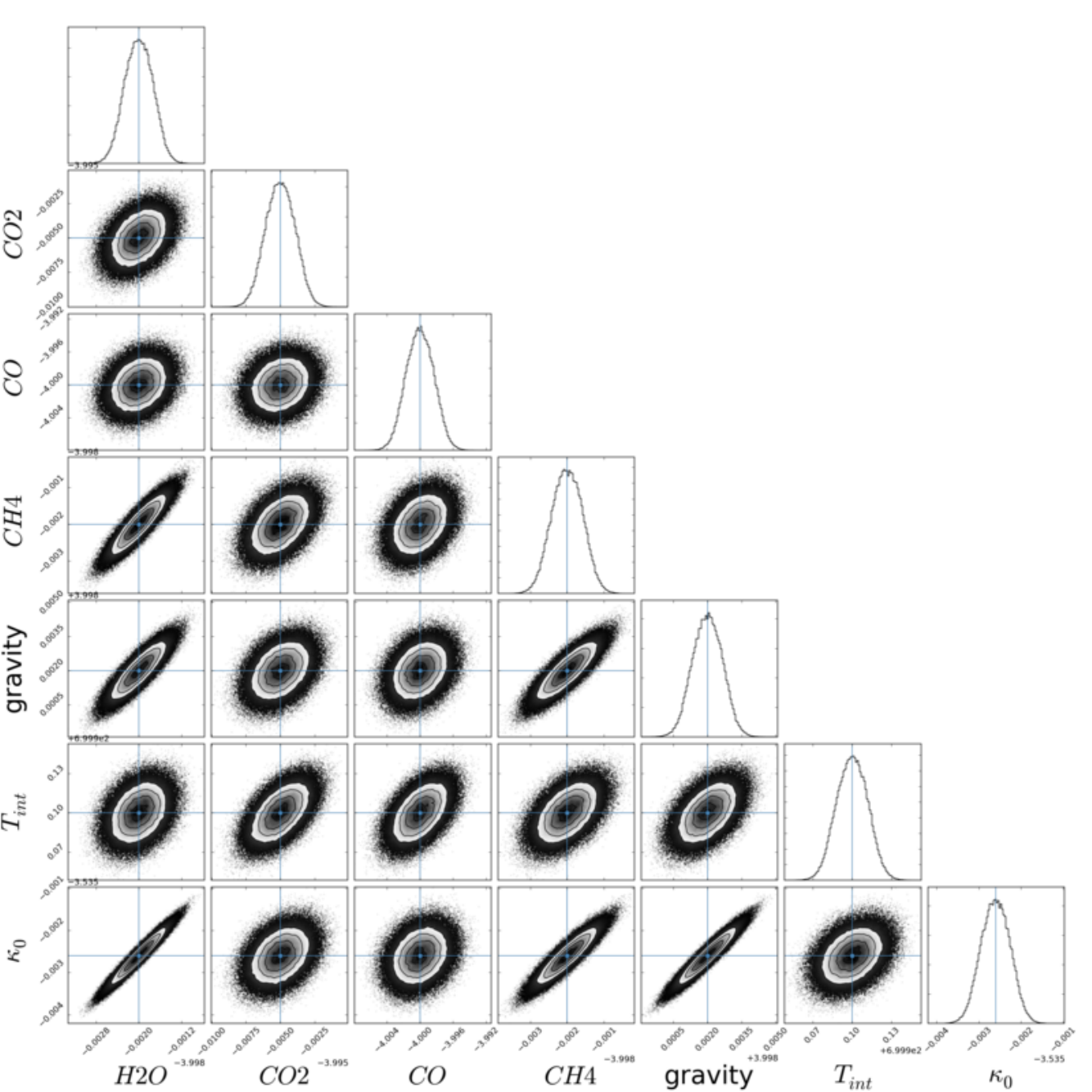}
\includegraphics[width=0.65\columnwidth]{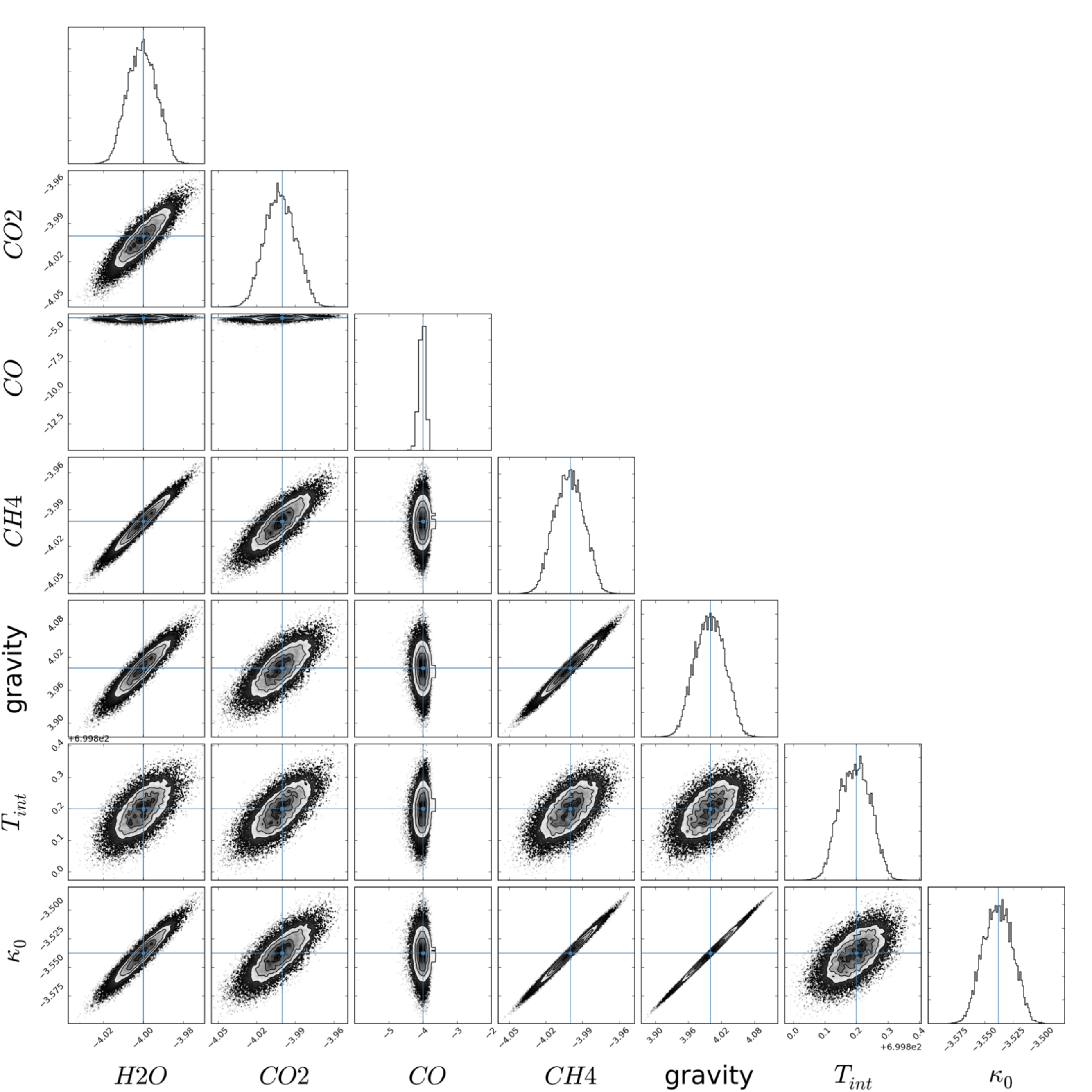}
\includegraphics[width=0.65\columnwidth]{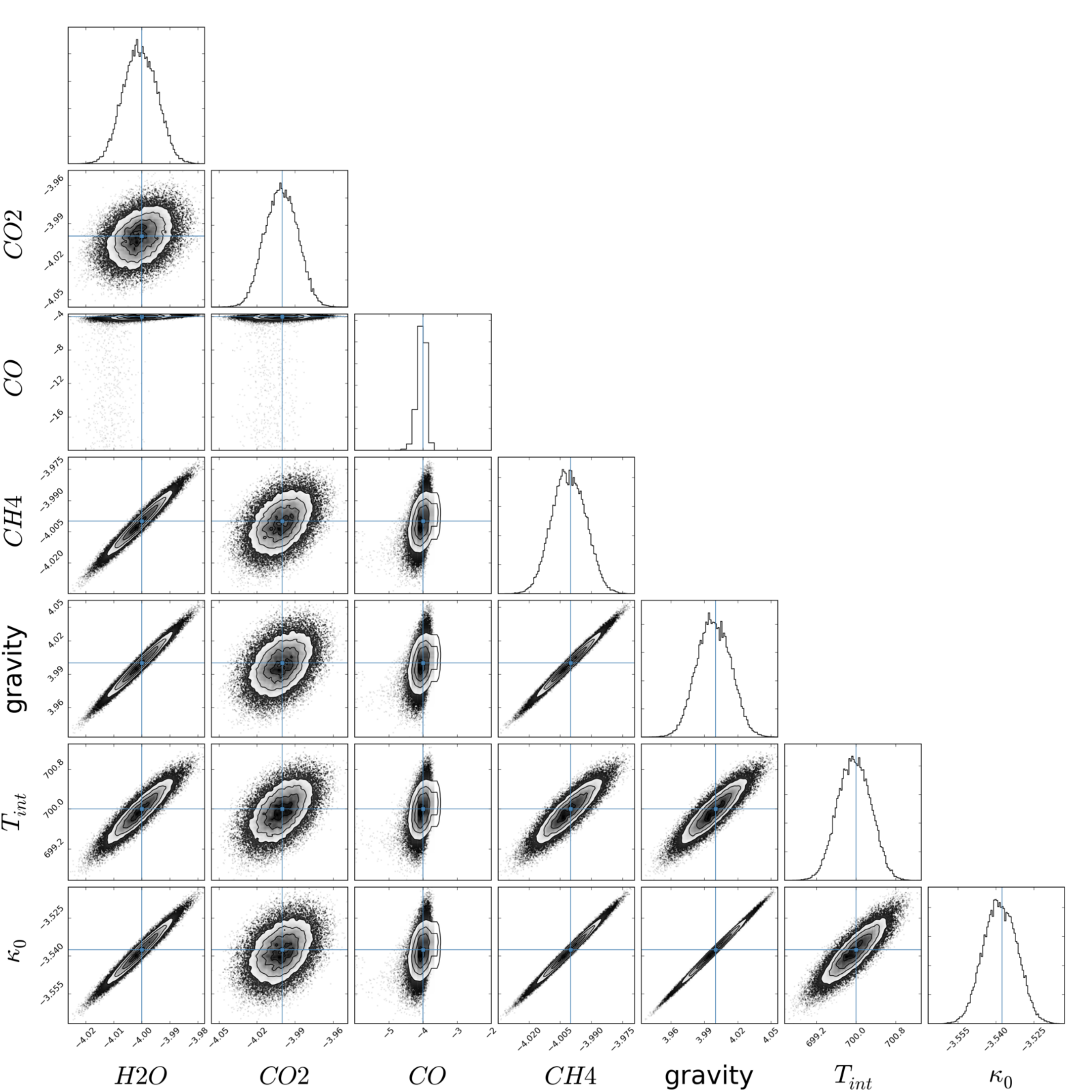}
\includegraphics[width=0.65\columnwidth]{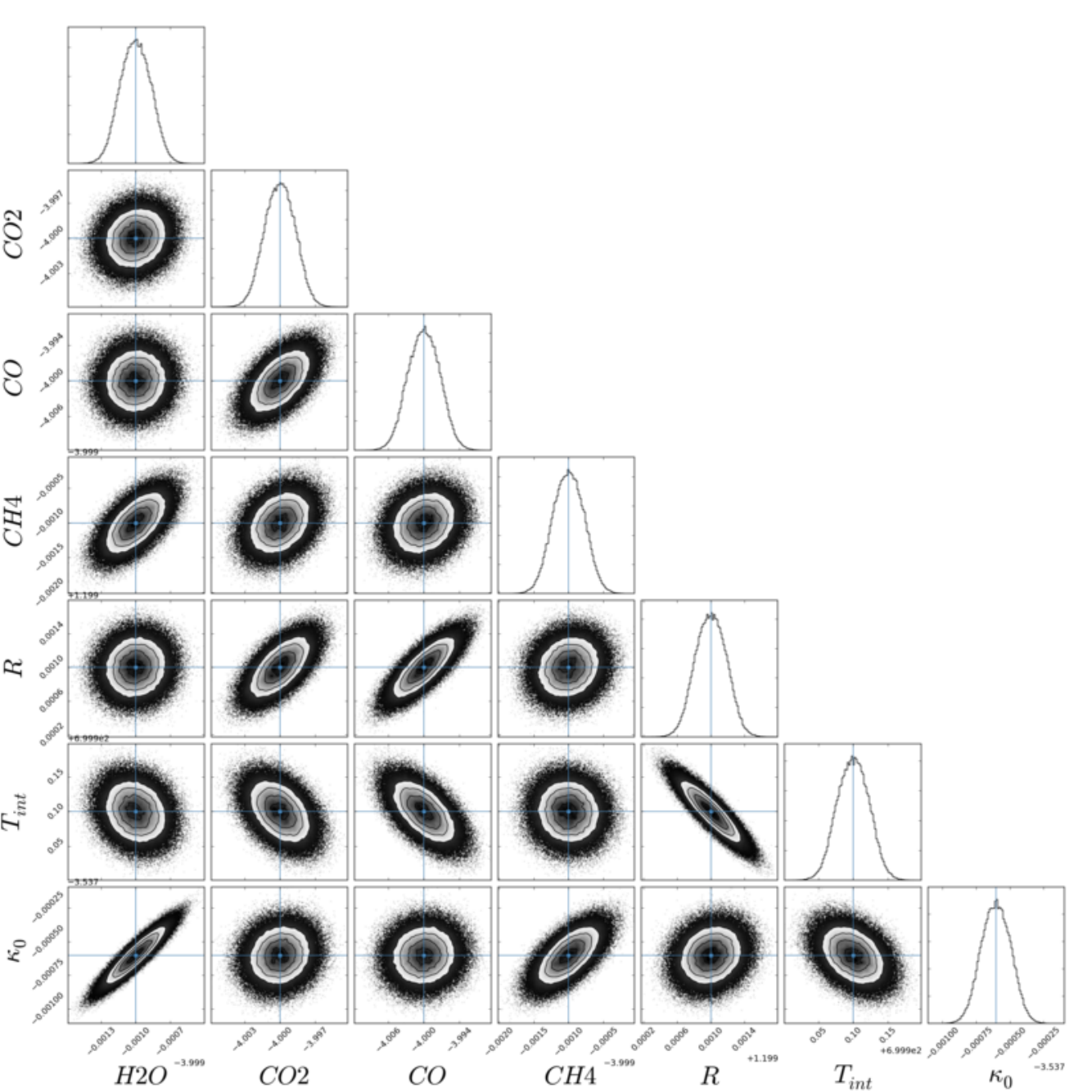}
\includegraphics[width=0.65\columnwidth]{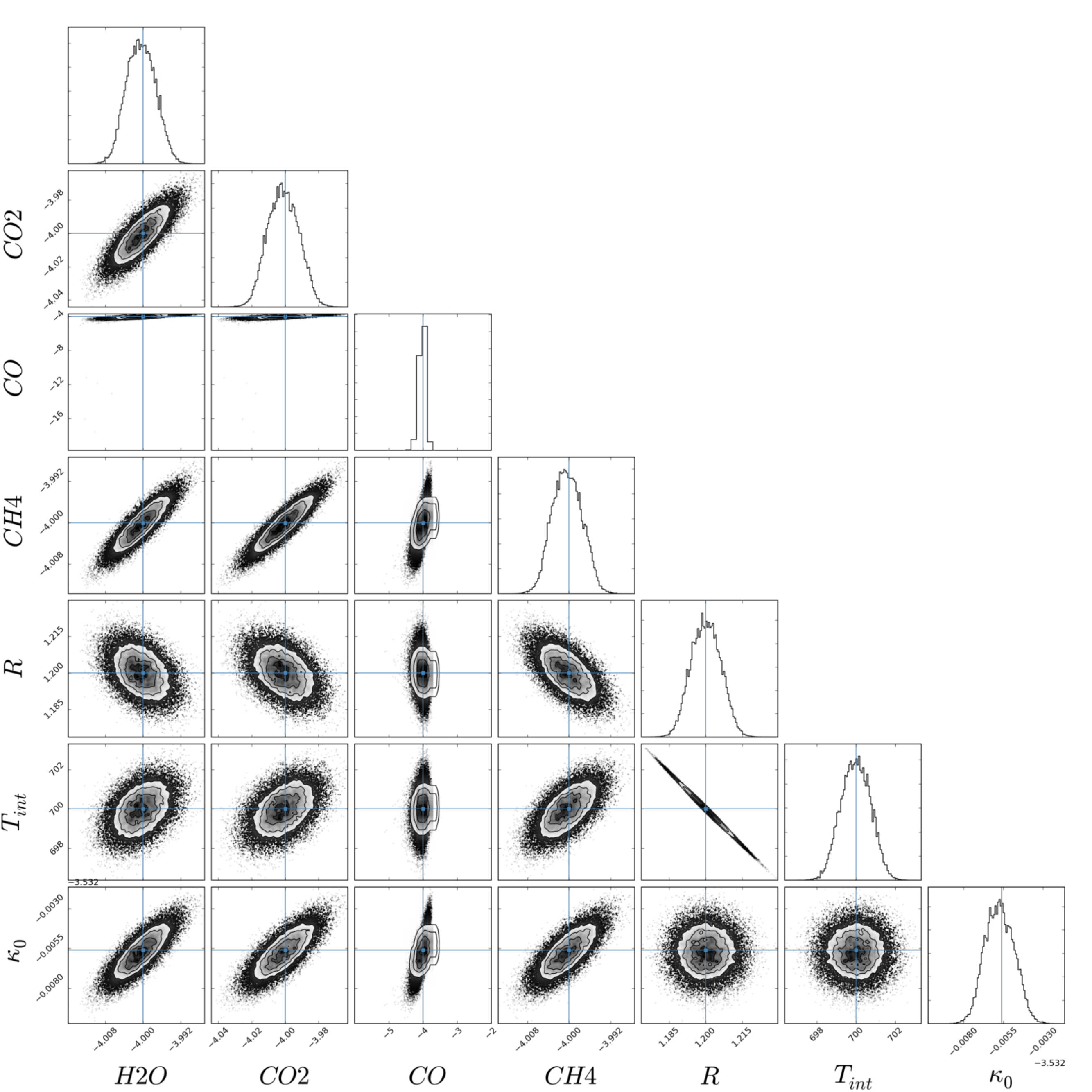}
\includegraphics[width=0.65\columnwidth]{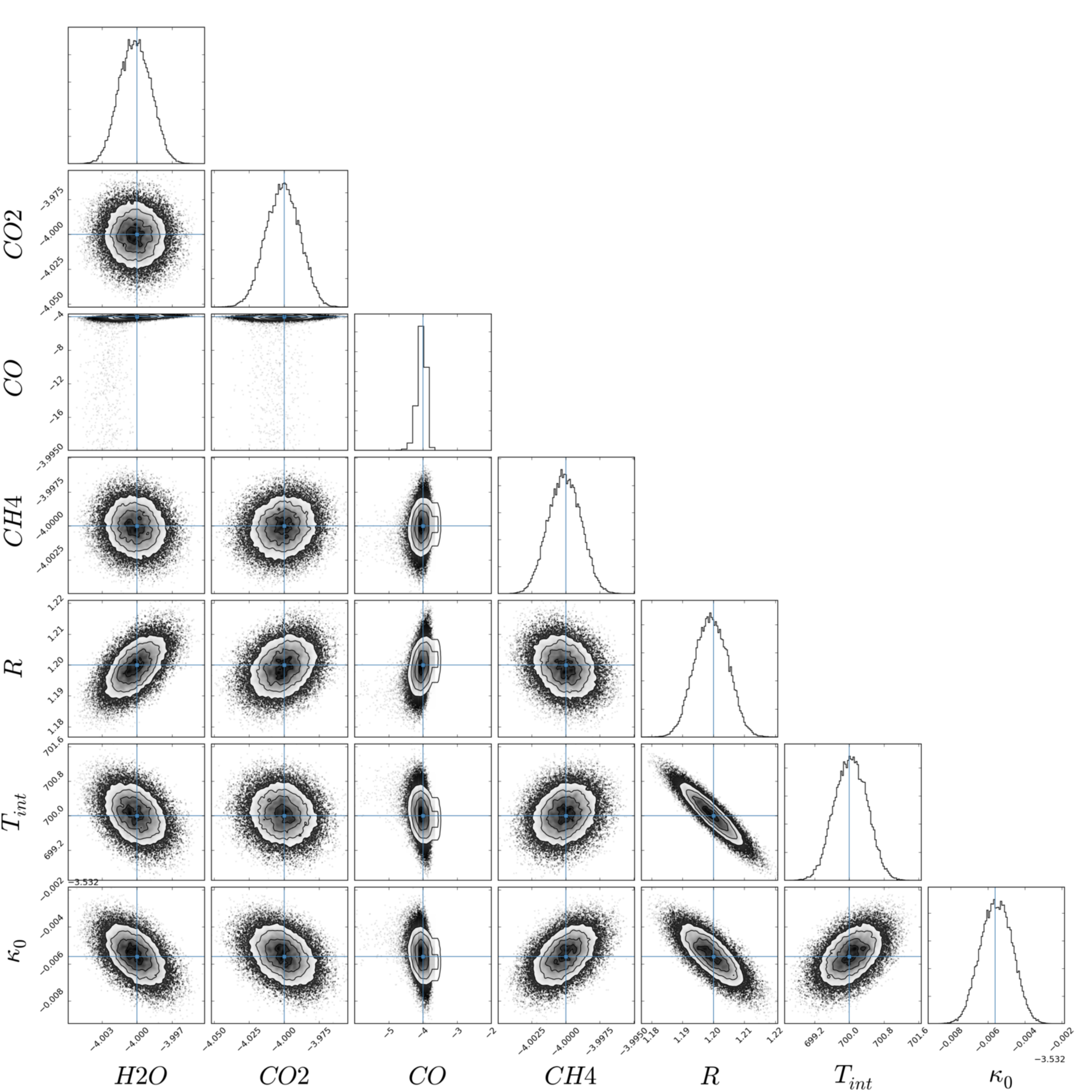}
\includegraphics[width=0.65\columnwidth]{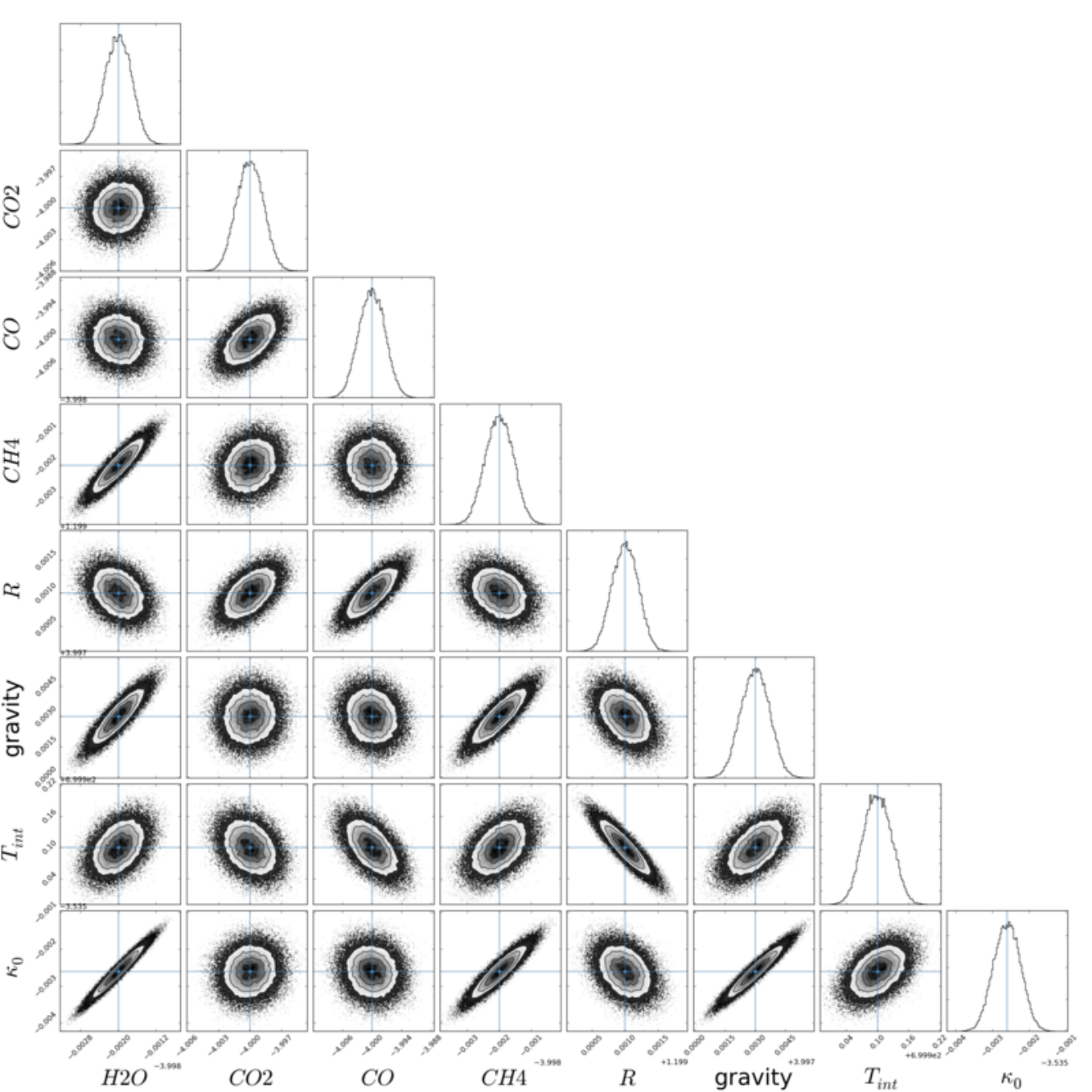}
\includegraphics[width=0.65\columnwidth]{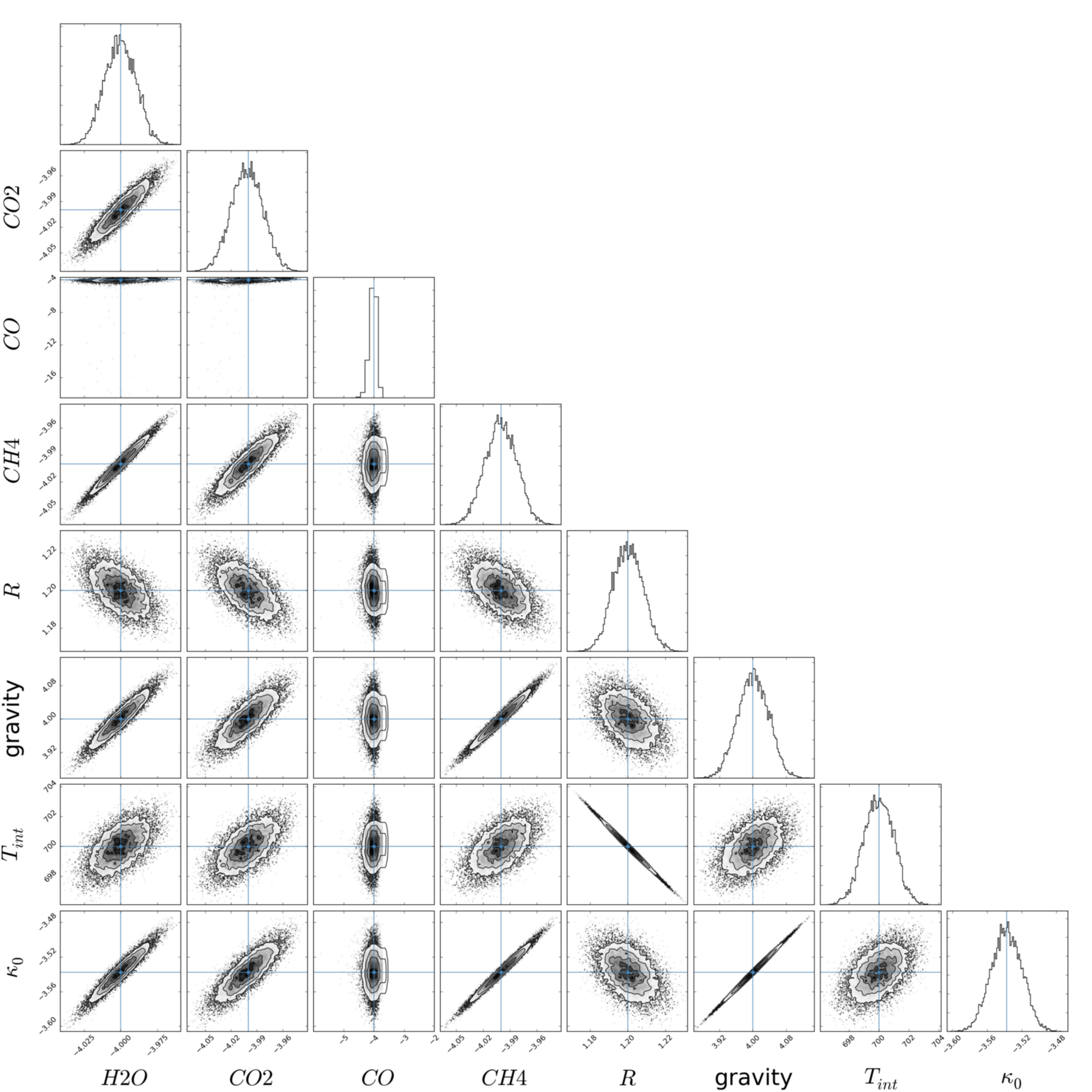}
\includegraphics[width=0.65\columnwidth]{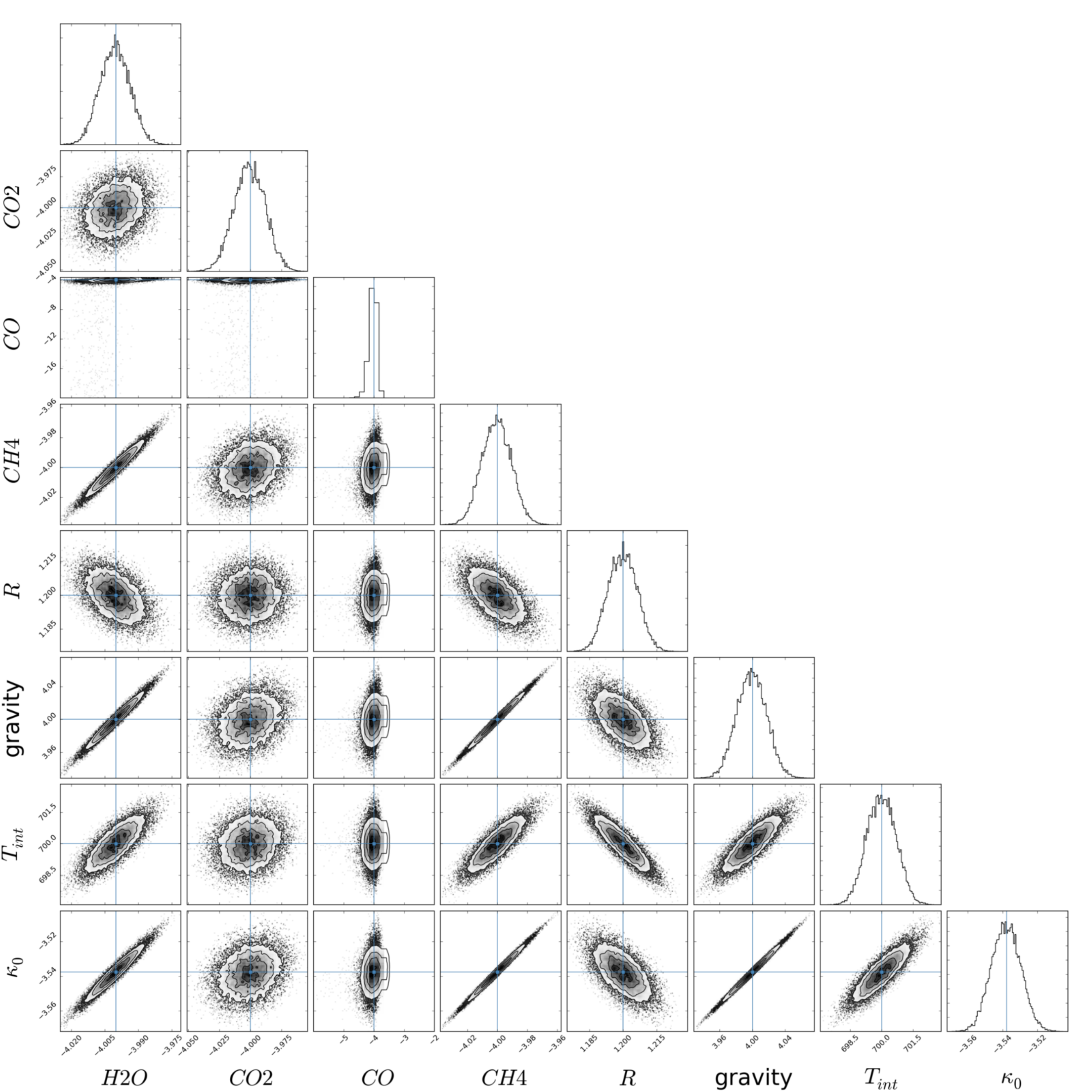}
\end{center}
\vspace{-0.1in}
\caption{Performing retrieval on a mock dataset, where the ``ground truth" is known (see text for input parameters).   We focus on performing retrieval with the UB model (see Table 1).  The columns represent the different wavelength coverages: 0.7--5 $\mu$m with 0.01 $\mu$m resolution (first column), HR 8799b-like data (second column) and HR 8799e-like data (third column).  The rows represent different assumptions for $R$ and $g$: fixed $R$ and $g$ (first row), fixed $R$ but $g$ is a fitting parameter (second row), fixed $g$ but $R$ is a fitting parameter (third row), both $R$ and $g$ are fitting parameters (fourth row).  The labels ``CO", ``CO2", ``CH4" and ``H2O" refer to the mixing ratios of carbon monoxide, carbon dioxide, methane and water, respective.  $T_{\rm int}$ is in units of K.  The labels ``$\kappa_0$" and ``gravity" are for $\log{\kappa_0}$ and $\log{g}$ in mks and cgs units, respectively.}
%\vspace{0.1in}
\label{fig:mock}
\end{figure*}

\begin{figure}%[!h]
\begin{center}
%\vspace{-0.2in}
\includegraphics[width=1.05\columnwidth]{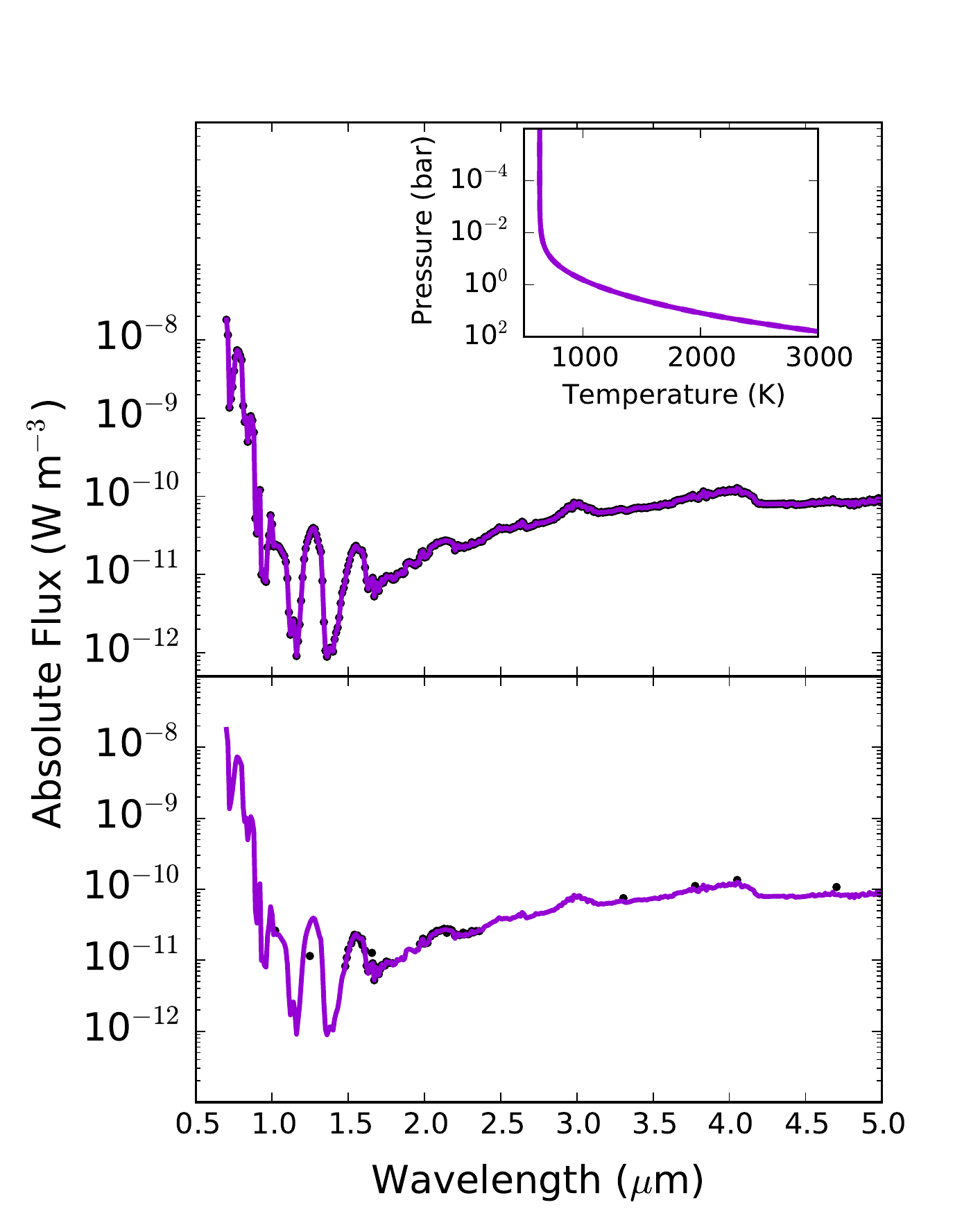}
\includegraphics[width=1.05\columnwidth]{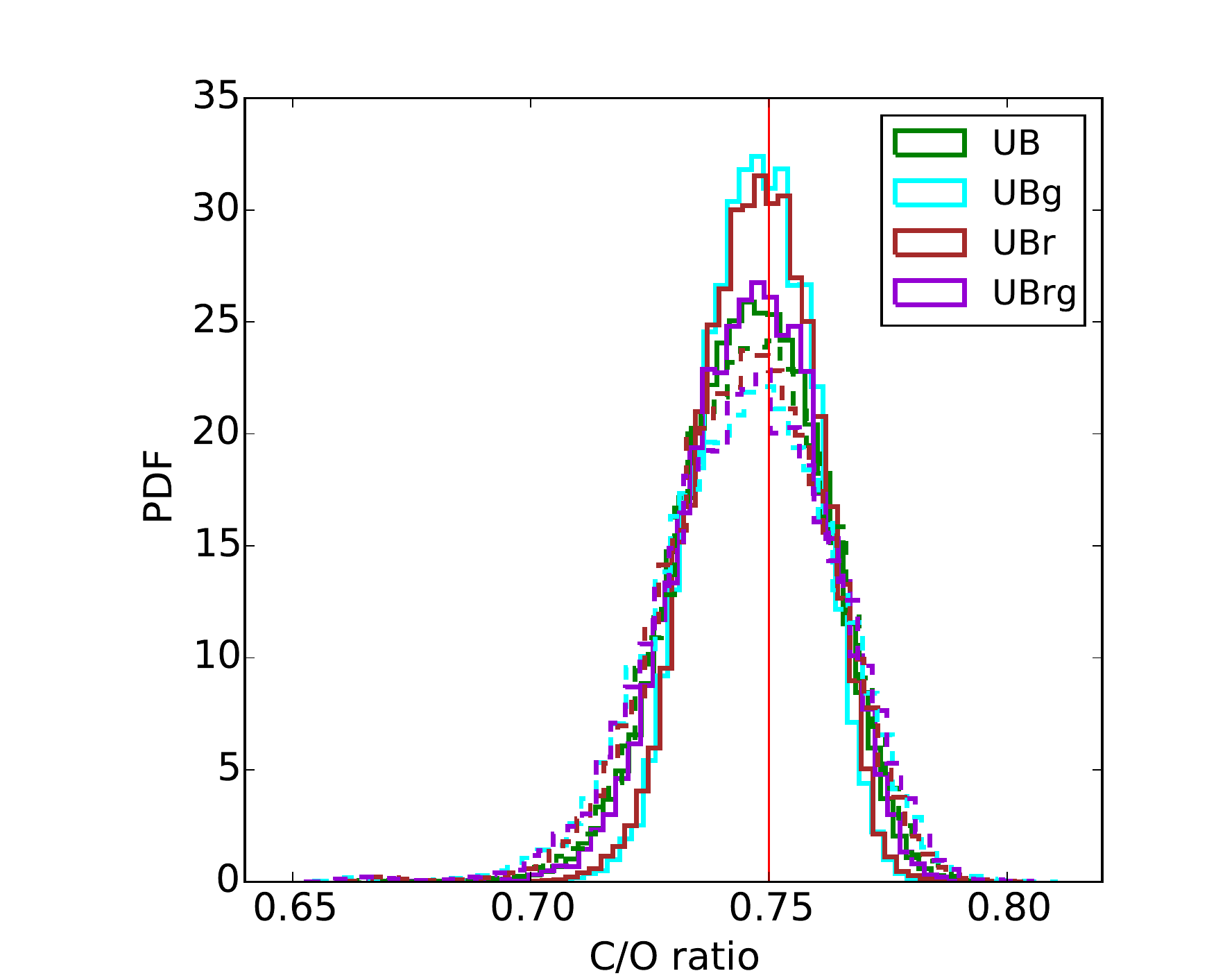}
\end{center}
%\vspace{-0.1in}
\caption{Further results from the retrievals on the mock dataset.  The top panel shows the mock dataset at full resolution (0.01 $\mu$m), and also with HR 8799b-like data coverage (circles).  The mock dataset and best-fit spectrum overlap exactly (to within the resolution of the plot).  The insert shows the retrieved temperature-pressure profile.  The bottom panel shows the retrieved posterior distributions of C/O assuming different models (see Table 1).  The solid and dotted curves are for HR 8799b-like and HR 8799e-like data coverage.  The broader posterior distributions of C/O associated with HR 8799e-like data coverage are primarily due to the lack of K band data.}
%\vspace{0.1in}
\label{fig:mock_co}
\end{figure}

\subsection{Radiative Transfer Scheme}

\begin{figure*}%[!h]
\begin{center}
%\vspace{0.2in}
\includegraphics[width=2\columnwidth]{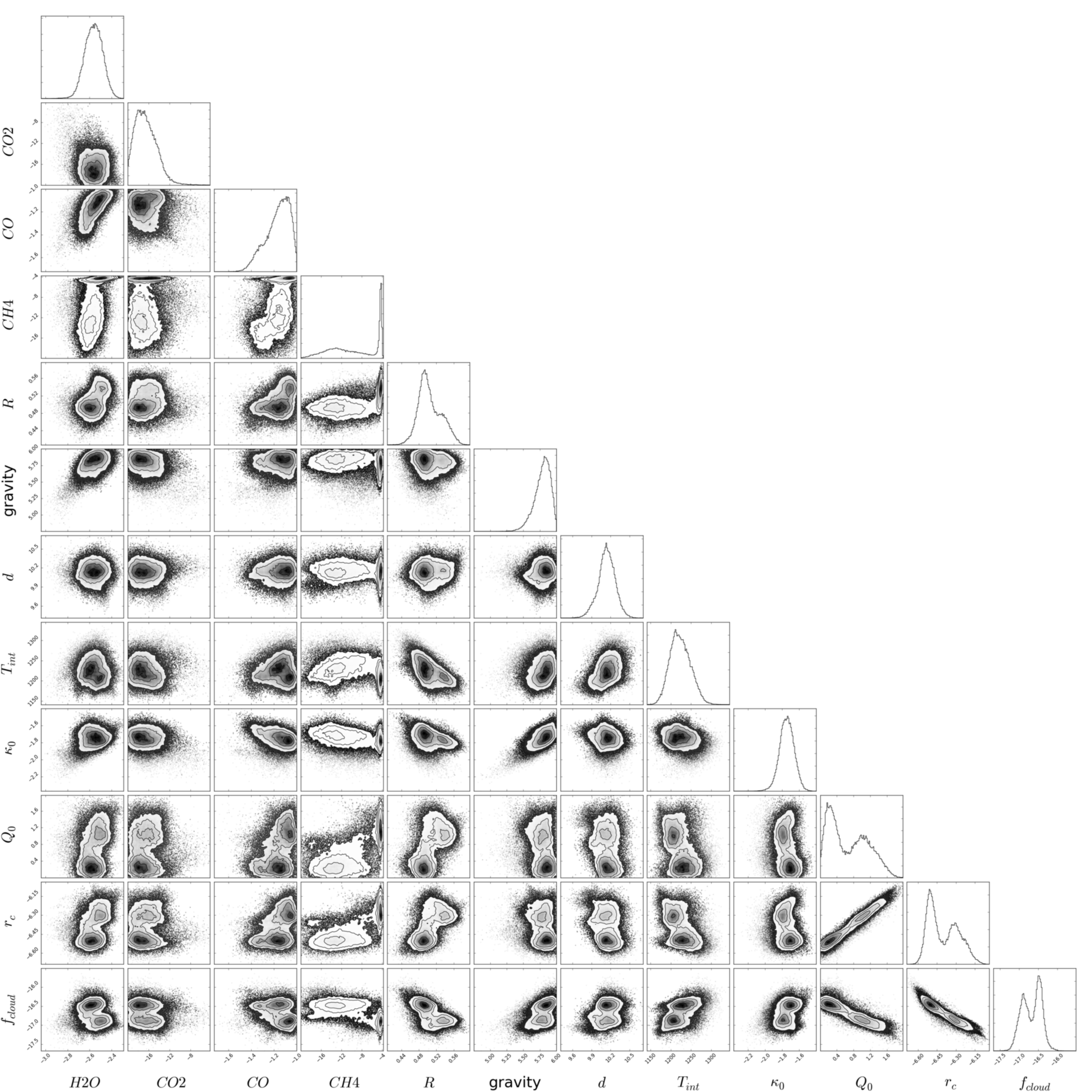}
\end{center}
%\vspace{-0.2in}
\caption{Montage of posterior distributions from performing retrieval on the measured spectrum of HR 8799b and allowing the radius, surface gravity and mean molecular weights to be uniform or log-uniform priors.  The retrieved value of the radius ($R \approx 0.5 R_{\rm J}$) is unphysical (see text for discussion).  The labels ``CO", ``CO2", ``CH4" and ``H2O" refer to the mixing ratios of carbon monoxide, carbon dioxide, methane and water, respectively.  $R$ and $T_{\rm int}$ are in units of $R_{\rm J}$ and K, respectively.  The labels ``$\kappa_0$" and ``gravity" are for $\log{\kappa_0}$ and $\log{g}$ in mks and cgs units, respectively. }
%\vspace{0.1in}
\label{fig:radius}
\end{figure*}

With the cross sections and temperature-pressure profiles in hand, one may compute the optical depth and hence the transmission function for each layer of the model atmosphere.  To propagate fluxes through the atmosphere and thus obtain the synthetic spectrum, we need a radiative transfer scheme.  Beer's law\footnote{Also known as the Beer-Lambert-Bouguer law.} is the simplest example of such a scheme, where incident radiation through a passive medium is exponentially attenuated.  A more sophisticated radiative transfer scheme needs to account for both the fluxes incident upon a layer and the thermal emission associated with the layer itself, since each layer has a finite temperature.  To this end, we use equation (B4) of \cite{hml14}, 
\begin{equation}
F_{\uparrow_{j+1}} = F_{\uparrow_j} {\cal T} + \pi B \left( 1 - {\cal T} \right),
\label{eq:flux}
\end{equation}
where the fluxes are computed at the $j$- and $(j+1)$-th interfaces.  The Planck function ($B$) is evaluated within each layer.  The transmission function is given by equation (B5) of \cite{hml14}
\begin{equation}
{\cal T} = \left( 1 - \Delta \tau \right) \exp{\left( -\Delta \tau \right)} + \left( \Delta \tau \right)^2 {\cal E}_1,
\end{equation}
with ${\cal E}_1(\Delta \tau)$ being the exponential integral of the first order.  The optical thickness of each layer is given by $\Delta \tau$.  Appendix \ref{append:e1} describes an analytical fitting formula for ${\cal E}_1$ that is highly accurate and allows for the computation to be significantly sped up.

We use equation (\ref{eq:flux}) to propagate the boundary condition at the bottom of the atmosphere (i.e., the internal/interior heat flux), which is the Planck function with a temperature given by the temperature-pressure profile at the bottom boundary.  The outgoing flux at the top of the atmosphere is then the synthetic spectrum.

We emphasize equation (\ref{eq:flux}) is an exact solution of the radiative transfer solution in the limit of isothermal layers and pure absorption. It is an improvement over using approximate solutions (e.g., two-stream solutions) and allows us to implement a radiative transfer scheme without taking any approximations besides assuming pure absorption.  Equation (\ref{eq:flux}) is equivalent to the approach of \cite{line16}, who used four-point Gaussian quadrature to account for angle-dependent flux propagation.  In our exact solution, the integration over angle has been performed analytically and is encapsulated in the exponential integral of the first order.  We gain computational efficiency both by bypassing the need for performing Gaussian quadrature and also by evaluating ${\cal E}_1$ using an analytical approach (Appendix \ref{append:e1}).  The overall accuracy is relegated to the number of discrete layers used.

The radius of the exoplanet ($R$) only appears as a scaling factor between the observed flux ($F_{\rm obs}$) and the flux escaping from the top of the atmosphere ($F_{\rm TOA}$),
\begin{equation}
F_{\rm obs} = \left( \frac{R}{d} \right)^2 F_{\rm TOA},
\end{equation}
where $d$ is the distance between the observer and the object.  The HR 8799 system is located at $39.4 \pm 1.0$ pc \citep{van07}, but the measured fluxes are usually reported as if it were located at $d=10$ pc (i.e., absolute fluxes).  

\subsection{Cloud Model}

\begin{figure}%[!h]
\begin{center}
%\vspace{0.2in}
\includegraphics[width=\columnwidth]{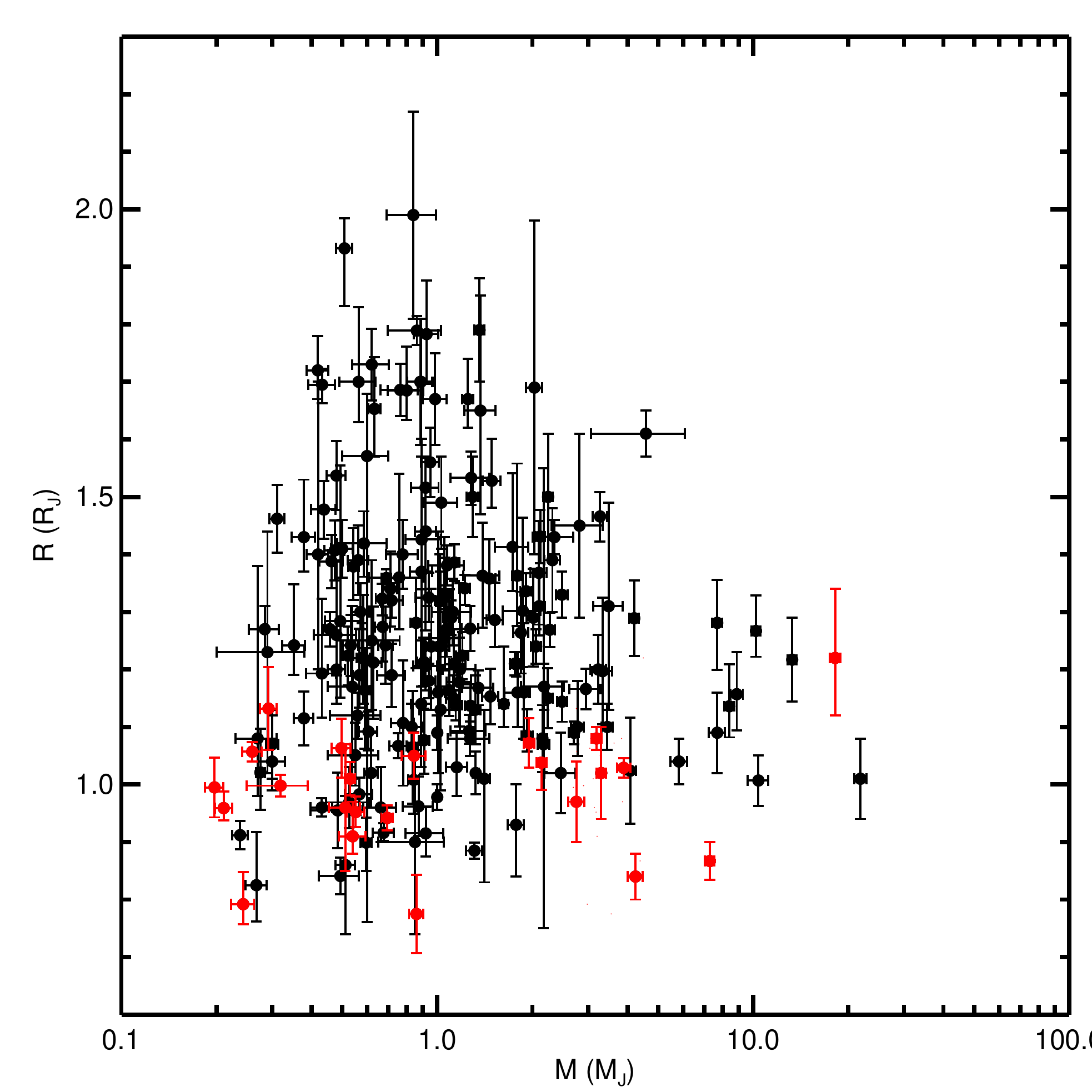}
\includegraphics[width=\columnwidth]{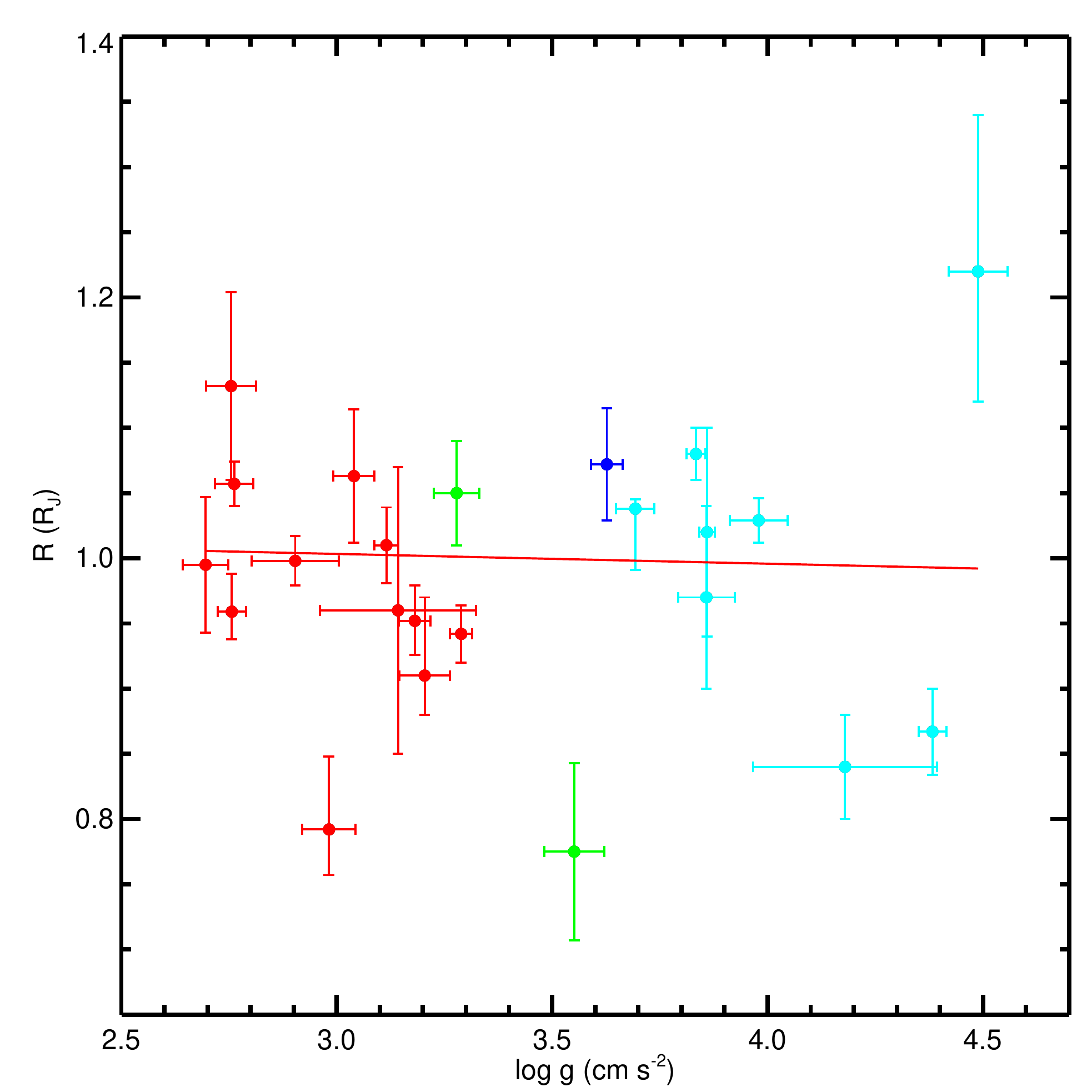}
\end{center}
%\vspace{-0.2in}
\caption{Top panel: transit radius versus surface gravity of a sample of transiting Jupiter-sized exoplanets around main-sequence stars (black data points).  The red data points are the sub-sample of transiting exoplanets with zero-albedo equilibrium temperatures below 1000 K.  The single outlier is Kep-447b (see text).  Bottom panel: the same sub-sample, but color-coded by mass.  The red, green, blue and cyan points are for $<0.8 M_{\rm J}$, $0.8 < M < 1.2 M_{\rm J}$, $1.2 < M < 2 M_{\rm J}$ and $>2 M_{\rm J}$, respectively.  Data taken from \texttt{http://www.exoplanets.org} \citep{han14}.}
%\vspace{0.1in}
\label{fig:radii}
\end{figure}

\begin{figure*}%[!h]
\begin{center}
\vspace{-0.1in}
\includegraphics[width=2.\columnwidth,trim = {3cm 5cm 3cm 2cm}]{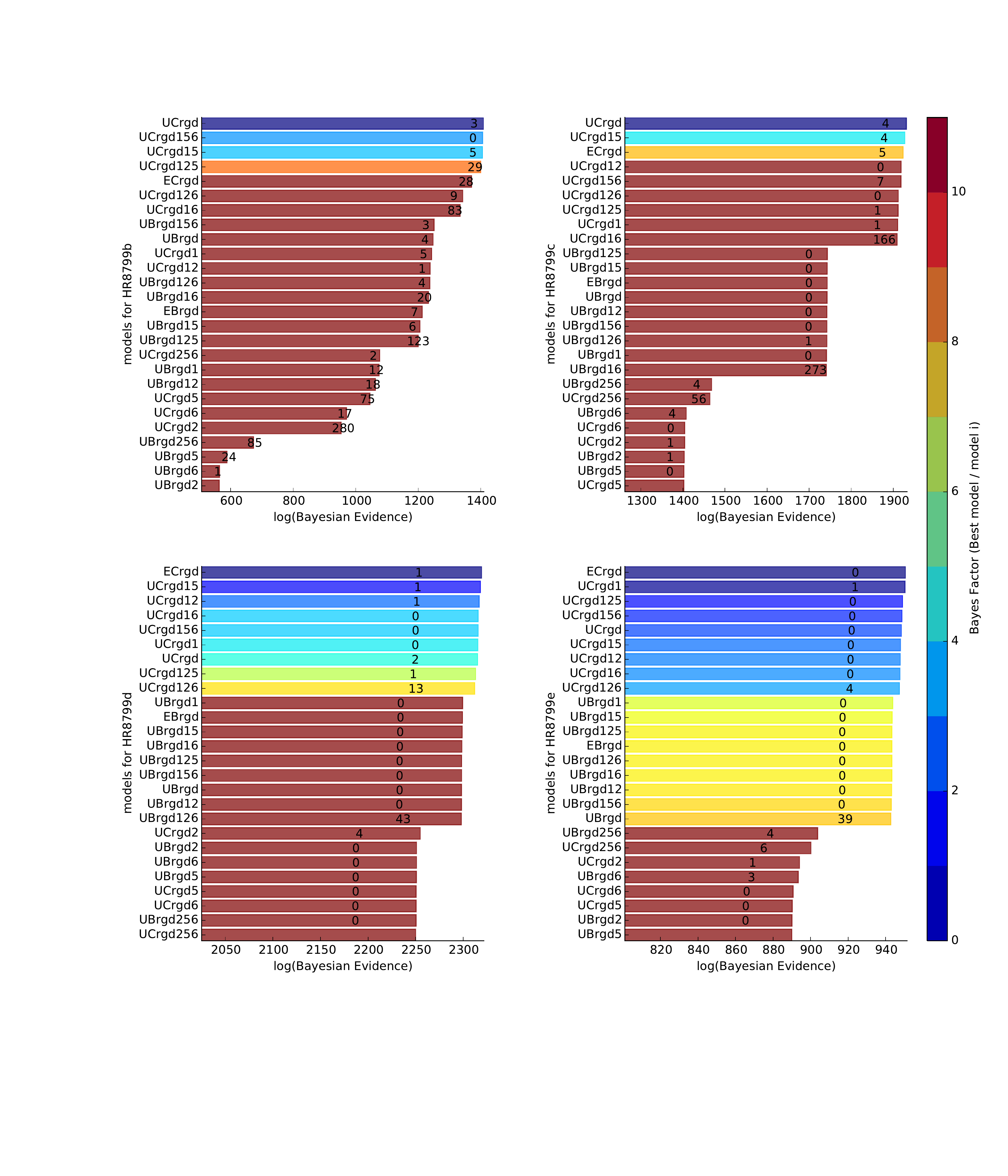}
\end{center}
\vspace{-0.2in}
\caption{ Bayes factors from a suite of models for each HR 8799 exoplanet.  See Table \ref{tab:models} for an explanation of the shorthand notation used to mark each model.  All of the models assume Gaussian priors on $R$ and $\log{g}$ (see Table 2).  For HR 8799b and c, the Bayesian evidence clearly favors cloudy models with non-equilibrium (unconstrained) chemistry.  For HR 8799e, the lack of K band spectroscopy implies that none of the models are strongly favored.  The number associated with each histogram is the logarithm of the Bayes factor between the model in question and its neighbor below.  The color bar shows the logarithm of the Bayes factor between the model in question and the best model, which is the model placed at the top of each panel.}
%\vspace{0.2in}
\label{fig:evidence}
\end{figure*}

The need for a cloud model is motivated by previous suggestions that the atmospheres of the HR 8799 exoplanets are cloudy \citep{barman11,madhu11,marley12}, and also by the finding that each cloud configuration essentially corresponds to a different mass-radius relationship \citep{burrows11,lee13}.  Our cloud model is based on the notion that, while cloud formation is challenging to model from first principles (e.g., \citealt{helling06}), once clouds do form it is somewhat easier to describe their effects on the synthetic spectrum, since this derives from our knowledge of classical optics and Mie theory \citep{pierrehumbert}.

Following \cite{lee13}, we consider the presence of clouds to add an extra contribution to the optical depth,
\begin{equation}
\Delta \tau_{\rm c} = Q_{\rm ext} \pi r_{\rm c}^2 n_{\rm cloud} ~\Delta z = Q_{\rm ext} \pi r_{\rm c}^2 f_{\rm cloud} \frac{\Delta P}{\bar{m} g},
\end{equation}
where $Q_{\rm ext}$ is the extinction efficiency, $r_{\rm c}$ is the radius of the (spherical) particles, $n_{\rm cloud}$ is the number density of clouds and $\Delta z = \Delta P / n \bar{m} g$ is the spatial thickness of the layer.  The cloud mixing ratio is $f_{\rm cloud} = n_{\rm cloud}/n$ and it is this quantity that we set a prior on (see Table 2 for its range of values). We assume the cloud to be uniformly distributed throughout the atmosphere.

In a departure from the approach of \cite{lee13}, we do not use a specific composition of cloud (e.g., enstatite).  Specifically, we adopt their approximate fitting formula (listed in the appendix of \citealt{lee13} but not used in their analysis),
\begin{equation}
Q_{\rm ext} = \frac{5}{Q_0 x^{-4} + x^{0.2}},
\end{equation}
where $x \equiv 2 \pi r_{\rm c}/\lambda$ and $\lambda$ is the wavelength.  When the particles are small ($x \ll 1$), we recover Rayleigh scattering: $Q_{\rm ext} \propto \lambda^{-4}$.  Large particles ($x \gg 1$) produce a roughly constant $Q_{\rm ext}$.  By contrast, \cite{bs13} assume their clouds to be described by only one number, which is the cloud-top pressure. Their model carries the implicit assumption that the cloud particles are large compared to the range of wavelengths examined.

The dimensionless quantity $Q_0$ serves as a proxy for the cloud composition.  Refractory species (e.g., silicates) have $Q_0 \approx 10$, while volatile species (e.g., ammonia, methane, water) have $Q_0 \approx 40$--80.  By using $Q_0$ as a fitting parameter in the retrieval, we can constrain the composition of the clouds.  The other fitting parameters in our cloud model are $r_{\rm c}$ and $f_{\rm cloud}$.

Since we do not self-consistently treat the cloud physics and gaseous chemistry, the caveat is that our retrieved C/O values are representative of only the gaseous component of the atmosphere.  It is conceivable that the true C/O values, which must account for the material sequestered in the cloud particles, are different.

\subsection{Data Selection: Spectra of HR 8799b, c, d and e}

The spectra and photometric data points of the HR 8799b, c, d and e exoplanets have been taken from \cite{bonnefoy16} and \cite{zurlo16}.  The new SPHERE data were presented in \cite{zurlo16}, while \cite{bonnefoy16} unified all of the previous data of the four exoplanets.  Specifically, we use the data from Figure 4 of \cite{bonnefoy16}.

To compute the flux in a photometric waveband, we simply integrate the synthetic spectrum over the range of wavelengths of the filter and assume a Heaviside function with a value of unity throughout.  Unlike \cite{lee13}, we do not apply filter functions with non-unity values to our synthetic spectrum, because this correction has already been done en route to reporting the observed fluxes in \cite{bonnefoy16} and \cite{zurlo16}.  It is unclear what has been done in previous studies.  \cite{madhu11} display filter functions in their Figure 1, but do not describe whether these filter functions were applied to their synthetic spectra.  \cite{line13} state that, ``For the broadband points we simply integrate the flux from the high-resolution model spectrum with the appropriate filter function for that point," but do not provide quantitative descriptions of their filter functions.  It cannot be ruled out that these filter functions have values of unity throughout.  For the spectroscopic data points, we do not convolve the synthetic spectrum with the instrument's response function, because the impact is minor for low-resolution spectra.

\section{Tests}
\label{sect:tests}

\begin{figure}%[!h]
\begin{center}
%\vspace{-0.2in}
\includegraphics[width=1.\columnwidth]{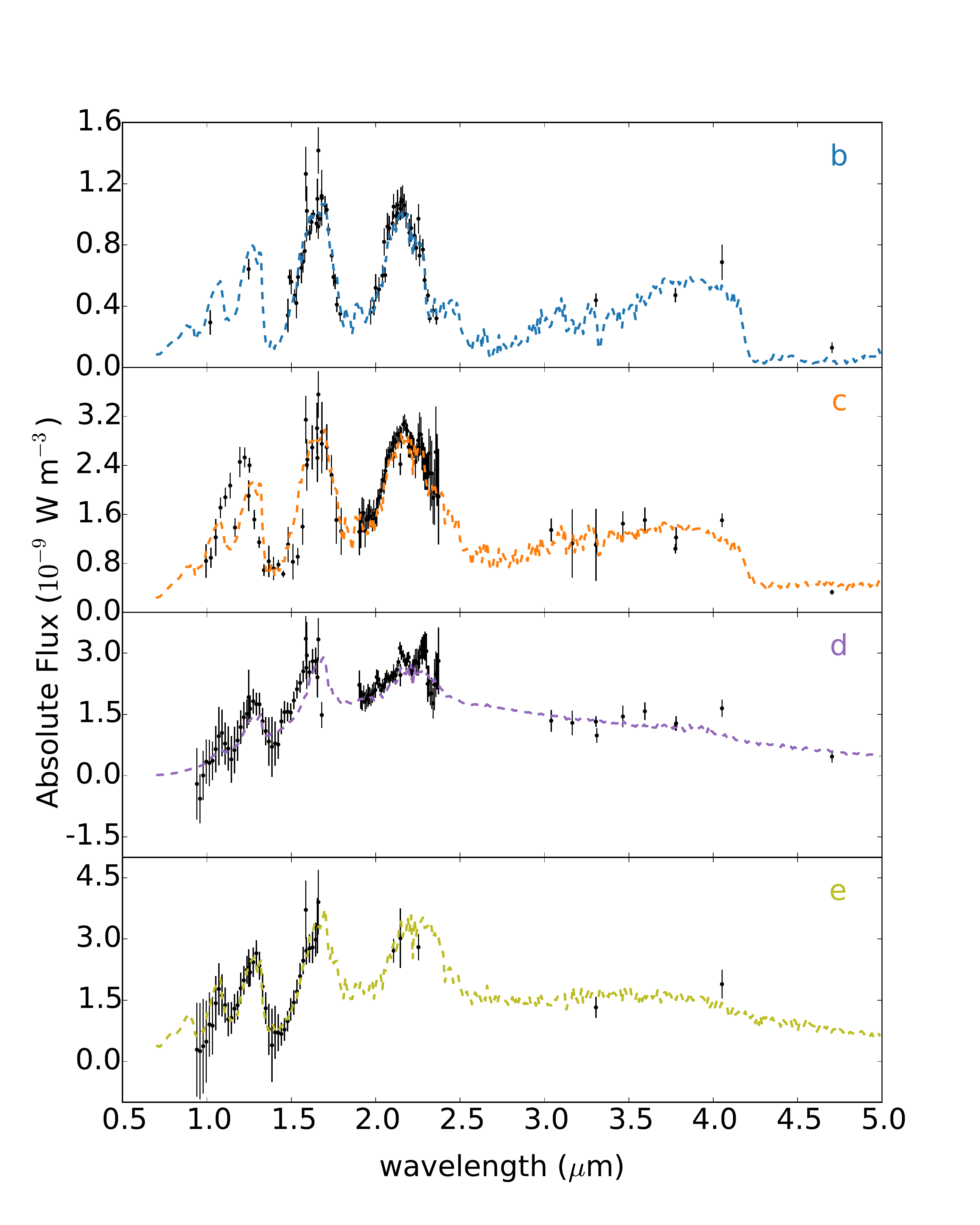}
\includegraphics[width=1.\columnwidth]{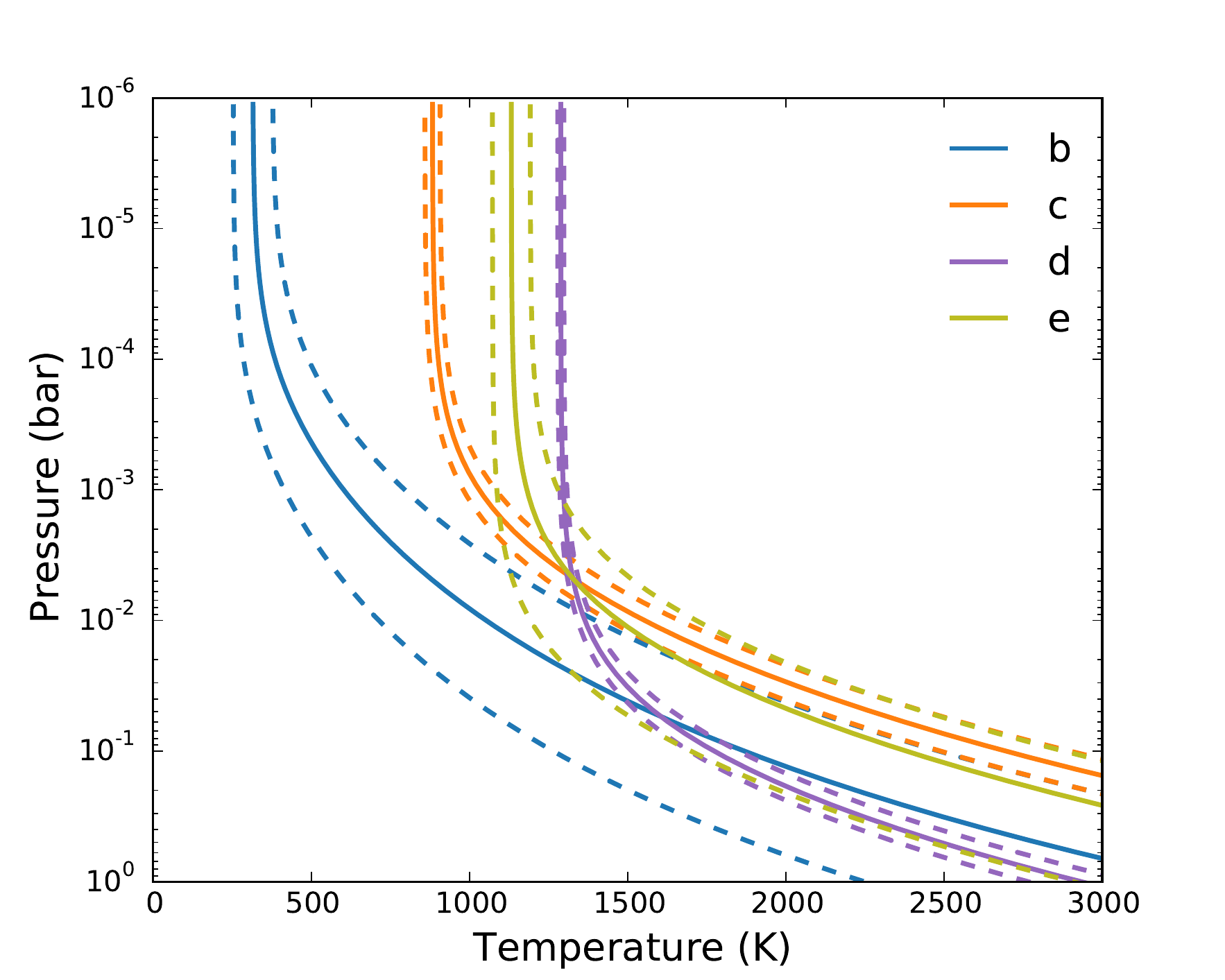}
\end{center}
%\vspace{-0.5in}
\caption{Best-fit spectra and temperature-pressure profiles for HR 8799b, c, d and e.}
%\vspace{0.2in}
\label{fig:spectra}
\end{figure}

Before analyzing the measured spectra of the HR 8799b, c, d and e directly imaged exoplanets, we subject \texttt{HELIOS-R} to a battery of tests.

\subsection{Number of Atmospheric Layers}

The number of layers used in a one-dimensional model atmosphere is a critical but often overlooked or unexplored detail.  We wish to quantify the mean and maximum errors associated with assuming a specific number of model layers.  We use the measured spectrum of HR 8799b as an illustration.  We consider an ensemble of $10^3$ cloudfree models with unconstrained chemistry. For each model, we randomly select our parameter values: 2 parameters for the temperature-pressure profile, 4 parameters for the mixing ratios and 1 for the surface gravity.  The range of parameter values used is listed in Table 2.  No model selection is performed for this test.  We consider forward models with both isothermal and non-isothermal layers.  For the latter, we use equation (B6) of \cite{hml14}.  

For each of the $10^3$ models, the spectrum computed with 10,000 non-isothermal layers is used as a reference.  We then compute coarser models with between 10 and 8000 isothermal or non-isothermal layers and calculate the fractional error on the synthetic spectrum compared to the reference model.  In Figure \ref{fig:layers}, we show both the mean and maximum errors associated with the synthetic spectrum.  With 100 layers, we see that models with isothermal and non-isothermal layers have the same mean and maximum errors of about 2.5\% and 8\%, respectively.  For the rest of the paper, we will use 100 isothermal layers.  For comparison, \cite{ms09}, \cite{lee13} and \cite{line13} used 100, 43 and 90 layers, respectively.

\subsection{Validating the Forward Model}

We previously developed a self-consistent radiative transfer code named \texttt{HELIOS}, which solves the radiative transfer equation in tandem with the first law of thermodynamics to obtain one-dimensional model atmospheres in radiative equilibrium \citep{malik17}.  \texttt{HELIOS} was validated against the radiative transfer model of \cite{mf10}.  In the limit of pure absorption, we also demonstrated that the two-stream and exact solutions produce excellent agreement if the diffusivity factor is set to 2 \citep{hml14,malik17}.  

The forward model of \texttt{HELIOS} uses the same equation as \texttt{HELIOS-R}, but was implemented independently by the first author of each study.  Here, we compare the forward models of \texttt{HELIOS-R} and \texttt{HELIOS} to verify that our implementation is bug-free.  In Figure \ref{fig:validation}, we constructed a cross section function consisting purely of water and used the k-distribution method to compute the fluxes.  \cite{malik17} used instead an opacity function, but also consisting purely of water and using the k-distribution method as well.  The k-distribution tables were constructed using a resolution of $10^{-5}$ cm$^{-1}$ evenly distributed across wavenumber (not shown).  Other assumptions include a hydrogen-dominated atmosphere ($\mu=2$), a water mixing ratio of $10^{-3}$ and a surface gravity of $\log{g} = 3.3$ in cgs units ( $\approx $19.5 m s$^{-2}$).  We then assumed an input temperature-pressure profile, as shown in the insert of Figure \ref{fig:validation}, in tandem with the k-distribution tables to compute the synthetic spectrum using both \texttt{HELIOS-R} and \texttt{HELIOS}.  The excellent agreement validates our implementation of the forward model.

\subsection{Retrieval on a Mock Dataset}
\label{subsect:mock}

A useful test is to create a mock dataset, where we know what the ``ground truth" is concerning the synthetic spectrum, temperature-pressure profile, molecular abundances, surface gravity, etc.  We assume a cloudfree model with unconstrained chemistry, which has the following input parameters:
\begin{equation}
\begin{split}
&X_{\rm CO}= X_{\rm CO_2}=X_{\rm CH_4}=X_{\rm H_2O}=10^{-4}, \\
&R = 1.2 R_{\rm J}, \log{g} = 4.0 \mbox{ (cgs)}, \\
&T_{\rm int} = 700 \mbox{ K}, ~\kappa_0 = 2.9 \times 10^{-4} \mbox{ m}^2 \mbox{ kg}^{-1}, 
\end{split}
\end{equation}
where $R_{\rm J}$ is the radius of Jupiter.  Using this setup, we create 3 mock datasets: a full mock spectrum from 0.7--5$\mu$m with 0.01 $\mu$m resolution, HR8799b-like and HR8799e-like data coverage. We assume this mock object to be located $d=10$ pc away.

Such a test serves three purposes.  First, if $R$ and $g$ are fixed to their input values (and excluded from being fitting parameters in the retrieval), then it is a test of the ability of our nested sampling algorithm to correctly recover the molecular abundances and temperature-pressure profile.  Second, if we now include $g$ and $R$ as fitting parameters, it allows us to study the degeneracies associated with our ignorance of the surface gravity and/or radius.  Third, by adapting and degrading the mock spectrum to the data resolution and spectral coverage of HR 8799b and HR 8799e, we may study the effects of incomplete or sparse data on the retrieved molecular abundances.  The key difference between the currently available data for HR 8799b and HR 8799e is that the latter does not have K band spectroscopy.

Figure \ref{fig:mock} shows the outcomes of these tests. When $R$ and $g$ are fixed to their input values, \texttt{HELIOS-R} correctly recovers the input values of the mixing ratios and $T$-$P$ profile parameters from the full mock spectrum (first row, first column).  Surprisingly, our ability to recover these input values appear to be insensitive to whether the mock spectrum is degraded or not (first row, second and third columns), if $R$ and $g$ are known.  

When the radius is implemented as a uniform prior, its value is correctly recovered, although the posterior distributions of the other fitting parameters become a little broader (third row of Figure \ref{fig:mock}).  With HR 8799e-like data coverage, we see clear signs of degeneracies being introduced into the posterior distributions.  It suggests that the K band spectrum contains important information on the molecular abundances, an issue we will explore further in \S\ref{subsect:main}.

Allowing the surface gravity to be a fitting parameter has more serious consequences, as it introduces degeneracies into all of the other fitting parameters (second row of Figure \ref{fig:mock}).  Even full data coverage does not lift these degeneracies (second row, first column).  It suggests that an informative prior needs to be set on the surface gravity. 
  
Surprisingly, the retrieved posterior distribution of C/O appears to be robust to the different model assumptions (Figure \ref{fig:mock_co}).  It suggests that the C/O is a robust outcome of the retrieval.

Overall, these exercises teach us that wavelength coverage and spectral resolution are generally not as important as knowledge of the surface gravity, although the K band spectrum appears to encode crucial information on the molecular abundances.  In \S\ref{subsect:priors}, we will argue for setting Gaussian priors on $\log{g}$ as well as $R$ when analyzing real data from the HR 8799 exoplanets.

\section{Results}
\label{sect:results}

\subsection{Setting Priors on Radius and Surface Gravity}
\label{subsect:priors}

The strongest demonstration of why our assumptions for the prior distributions of input parameters are important comes from examining a model where the radius and surface gravity are implemented as uniform priors in the retrieval.  Specifically, we perform a retrieval on the measured spectrum of HR 8799b using model UBrg in Figure \ref{fig:radius}, where $R$ and $g$ are specified as uniform priors.  We see that the retrieved solution is $R \approx 0.5 R_{\rm J}$, which is physically unreasonable.  The surface gravity takes on unphysical values of $\log{g} \approx 5.5$--6.  As we have learned from the mock-retrieval exercises in \S\ref{subsect:mock}, these difficulties stem from specifying the radius and surface gravity as unconstrained fitting parameters.

We now discuss why the values for $R$ are physically unreasonable.  There are indirect arguments for why retrieved solutions with radii well below a Jupiter radius should be rejected.  First, brown dwarfs and low-mass stars with masses between 20 and 100 Jupiter masses have transit radii that are at least $0.8 R_{\rm J}$ (see \citealt{burrows11} and references therein), including CoRoT-3b, which is a low-mass brown dwarf with a dynamical mass of $M=21.66 \pm 1.0 ~M_{\rm J}$ and a transit radius of $R = 1.01 \pm 0.07 ~R_{\rm J}$ \citep{deleuil08}.  Second, a review of the data for all of the transiting Jupiter-like exoplanets reveals also that objects with radii below $0.8 R_{\rm J}$ do not exist (Figure \ref{fig:radii}).  When a cut is made to only include objects with zero-albedo equilibrium temperatures below 1000 K (to exclude objects that are ``inflated" by some unknown mechanism related to stellar heating, e.g., \citealt{ds11}), we find that the radii are bound between 0.8 and $1.2 R_{\rm J}$.  The single outlier with $R = 1.65^{+0.59}_{-0.56} ~R_{\rm J}$ is Kep-447b, which has an extremely grazing transit \citep{lillo15} that may render its radius measurement unreliable.  Third, objects with a mass of Jupiter (or higher) are partially degenerate and it is theoretically challenging to get their radius to be less than that of Jupiter's \citep{burrows93}.

While it may be tempting to fix our model radius at between 0.8 and $1.2 ~R_{\rm J}$, we should be reminded of the fact that these radii are measured for $>1$ Gyr-old objects, whereas the HR 8799 exoplanets are estimated to be $\sim 10$--100 Myr old.  Guided by evolutionary models \citep{mordasini12,sb12}, we set $R = 1.2 \pm 0.1 ~R_{\rm J}$ as a Gaussian prior of our retrievals.  The uncertainty of $0.1 R_{\rm J}$ is the full-width at half-maximum of the Gaussian.  We note that \cite{moses16} assume a fixed value of $R=1.2 ~R_{\rm J}$ for their self-consistent model of HR 8799b.

The bottom panel of Figure \ref{fig:radii} is also revealing, as it shows the measured surface gravities of transiting Jupiter-sized exoplanets to be hovering around $\log{g} \approx 4$ for objects with masses of $M>2 M_{\rm J}$, where $M_{\rm J}$ is the mass of Jupiter.  Since we expect the HR 8799 exoplanets to have radii that are slightly larger than Jupiter's, we expect their surface gravities to also be $\log{g} \approx 4$.  Surface gravities of $\log{g} \approx 4.5$--5.0 are only appropriate when one crosses over into the brown dwarf regime ($\gtrsim 13 M_{\rm J}$), e.g., CoRoT-3b has $\log{g} = 4.72 \pm 0.07$.  The photometric masses of HR 8799b, c, d and e are less than half that of CoRoT-3b \citep{marois08,marois10}.  Based on the evolutionary calculations of \cite{mc14}, who estimated $M \approx 4$--13 $M_{\rm J}$ for the HR 8799 exoplanets, we set a Gaussian prior of $\log{g} = 4.1 \pm 0.3$ on the surface gravity (taking into account $R = 1.2 \pm 0.01 ~R_{\rm J}$).  This range of surface gravities is somewhat higher than the $\log{g} = 3.5 \pm 0.5$ values considered by \cite{barman15}.

In summary, we find that what we assume for the prior distributions of the input quantities is critical to the outcome of the retrieval.  Uniform or log-uniform priors may not always be the best choice as they may lead to unphysical or even nonsensical outcomes.  Gaussian priors are better choices in these instances, but only when they are guided by physics.  We find our retrievals to be physically meaningful only when Gaussian priors are set on the radius and surface gravity, which is a departure from the HR 8799b analysis of, e.g., \cite{lee13}.

\subsection{Model Selection Using Bayesian Evidence}

Traditionally, model selection is performed manually by the modeler or theorist.  One starts with a set of assumptions, computes forward and arrives at a prediction for the thermal structure and synthetic spectrum.  These assumptions include chemical equilibrium or disequilibrium, a value for the strength of atmospheric mixing, the number of atoms and molecules included in the model, the metallicity and C/O, etc.  Other assumptions are more closely related to technique, e.g., the approximate or limiting form of the radiative transfer equation being solved.  

Like all of the other previous studies involving both forward modeling and retrieval, we inevitably make a set of both physical and technical assumptions.  However, we use our nested sampling approach to go a step further: we compute the Bayesian evidence for models with and without equilibrium chemistry.  We then compare them in order to formally quantify whether equilibrium chemistry is a warranted assumption.  Instead of assuming a fixed set of cloud parameters for each retrieval, as was done by \cite{lee13}, we allow our cloud model to be part of the retrieval and also compare its Bayesian evidence to a retrieval that assumes a cloudfree atmosphere.  In these ways, we allow model selection based on the Bayesian evidence to inform us on whether the atmosphere is cloudy or cloudfree and in chemical equilibrium or disequilibrium.

Figure \ref{fig:evidence} shows a montage of all of the models tested for all four HR 8799 exoplanets.  Table \ref{tab:models} explains what the labels of the models correspond to.  \textit{For HR 8799b, c and d, we see that the Bayesian evidence favors model atmospheres that are not in chemical equilibrium and are cloudy.}  For HR 8799e, the relative lack of data, compared to the other HR 8799 exoplanets, means that we are unable to strongly select between the different models.

Figure \ref{fig:spectra} shows the best-fit spectra.  Our retrieval procedure generally manages to find good fits to the data, except for the band-head near 1 $\mu$m for HR 8799c.  We speculate that this mismatch could be due to the influence of an additional molecule we have not included in our analysis, but we deem it beyond the scope of the present paper to identify it.  In the Appendix, Figure \ref{fig:spectra2} elucidates the effects of using ExoMol methane and water versus HITRAN methane and HITEMP water.

For the rest of this paper, we will discuss the retrieved properties of the HR 8799b, c and d exoplanets based on the best-fit models only.  For HR 8799e, we will discuss results from the model with equilibrium chemistry and that includes all four molecules in the retrieval.  In Figures \ref{fig:retrieval1}, \ref{fig:retrieval2}, \ref{fig:retrieval3} and \ref{fig:retrieval4} of Appendix \ref{append:full}, we provide the full posterior distributions of the best models for all exoplanets for completeness.  

\subsection{Retrieving the Cloud Properties and Inferring $K_{\rm zz}$}

\begin{figure}%[!h]
\begin{center}
%\vspace{0.2in}
\includegraphics[width=\columnwidth]{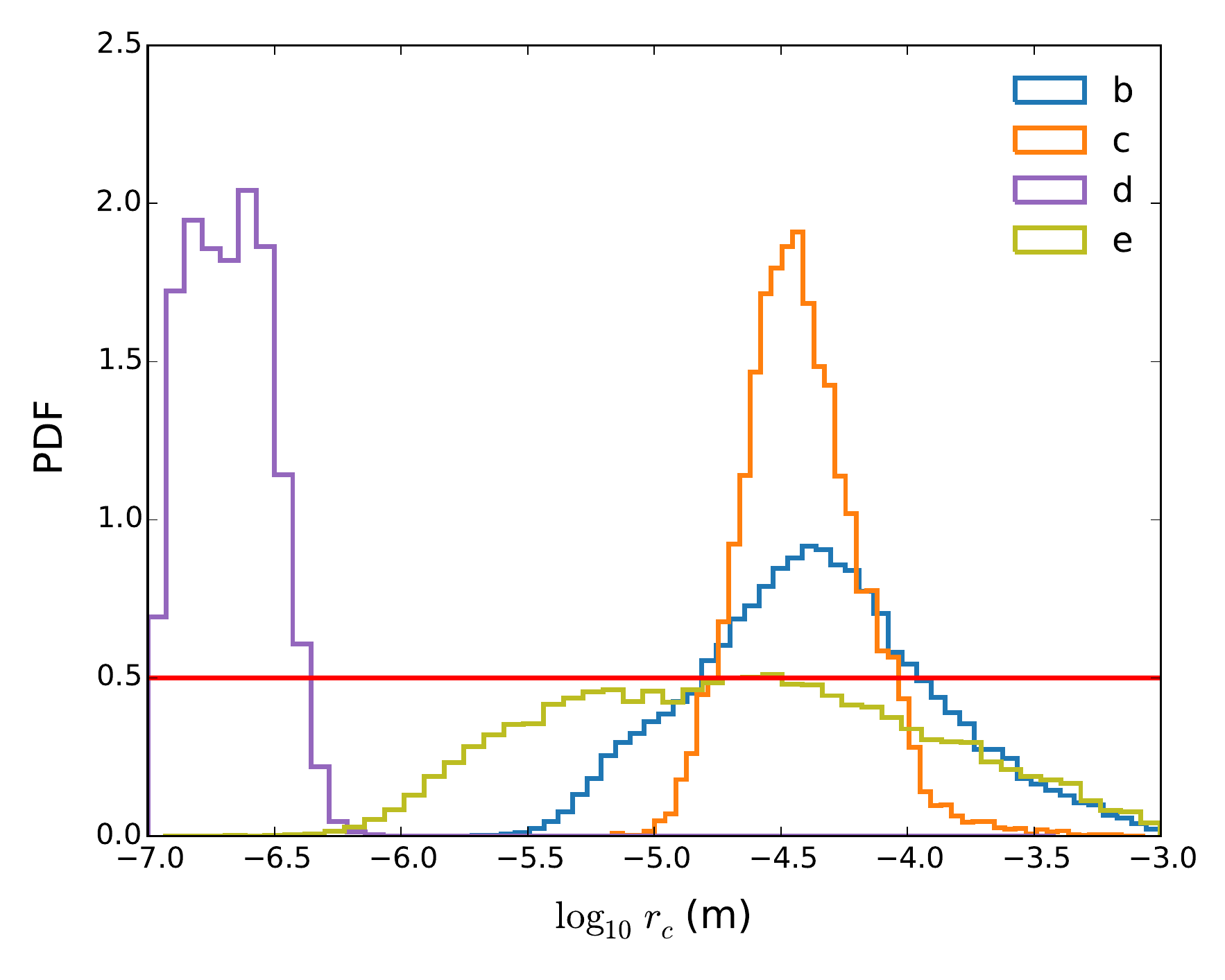}
\includegraphics[width=\columnwidth]{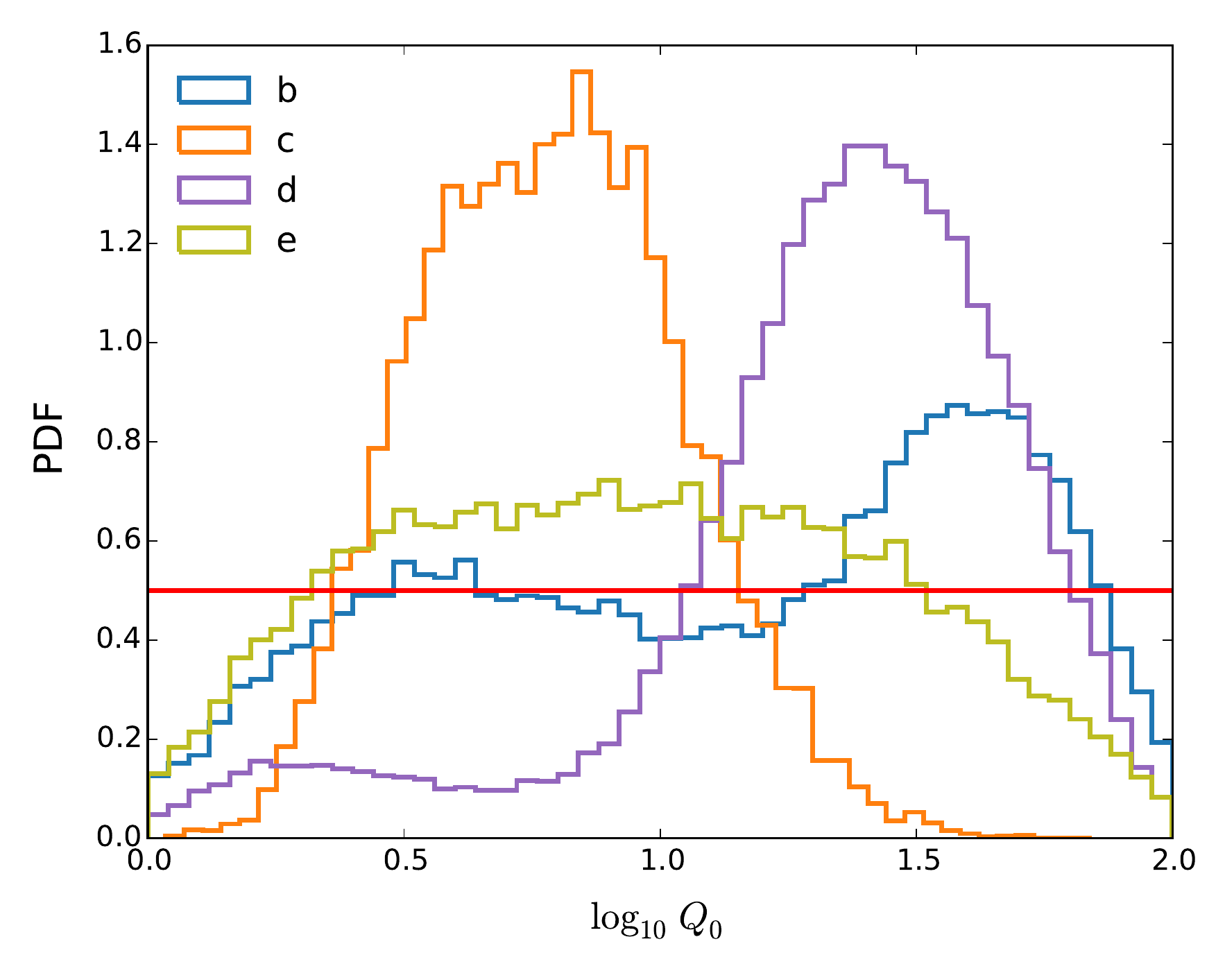}
\includegraphics[width=1.07\columnwidth]{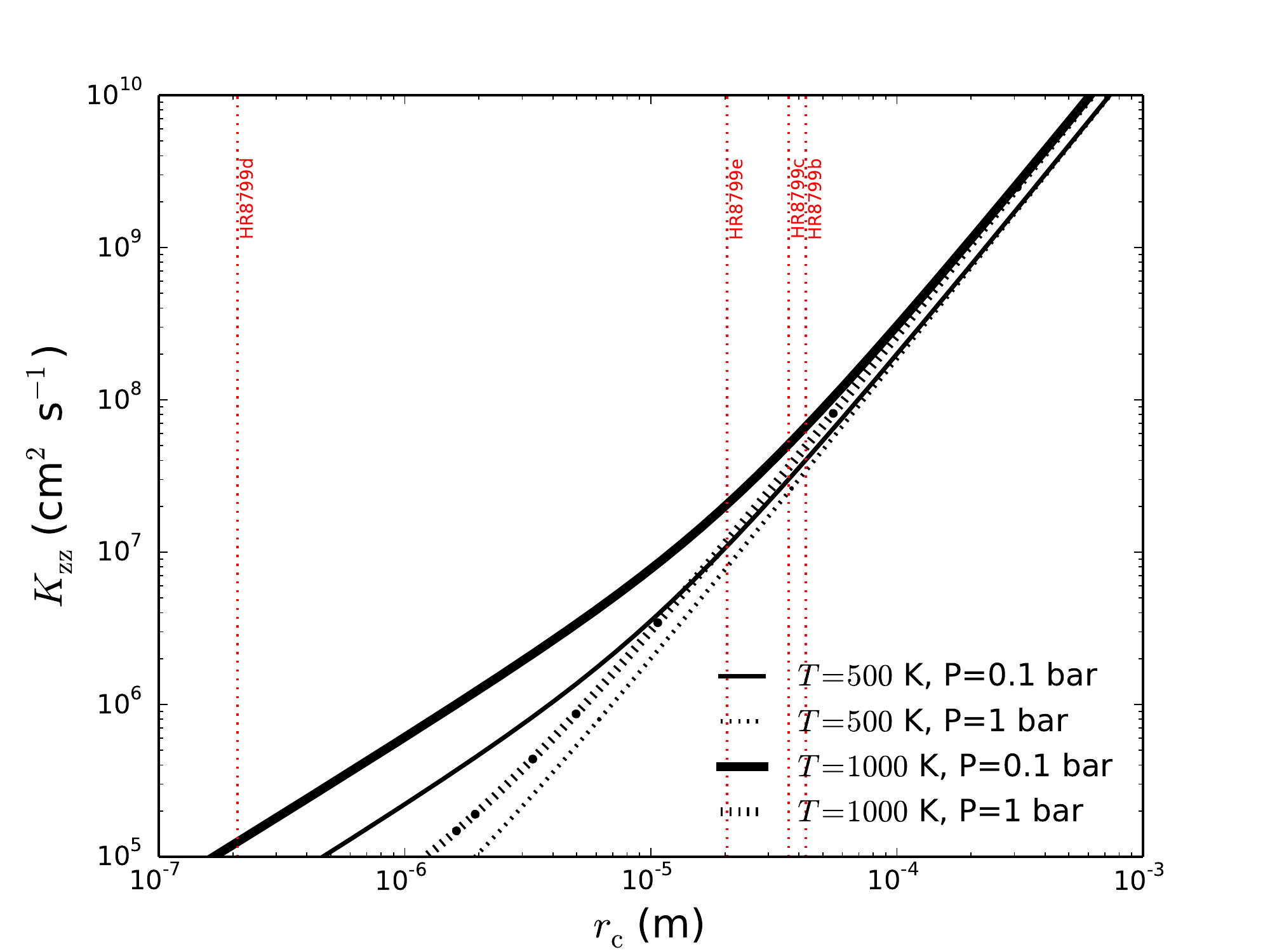}
\end{center}
%\vspace{-0.2in}
\caption{The top and middle panels show the retrieved cloud particle radius and composition parameter for the HR 8799b, c, d and e directly imaged exoplanets, where the flat line is the prior.  The bottom panel shows the eddy diffusion coefficient and demonstrates that its inferred value corresponding to the peak value of the posterior distribution for $r_{\rm c}$ are not unreasonable and broadly consistent with the assumed $K_{\rm zz}$ values in previous studies.  PDF stands for ``probability density function".}
%\vspace{0.1in}
\label{fig:clouds}
\end{figure}

Figure \ref{fig:clouds} shows the retrieved posterior distributions of the cloud particle radius ($r_{\rm c}$) and composition parameter ($Q_0$).  Unsurprisingly, the retrieved values of $Q_0$ span a broad enough range (3 to 4 orders of magnitude) that they are uninformative with regards to distinguishing between different compositions, consistent with the expectation that the absorption and scattering properties of the cloud are mainly determined by the particle size and less by the composition \citep{hd13}.

The inferred values of $r_{\rm c}$ span a broad range and lie between about 1 and 100 $\mu$m.  The presence of these cloud particles implies that they are being held aloft by atmospheric motion.  Since these exoplanets are not being heavily irradiated (unlike for hot Jupiters), we can safely assume that the underlying mechanism driving this motion is convection \citep{burrows97,chabrier00,baraffe02} and estimate approximate values for the associated ``eddy diffusion coefficient", which we denote by $K_{\rm zz}$.  We use equations (15) and (17) of \cite{spiegel09}, as well as equations (6) and (8) of \cite{hd13}, to calculate the terminal speed associated with a particle of radius $r_{\rm c}$, which we denote by $v_{\rm terminal}$.  The eddy diffusion coefficient is roughly
\begin{equation}
K_{\rm zz} \sim 0.1 v_{\rm terminal} H,
\end{equation}
where the pressure scale height is $H=k_{\rm B}T / \tilde{m} g$ and $k_{\rm B}$ is the Boltzmann constant.  We follow the prescription of \cite{smith98} and use $0.1H$ as the characteristic length scale, which is more conservative than what was assumed in \cite{lee13}.  We note that the preceding expression for $K_{\rm zz}$ has no dependence on $g$, as it appears in the numerator of $v_{\rm terminal}$ and the denominator of $H$.  We assume the intrinsic density of the particles to be 3 g cm$^{-3}$.

In Figure \ref{fig:clouds}, we see that $K_{\rm zz}$ spans a broad range of values from $\sim 10^5$ cm$^2$ s$^{-1}$ to $\sim 10^{10}$ cm$^2$ s$^{-1}$ as $r_{\rm c}$ increases from 1 $\mu$m to 1 mm.  The deviation in the curves between $P=0.1$ and 1 bar arises from the Cunningham-Millikan-Davies ``slip factor correction" kicking in when the mean free path for collisions between the hydrogen molecules becomes comparable to the cloud particle radius.  If we place the retrieved values of $r_{\rm c}$ corresponding to the peak of each posterior distribution on the plot, we infer $K_{\rm zz} \sim 10^5$--$10^8$ cm$^2$ s$^{-1}$, in agreement with \cite{barman15}.  \cite{madhu11} assume $K_{\rm zz} = 10^2$--$10^6$ cm$^2$ s$^{-1}$, while \cite{barman11} and \cite{marley12} assume $K_{\rm zz} = 10^4$ cm$^2$ s$^{-1}$.

\subsection{Retrieving C/O, C/H and O/H for the HR 8799b, c, d and e Exoplanets and Implications for Planet Formation}
\label{subsect:main}

\begin{figure}%[!h]
\begin{center}
%\vspace{0.2in}
\includegraphics[width=\columnwidth]{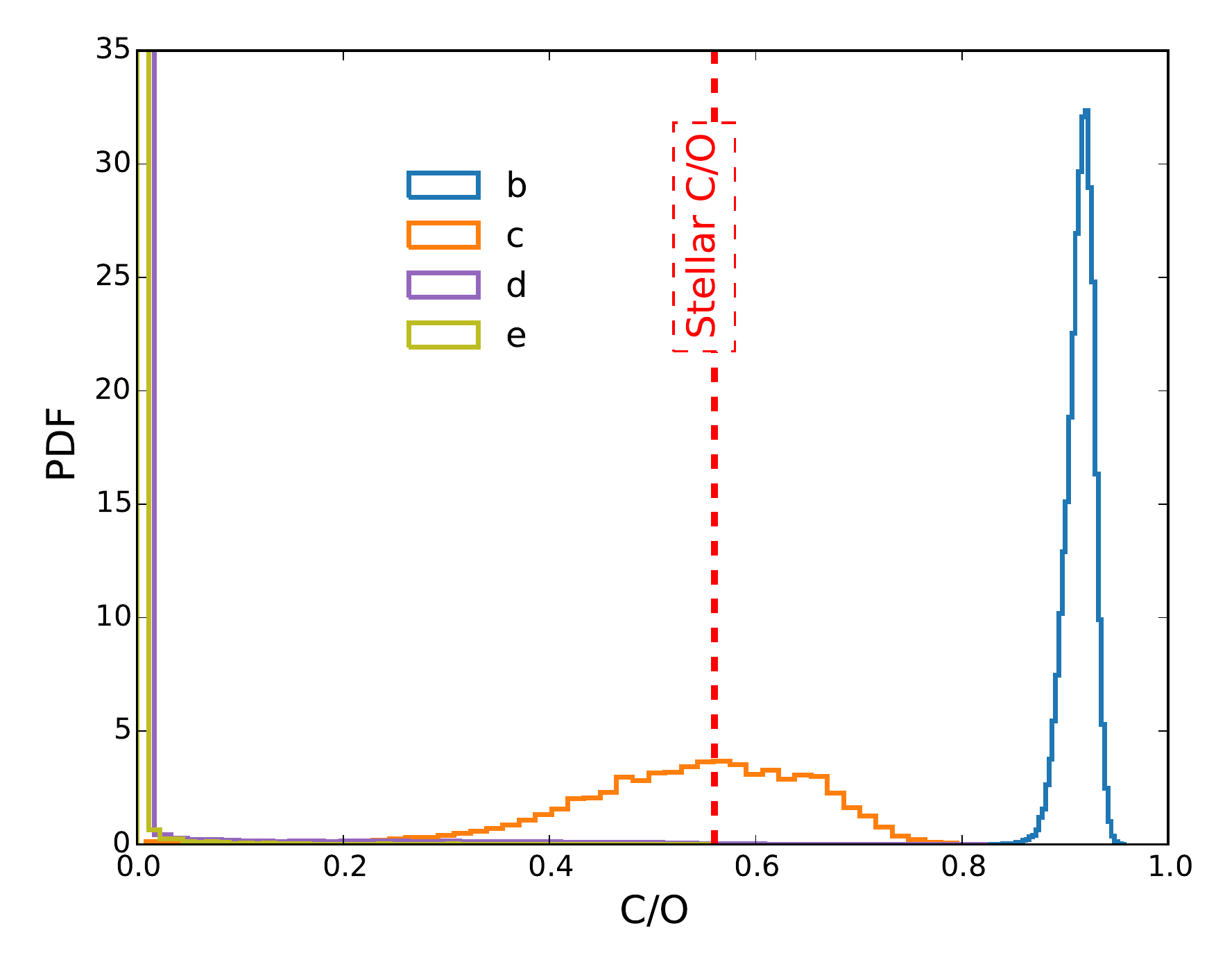}
\end{center}
%\vspace{-0.2in}
\caption{Retrieved C/O values for the HR 8799b, c, d and e directly imaged exoplanets.  The stellar C/O value is about 0.56.  PDF stands for ``probability density function".}
%\vspace{0.1in}
\label{fig:co}
\end{figure}

\begin{figure}%[!h]
\begin{center}
%\vspace{0.2in}
\includegraphics[width=\columnwidth]{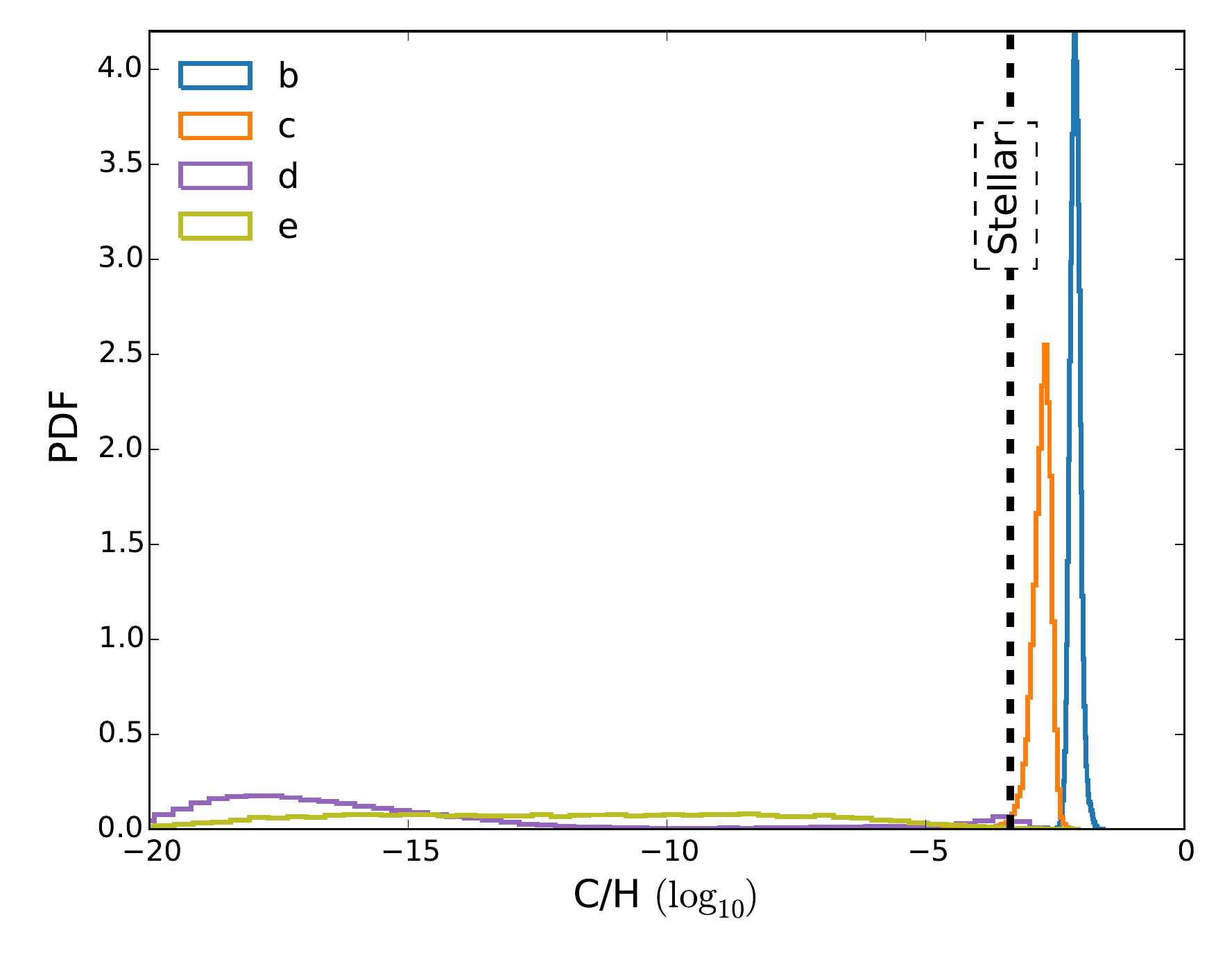}
\end{center}
%\vspace{-0.2in}
\caption{Retrieved C/H values for the HR 8799b, c, d and e directly imaged exoplanets.  The stellar C/H value is about $4.3 \times 10^{-4}$.  PDF stands for ``probability density function".}
%\vspace{0.1in}
\label{fig:ch}
\end{figure}

\begin{figure}%[!h]
\begin{center}
%\vspace{0.2in}
\includegraphics[width=\columnwidth]{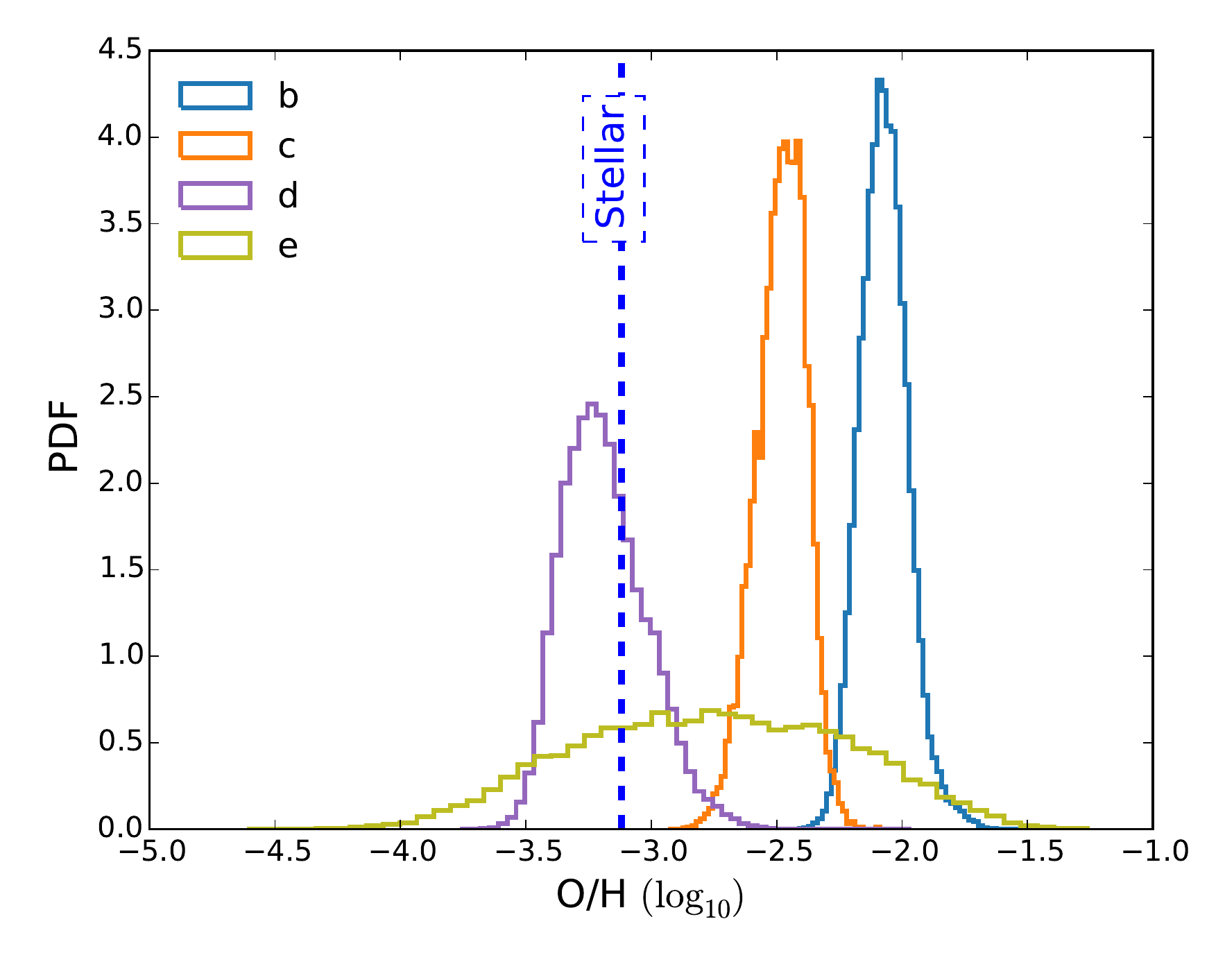}
\end{center}
%\vspace{-0.2in}
\caption{Retrieved O/H values for the HR 8799b, c, d and e directly imaged exoplanets.  The stellar O/H value is about $7.6 \times 10^{-4}$.  PDF stands for ``probability density function".}
%\vspace{0.1in}
\label{fig:oh}
\end{figure}

\subsubsection{The Star of HR 8799}

We refer to the ``metallicity" as the set of elemental abundances with atomic mass numbers that are larger than that of hydrogen and helium.  In our current study, these would be $f_{\rm C} \equiv \mbox{C/H}$ and $f_{\rm O} \equiv \mbox{O/H}$.  For comparison, their values in the solar photosphere are $f_{\rm C} \approx 3 \times 10^{-4}$ and $f_{\rm O} \approx 6 \times 10^{-4}$, such that $\mbox{C/O} \approx 0.5$ \citep{lodders03}.  For the star of the HR 8799 system, \cite{sada06} has found that
\begin{equation}
\mbox{C/H}_\star \approx 4.3 \times 10^{-4}, ~\mbox{O/H}_\star \approx 7.6 \times 10^{-4}, ~\mbox{C/O}_\star \approx 0.56.
\label{eq:star}
\end{equation}

\subsubsection{Retrieved C/O, C/H and O/H Values}

Given the interest in the possibility of carbon-rich exoplanets \citep{gaidos00,ks05}, our retrieval analysis yields the posterior distributions of C/O, C/H and O/H for the atmospheres of HR 8799b, c, d and e in Figures \ref{fig:co}, \ref{fig:ch} and \ref{fig:oh}, respectively, which we then compare to the values for the star listed in equation (\ref{eq:star}).  A caveat is that the retrieved values are only for the gaseous phase and the true C/O ratio may be hidden in a condensed phase such as graphite \citep{moses13}.  The retrieved posterior distributions of C/O and C/H for HR 8799e are not as definitive as for the other three exoplanets, because its K band spectrum has not been measured.

\subsubsection{Locations of Snowlines/Icelines}

\begin{figure}%[!h]
\begin{center}
%\vspace{0.2in}
\includegraphics[width=\columnwidth, trim={2cm 2cm 2cm 1cm}]{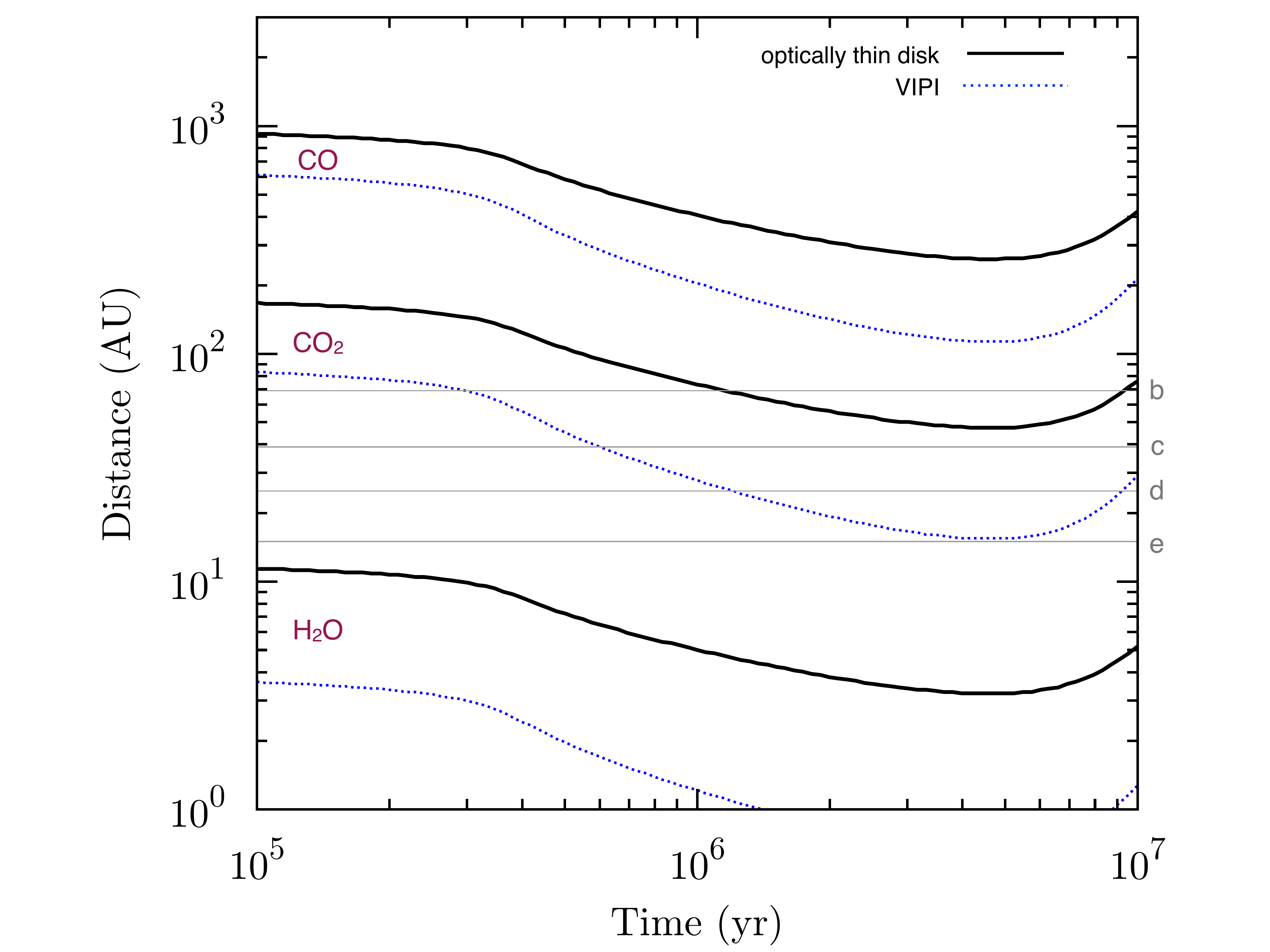}
\end{center}
%\vspace{-0.2in}
\caption{Positions of the CO, CO$_2$ and H$_2$O snowlines or icelines as functions of the stellar age of the HR 8799 system.  The solid curves are calculated assuming an optically thin disk.  The dotted curves are calculations from a vertically isothermal, passively irradiated (VIPI) disk.}
%\vspace{0.1in}
\label{fig:snowlines}
\end{figure}

\cite{konopacky13} have previously estimated that the H$_2$O, CO$_2$ and CO snowlines or icelines are located at about 10, 90 and 600 AU, respectively.  We wish to point out that the iceline locations depend on the formation history of the HR 8799 exoplanets.

In Figure \ref{fig:snowlines}, we show calculations of the locations of the CO, CO$_2$ and H$_2$O icelines as functions of the age of the HR 8799 system.  We consider two scenarios: an optically thin disk and a vertically isothermal, passively irradiated disk.  For the optically thin disk, the temperatures are simply the zero-albedo equilibrium temperatures at a given distance from the star informed by the Pisa stellar evolution models \citep{tog11}.  By ``passively irradiated", we mean that viscous heating associated with turbulence is neglected \citep{cg97}.  Both models consider the evolution of stellar heating as the star ages.  We expect more sophisticated calculations that involve temperature gradients, photoevaporation and viscous heating to produce iceline curves that are intermediate between these two scenarios.  The calculations are shown for $t=10^5$ to $10^7$ years, because this encompasses the gas-clearing phase of the protoplanetary disk.  Curiously, the CO$_2$ iceline sits between different pairs of HR 8799 exoplanets as its location evolves during the gas-clearing phase ($t \sim 10^6$ years), implying that a variation in the C/O, C/H and O/H values of these exoplanets may be a natural outcome of the planet formation process.

\subsubsection{Implications for Planet Formation}

Our findings have implications for planet formation, if we assume the retrieved C/O, C/H and O/H values to be representative of the bulk composition of each exoplanet.  \cite{oberg11} have previously elucidated the chemical signatures associated with the planet formation mechanism and history of an exoplanet.  If an exoplanet forms by gravitational instability, the zeroth-order expectation is that its C/O, C/H and O/H values mirror that of the star, unless late-time accretion occurred.  This is clearly at odds with our inferred values of C/O, C/H and O/H for the HR 8799b, c, d and e exoplanets.  

In the context of the core accretion formation mechanism, all four exoplanets should have C/O values that are enhanced above stellar, but below unity, if they formed in-situ and in between the water and carbon dioxide snowlines/icelines \citep{oberg11}.  Our retrieved values of C/O for HR 8799b and c are consistent with this scenario, whereas HR 8799d and e have sub-solar C/O values.  \cite{oberg11} have suggested that substellar C/O values are still consistent with core accretion if the late-time accretion of planetesimals has occurred to pollute the atmospheres.  The link between late-time planetesimal accretion and atmospheric composition has been emphasized by \cite{mordasini16}.  The HR 8799b and c exoplanets have super-stellar C/H and O/H values, which suggests that they accreted both carbon- and oxygen-rich ices.  The HR 8799d and e exoplanets, which reside closer to the star, have substellar C/H values but stellar to superstellar O/H values, which suggest the accretion only of oxygen-rich ices.  

Overall, our retrieved values of C/O, C/H and O/H appear to be consistent with the core accretion formation mechanism and inconsistent with gravitational instability without late-time accretion, as has been suggested by, e.g., \cite{kratter10}.

\subsubsection{Why Spectroscopy in the K Band is Crucial}

A lesson we have learned from our analysis is that spectroscopy in the K band is crucial for obtaining meaningful constraints on C/H and C/O, as it affects the ability of the retrieval approach to constrain the abundances of CO and/or CH$_4$.  The lack of K band spectroscopy for HR 8799e hampers our ability to make stronger statements on its C/H and C/O values.  These findings have implications for the design of future instruments on the European Extremely Large Telescope (ELT).  Furthermore, multiple wavebands should be monitored simultaneously in order to detect variability \citep{apai16}.

\section{Discussion}
\label{sect:discussion}

\subsection{Summary and Comparison to Previous Work}

\begin{figure}%[!h]
\begin{center}
%\vspace{0.2in}
\includegraphics[width=\columnwidth]{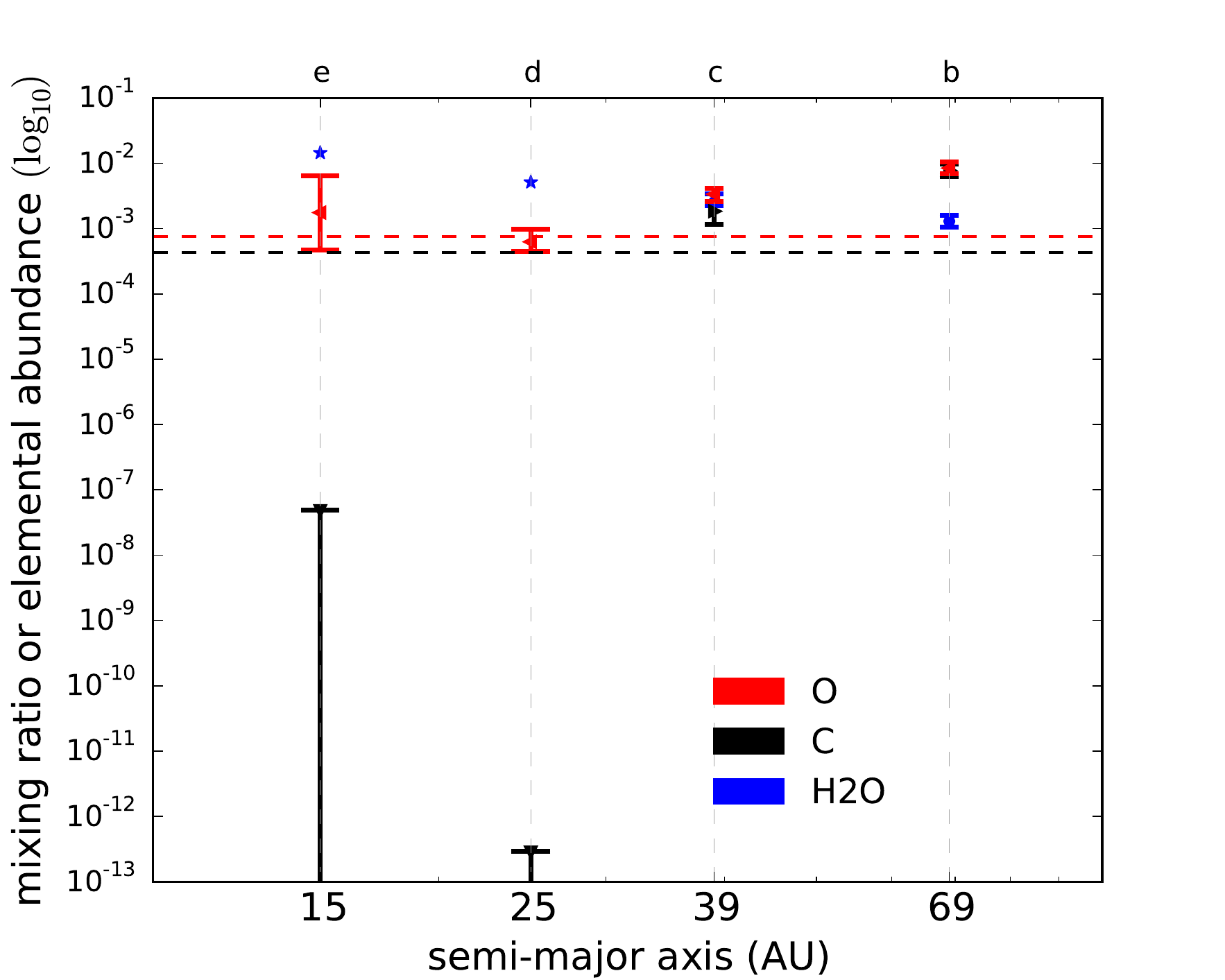}
\includegraphics[width=\columnwidth]{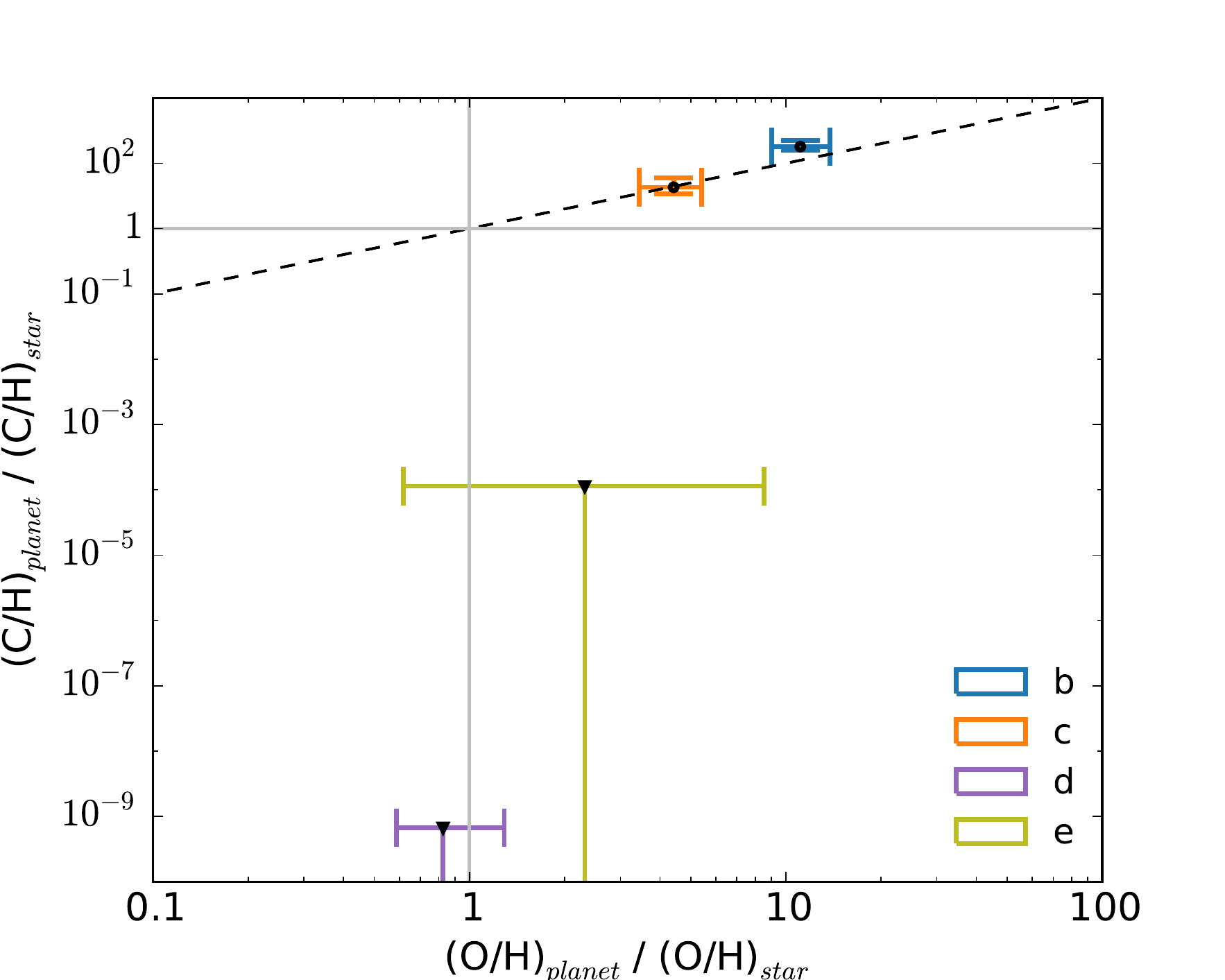}
\end{center}
%\vspace{-0.2in}
\caption{Summary of our main results.  The top panel shows the retrieved water mixing ratios and elemental abundances of carbon and oxygen for all four HR 8799 exoplanets. For HR 8799d and e, we show the water abundance in chemical equilibrium at 1 bar (represented by the blue stars).  For C/H and O/H, we also show the corresponding values of the HR 8799 star (horizontal dashed lines).  The bottom panel shows the exoplanetary elemental abundances normalized to their stellar values with the dashed line denoting parity.}
%\vspace{0.1in}
\label{fig:summary}
\end{figure}

We have presented the complete methodology for a nested sampling atmospheric retrieval code named \texttt{HELIOS-R}, which allows us to insert arbitrary prior distributions of parameters and also compute the full posterior distributions of the retrieved quantities.  In its current implementation, we used analytical formulae for the forward model, temperature-pressure profile and equilibrium chemistry, as well as a customized opacity calculator (\texttt{HELIOS-K}).  By computing the Bayesian evidence, we can compare models that assume equilibrium versus unconstrained chemistry and determine which scenario is favored by the data.

We apply \texttt{HELIOS-R} to the measured spectra of the HR 8799b, c, d and e directly imaged exoplanets.  We find that the outer HR 8799b and c exoplanets are enriched in carbon and have superstellar and stellar C/O values, respectively.  The inner HR 8799d and e exoplanets are diminished in carbon and C/O.  All four exoplanets are possibly enriched in oxygen relative to the star, which is a clear signature of late-time accretion of water-rich planetesimals.  Figure \ref{fig:summary} provides a summary of our findings.  We note that our retrieved water abundances are about 2 to 3 orders of magnitude higher than what was found by \cite{madhu14} for three hot Jupiters, although it should be noted that these authors do not include a cloud model in their retrievals.  The inclusion of a cloud model should worsen the discrepancy between these outcomes.  Our retrieved molecular abundances and C/O for HR 8799b are in broad agreement with \cite{lee13}, despite differences in our retrieval techniques.  Table 3 summarizes the properties of the four exoplanets inferred from the retrieval.

Our conclusions differ somewhat from previous studies, which reach a diversity of conclusions.  \cite{barman11} used self-consistent models to interpret the H and K band spectra of HR 8799b.  They infer $R = 0.75^{+0.17}_{-0.12} R_{\rm J}$ and $M=0.72^{+2.6}_{-0.6} M_{\rm J}$.  We deem this radius value to be unphysical for the reasons described in \S\ref{subsect:priors}.  \cite{marley12} also used self-consistent models and found that if the theoretical interpretation is made of the photometry alone, then the inferred radius for the HR 8799b exoplanet is 1.11 $R_{\rm J}$ but with a surface gravity of $\log{g}=4.75$, considerably higher than the $\log{g}=3.5 \pm 0.5$ value of \cite{barman11}.  \cite{madhu11} used self-consistent models\footnote{Strictly speaking, these are parametric models, because the cloud physics is not treated self-consistently with the gaseous chemistry and is instead parametrized.} with various cloud configurations to conclude that the HR 8799b, c and d exoplanets have masses of 2--12, 6--13 and 3--11 $M_{\rm J}$, respectively, and surface gravities $\log{g} \approx 4$.  In these studies, solar abundance is assumed.  The diversity of reported results from these studies already hint at the difficulty of using photometry and spectroscopy to infer the radius and mass of a directly imaged exoplanet from the traditional use of forward modeling.

\cite{barman15} performed a manual fitting of the H and K band spectra of HR 8799b and HR 8799c.  They first held the CO and CH$_4$ abundances fixed to their solar values, then fitted for the abundance of H$_2$O.  The bandheads involving CO and CH$_4$ are masked or excluded from the fit.  Next, the H$_2$O abundance is held at its best-fit value (and CH$_4$ is again held fixed at its solar value) and the abundance of CO is inferred.  The final step involves fitting for CH$_4$.  Such an approach is plausible as a first step, but does not explore the model degeneracies.  It is likely that the reported value of $\mbox{C/O}=0.61 \pm 0.05$ for HR 8799b has uncertainties that are under-estimated.  \cite{barman15} themselves remark that, ``The various sources of uncertainty in the models (are) not accounted for in the formal mole fraction error-bars."  Building on the work of \cite{barman15}, \cite{moses16} assumed fixed values for the equilibrium temperature, surface gravity, radius, C/O, metallicity and $K_{\rm zz}$, as well as a fixed temperature-pressure profile.  They explored thermo- and photochemical models of HR 8799b and produced synthetic spectra that somewhat match the measured spectrum (see their Figure 14).

\cite{lee13} analyzed the HR 8799b exoplanet and reported super-solar metallicities for their best-fits, consistent with the present study.  They considered two cloud models, where the monodisperse cloud particle radius is fixed manually and not formally included as part of the retrieval.  The cloud composition is also assumed to be enstatite, whereas we have allowed the cloud composition to be part of the retrieval.  The models of \cite{lee13} allowed for $R$ and $g$ to be uniform or log-uniform priors, whereas in the current study we have chosen $R$ and $\log{g}$ to be Gaussian priors.  Somewhat surprisingly, despite these differences, they retrieve a C/O value that is similar to what we find (see Figure \ref{fig:co}).  On the technical side, \cite{lee13} used the \texttt{NEMESIS} code, which implements non-linear optimal estimation (versus the nested sampling algorithm we implemented).  This technique, which is also used by \cite{barstow15}, assumes that the priors and posteriors are Gaussian and is unable to formally perform model selection via Bayesian evidence comparison.  \cite{lee13} also do not consider equilibrium chemistry in their comparison of models.  (See \citealt{line13} for a comparison of these optimization methods.)  Overall, \texttt{HELIOS-R} implements a number of improvements over \texttt{NEMESIS} that are more appropriate for the sparse data regime of exoplanetary atmospheres (compared to the remote sensing data of Solar System objects) and is able to more rigorously explore a broader range of parameter space.

\subsection{Opportunities for Future Work}

There are ample opportunities for future work.  Instead of unconstrained chemistry, disequilibrium chemistry may be described by some form of atmospheric mixing (e.g., eddy diffusion).  More molecules may be added to the analysis, including acetylene, ethylene and hydrogen cyanide, which are known to be spectroscopically active in the infrared at temperatures higher than for the photospheres of the HR 8799 exoplanets.  Ultimately, it is our hope that the collective body of work on atmospheric retrieval will stimulate and connect to work on disk chemistry (e.g., \citealt{cridland16}).  It will also be insightful to train \texttt{HELIOS-R} on a large sample of brown dwarf photometry and spectra, as \cite{line15} have done for two T dwarfs.

\acknowledgments
B.L., M.M., J.M., S.G., M.O. and K.H. thank the Center for Space and Habitability (CSH), the Swiss National Science Foundation (SNSF), the PlanetS National Center of Competence in Research (NCCR) and the MERAC Foundation for partial financial support.

\software{Corner \citep{fm16},
		HELIOS-K \citep{gh15},
		MultiNest \citep{feroz08,feroz09,feroz13},
		PyMultiNest \citep{buchner14},
		VULCAN \citep{tsai17}}
		
\appendix

\section{Analytical Formula for the Exponential Integral of the First Order}
\label{append:e1}

\begin{figure}%[!h]
\begin{center}
\vspace{0.2in}
\includegraphics[width=0.6\columnwidth]{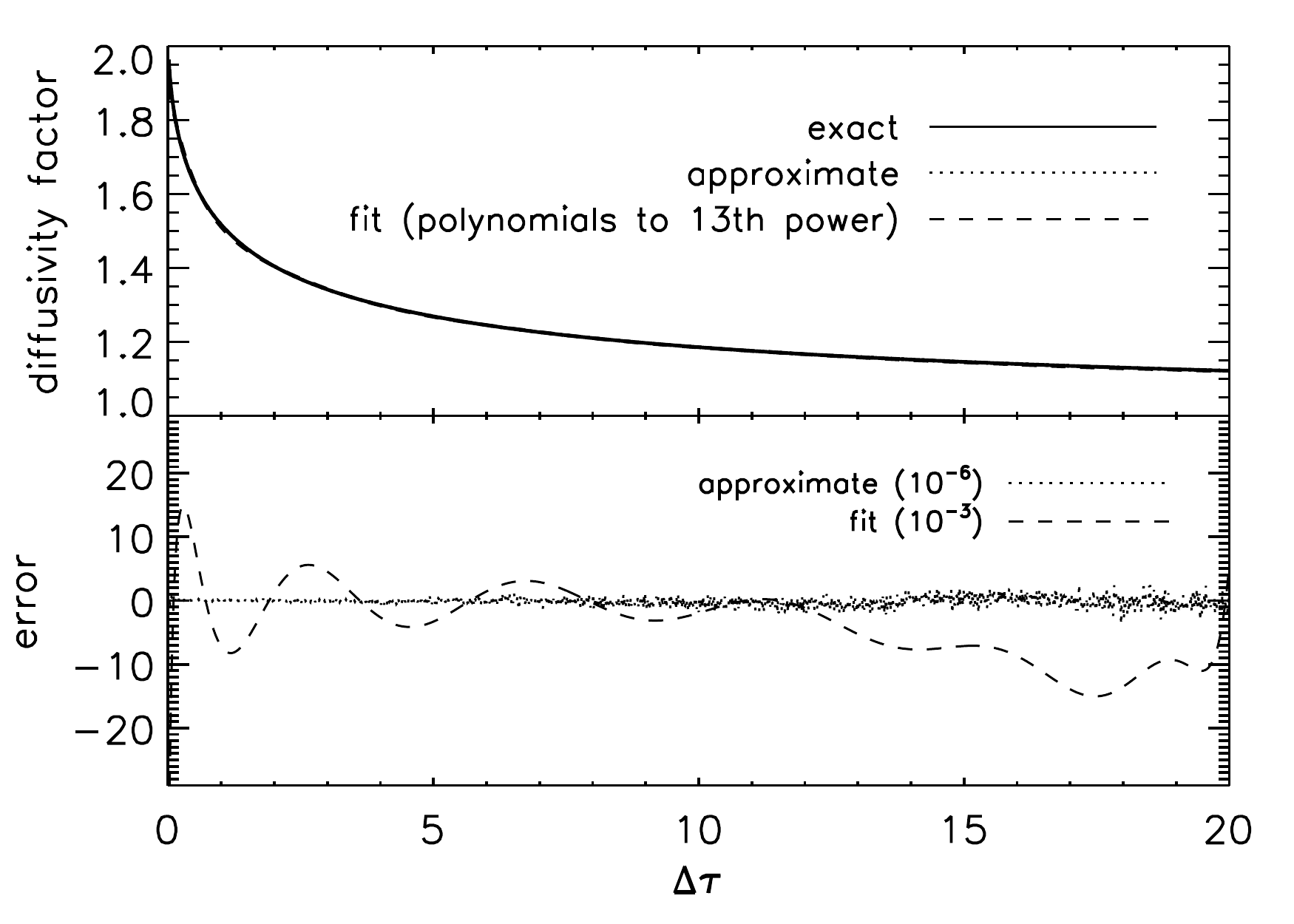}
\end{center}
%\vspace{-0.2in}
\caption{Calculations of the diffusivity factor using a canned routine (``exact") versus those performed using our fitting formulae for ${\cal E}_1$ in equation (\ref{eq:expint}) (``approximate").  The calculation labeled ``fit" is a 13th order polynomial fit to the exact solution, which performs poorly even at the $\sim 10^{-3}$ level.}
%\vspace{0.1in}
\label{fig:diffusivity}
\end{figure}

We may avoid the numerical integration of the exponential integral of the first order by using the approximate, but highly accurate, analytical formulae presented in \cite{abram},
\begin{equation}
{\cal E}_1 =
\begin{cases}
-\ln{\Delta \tau} + \sum_{j=0}^{5} {\cal A}_j \left( \Delta \tau \right)^j, & \Delta \tau \le 1, \\
\left(\Delta \tau\right)^{-1} \exp{\left( - \Delta \tau \right)} \frac{\sum_{j=0}^{4} {\cal B}_j \left( \Delta \tau \right)^{4-j}}{\sum_{j=0}^{4} {\cal C}_j \left( \Delta \tau \right)^{4-j}}, & \mbox{otherwise}.
\end{cases}
\label{eq:expint}
\end{equation}
The fitting coefficients ${\cal A}_j$, ${\cal B}_j$ and ${\cal C}_j$ are given in equations (5.1.53) and (5.1.56) of \cite{abram}, but we reproduce them here for convenience: ${\cal A}_0 = -0.57721566$, ${\cal A}_1 = 0.99999193$, ${\cal A}_2 = -0.24991055$, ${\cal A}_3 = 0.05519968$, ${\cal A}_4 = -0.00976004$ and ${\cal A}_5 = 0.00107857$; ${\cal B}_0 = {\cal C}_0 = 1$, ${\cal B}_1 = 8.5733287401$, ${\cal B}_2 = 18.059016973$, ${\cal B}_3 = 8.6347608925$, ${\cal B}_4 = 0.2677737343$, ${\cal C}_1 = 9.5733223454$, ${\cal C}_2 = 25.6329561486$, ${\cal C}_3 =  21.0996530827$ and ${\cal C}_4 = 3.9584969228$.  As originally stated by \cite{abram}, the formula involving ${\cal A}_j$ has a precision better than $2 \times 10^{-7}$, while that involving ${\cal B}_j$ and ${\cal C}_j$ is precise to better than $2 \times 10^{-8}$.  In Figure \ref{fig:diffusivity}, we check these claims by evaluating ${\cal E}_1$ using a canned routine (\texttt{expint} in \texttt{IDL}) and computing the diffusivity factor using
\begin{equation}
{\cal D} = -\frac{1}{\Delta \tau} \ln{\left[\left( 1 - \Delta \tau \right) \exp{\left(-\Delta \tau \right)} + \left( \Delta \tau \right)^2 {\cal E}_1\right]}.
\end{equation}
We label these calculations as ``exact".  The calculations labeled ``approximate" were performed using the fitting formulae in equation (\ref{eq:expint}).  We see that the error is better than $10^{-6}$.  By contrast, a 13th order polynomial fit to the exact solution incurs large errors ($\gtrsim 10^{-3}$).

\section{Full Posterior Distributions for best models of HR 8799b, c, d and e}
\label{append:full}

\begin{figure*}%[!h]
\begin{center}
%\vspace{0.2in}
\includegraphics[width=\columnwidth]{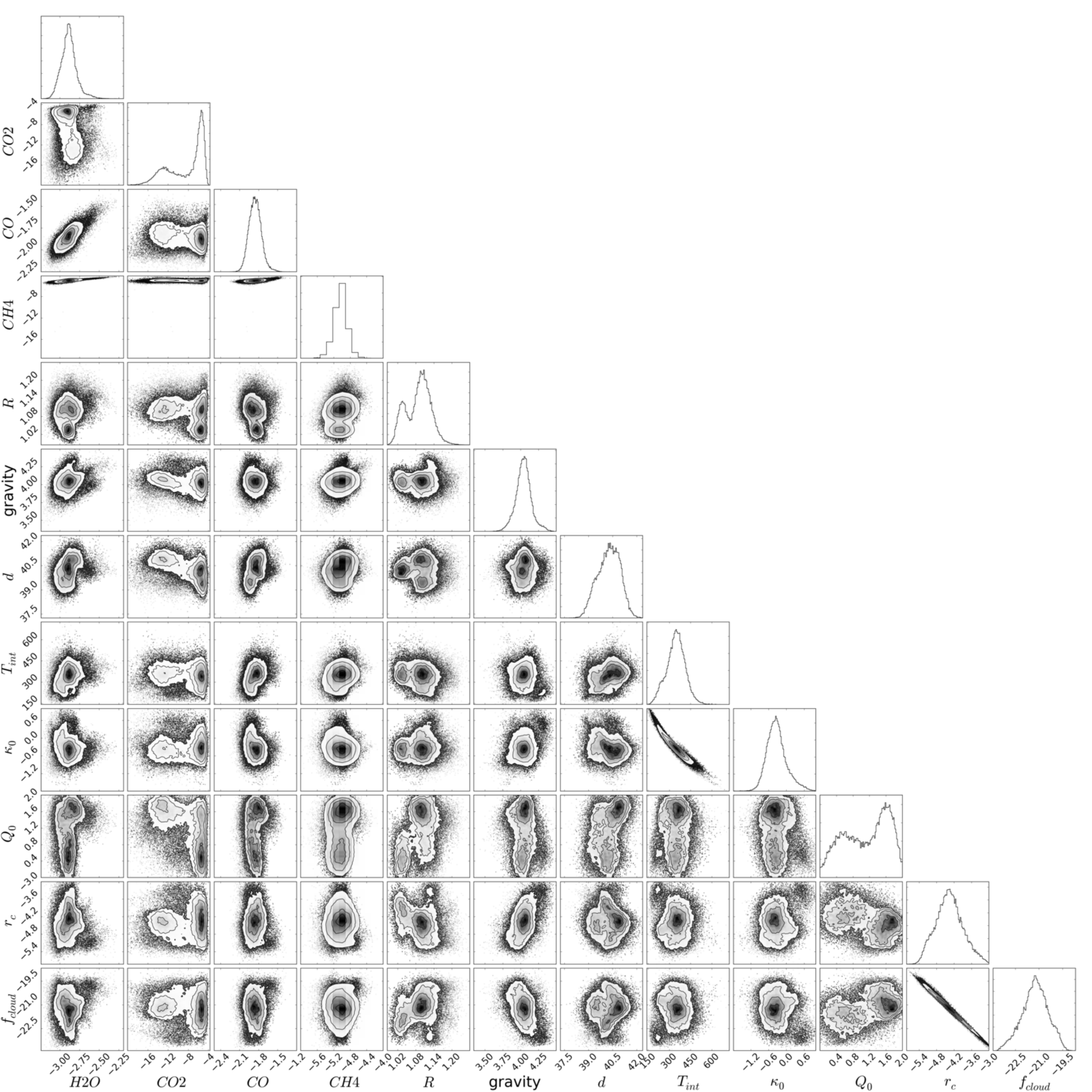}
\end{center}
%\vspace{-0.2in}
\caption{Montage of posterior distributions from the best-fit retrieval model of HR 8799b.}
%\vspace{0.1in}
\label{fig:retrieval1}
\end{figure*}

\begin{figure*}%[!h]
\begin{center}
%\vspace{0.2in}
\includegraphics[width=\columnwidth]{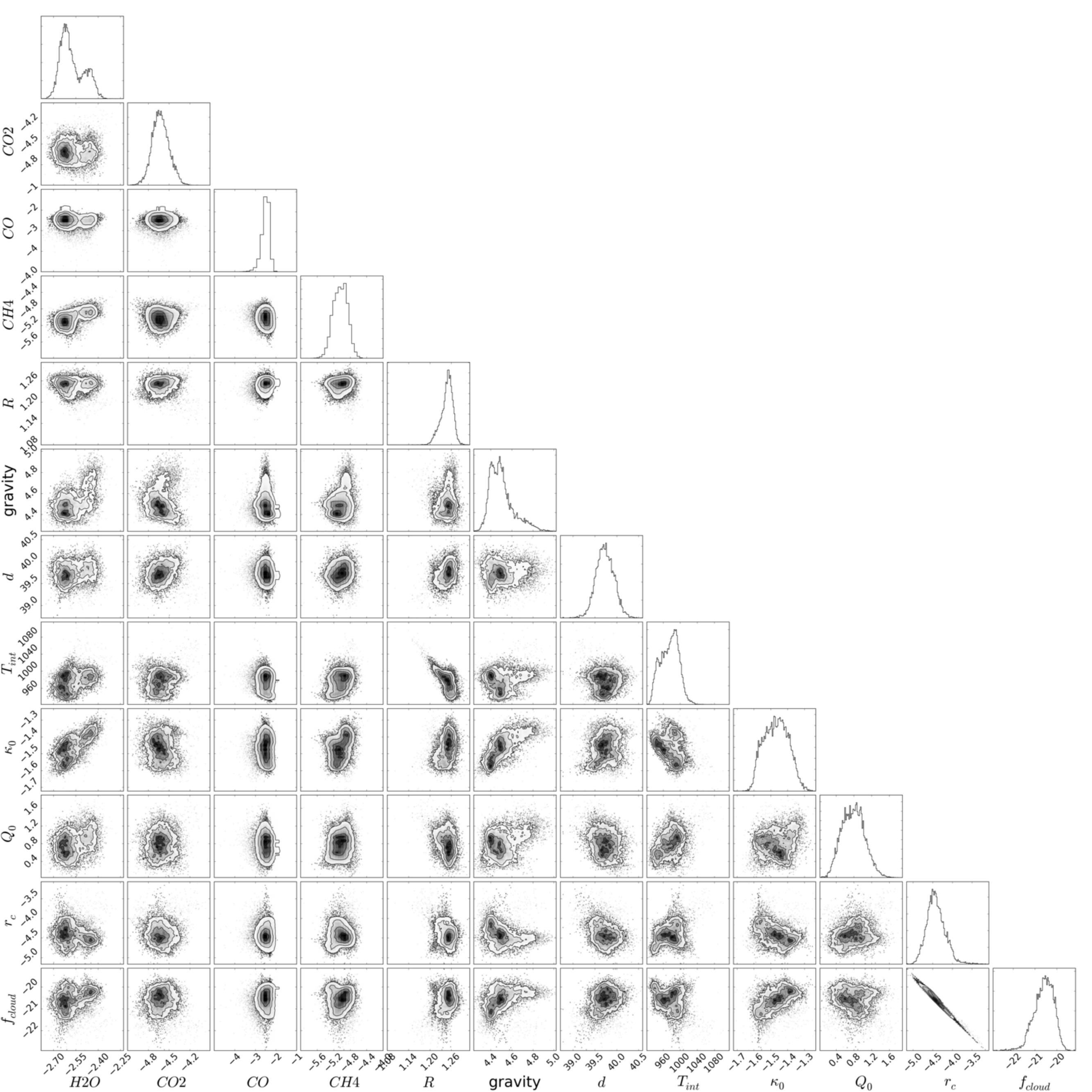}
\end{center}
%\vspace{-0.2in}
\caption{Same as Figure \ref{fig:retrieval1}, but for HR 8799c.}
%\vspace{0.1in}
\label{fig:retrieval2}
\end{figure*}

\begin{figure*}%[!h]
\begin{center}
%\vspace{0.2in}
\includegraphics[width=\columnwidth]{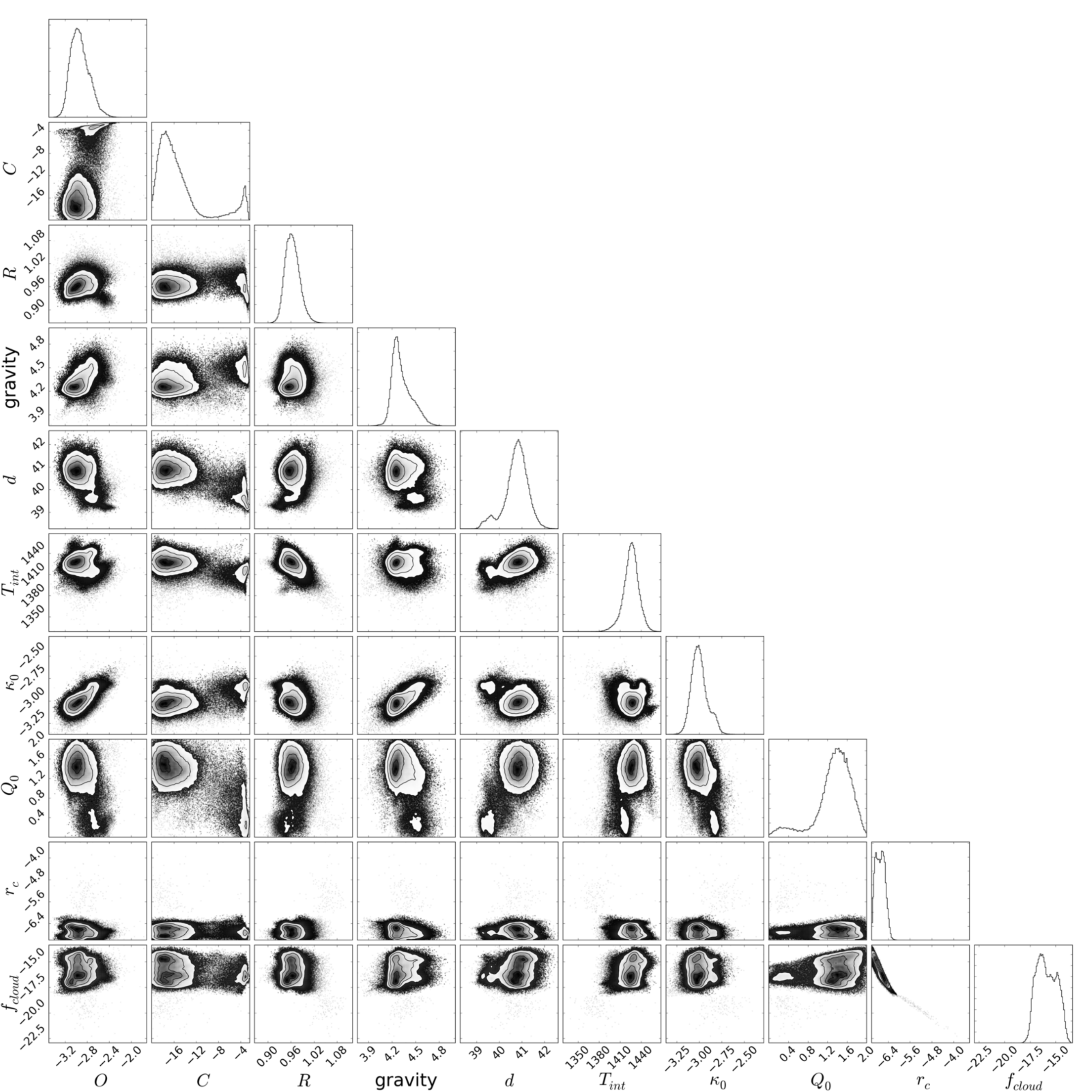}
\end{center}
%\vspace{-0.2in}
\caption{Same as Figure \ref{fig:retrieval1}, but for HR 8799d.}
%\vspace{0.1in}
\label{fig:retrieval3}
\end{figure*}

\begin{figure*}%[!h]
\begin{center}
%\vspace{0.2in}
\includegraphics[width=\columnwidth]{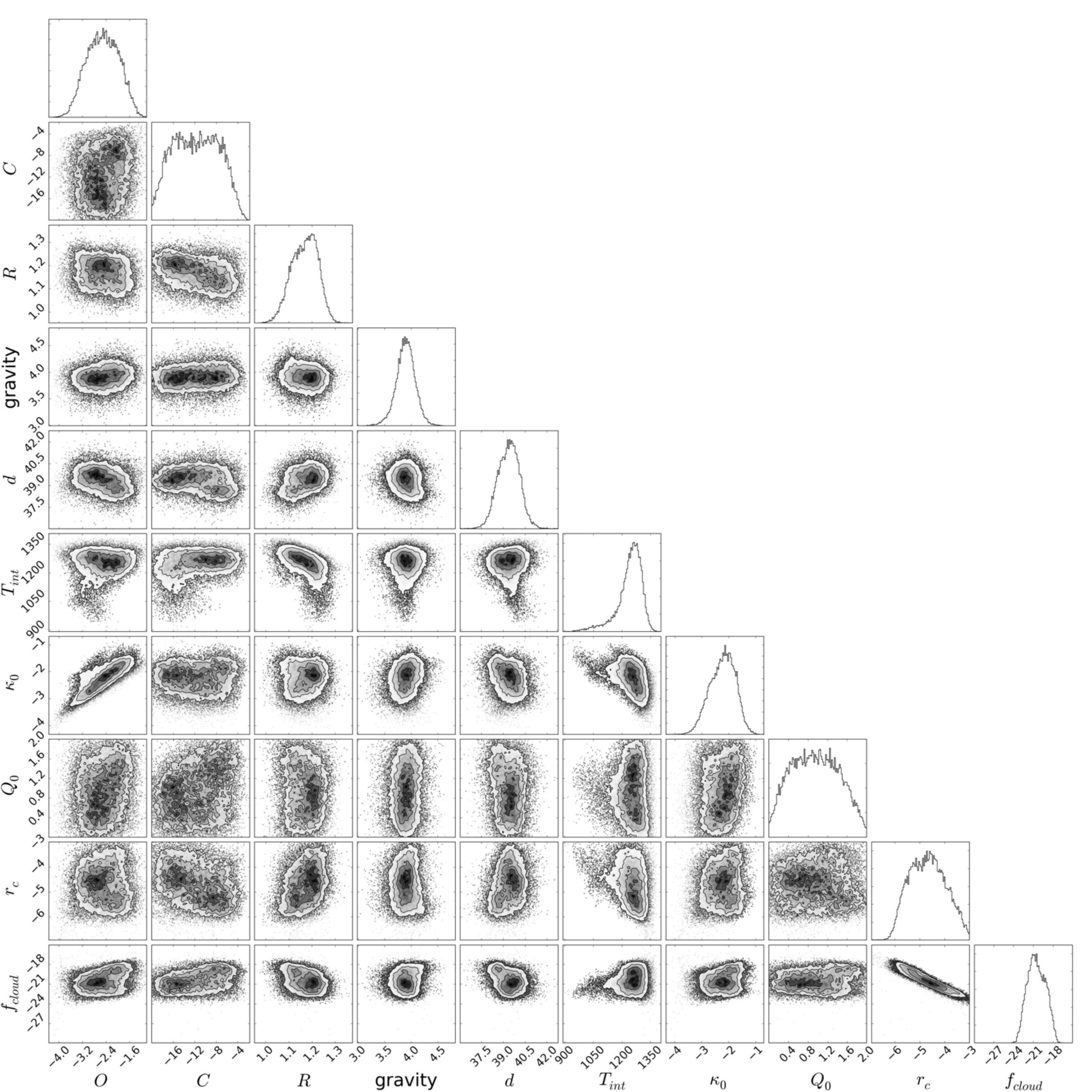}
\end{center}
%\vspace{-0.2in}
\caption{Same as Figure \ref{fig:retrieval1}, but for HR 8799e.}
%\vspace{0.1in}
\label{fig:retrieval4}
\end{figure*}

For completeness, we show in Figures \ref{fig:retrieval1}, \ref{fig:retrieval2}, \ref{fig:retrieval3} and \ref{fig:retrieval4} the full posterior distributions for our best models of HR 8799b, c, d and e, which elucidate the model degeneracies between each pair of parameters.

\begin{figure*}%[!h]
\begin{center}
%\vspace{0.2in}
\includegraphics[width=\columnwidth]{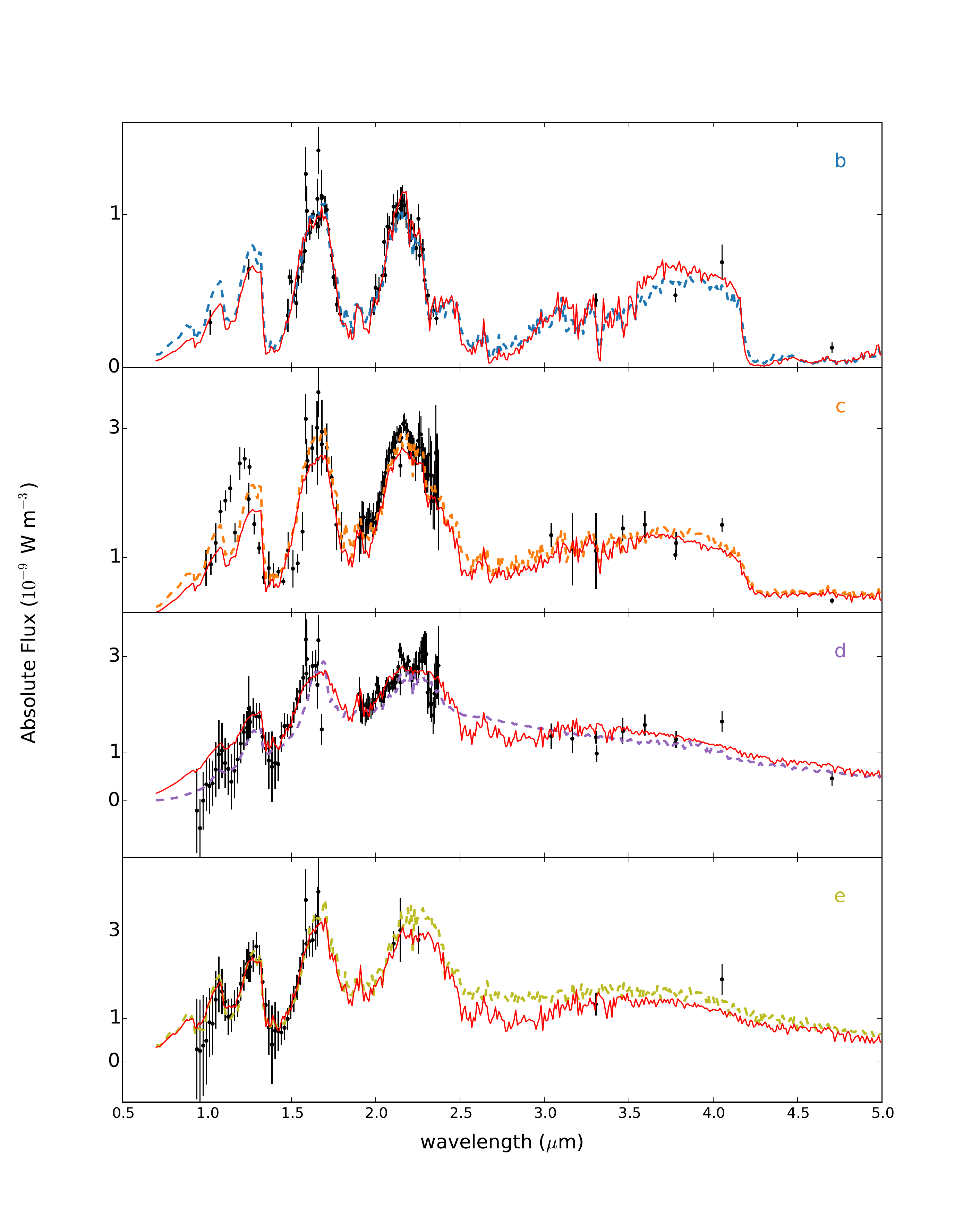}
\end{center}
\vspace{-0.2in}
\caption{Elucidating the effects of using different spectroscopic line lists.  The dashed curves in each panel show the retrievals using ExoMol data for water and methane.  The red, continuous curves use the retrieved parameters to produce model spectra but using HITEMP water and HITRAN methane (post-processing).}
%\vspace{0.1in}
\label{fig:spectra2}
\end{figure*}

%%% REFERENCES %%%

\label{lastpage}

\end{document}